\documentclass[useAMS,usenatbib]{mn2e}
\usepackage[final]{graphics}
\usepackage{epsfig}

\newcommand{\he}[1] {He\,{\sc #1}}
\newcommand{\hel}[2] {He\,{\sc #1}~$\lambda$#2}
\newcommand{\kms}{\mbox{$\mathrm{km~s^{-1}}$}}

\newcommand{\Line}[3]{\Ion{#1}{#2}~$\lambda$#3}

\newcommand{\Ion}[2]{#1{\,\scriptsize #2}}

\newcommand{\Ha}{\mbox{${\mathrm H\alpha}$}}
\newcommand{\Hb}{\mbox{${\mathrm H\beta}$}}
\newcommand{\Hg}{\mbox{${\mathrm H\gamma}$}}

\newcommand{\hl}{HL\,Aqr}
\newcommand{\bo}{BO\,Cet}
\newcommand{\vhya}{V393\,Hya}
\newcommand{\ah}{AH\,Men}
\newcommand{\voph}{V380\,Oph}
\newcommand{\lqpeg}{LQ\,Peg}
\newcommand{\ahp}{AH\,Pic}
\newcommand{\lnu}{LN\,UMa}

\newcommand{\vher}{V849\,Her}
\newcommand{\yes}{$\surd$}		
\newcommand{\no}{$\times$}		

\title[New SW Sextantis stars]{Spectroscopic search for new SW Sextantis stars in the 3--4 hour orbital period range -- I.}
\author[P. Rodr\'\i guez-Gil, L. Schmidtobreick, B.~T. G\"ansicke]{P. Rodr\'\i guez-Gil$^{1,2}$\thanks{E-mail:prguez@iac.es}, L. Schmidtobreick$^{2}$ and B. T. G\"ansicke$^{3}$\\
$^{1}$Instituto de Astrof\'\i sica de Canarias, V\'\i a L\'actea s/n, La Laguna, E-38205, Santa Cruz de Tenerife, Spain\\
$^{2}$European Southern Observatory, Casilla 19001, Santiago 19, Chile\\
$^{3}$Department of Physics, University of Warwick, Coventry CV4 7AL, UK}
\begin{document}
\date{Accepted 2006. Received 2006}
\pagerange{} \pubyear{2006}
\maketitle
\begin{abstract}
We report on time-resolved optical spectroscopy of ten non-eclipsing nova-like cataclysmic variables in the orbital period range between 3 and 4 hours. The main objective of this long-term programme is to search for the characteristic SW\,Sextantis behaviour and to eventually quantify the impact of the SW\,Sex phenomenon on nova-likes at the upper boundary of the orbital period gap.

Of the ten systems so far observed, HL Aqr, BO Cet, AH Men, V380 Oph, AH Pic, and LN UMa are identified as new members of the SW Sex class. We present improved orbital period measurements for HL Aqr ($P_\mathrm{orb} = 3.254 \pm 0.001$\,h) and V380 Oph ($P_\mathrm{orb} = 3.69857 \pm 0.00002$\,h). BO Cet and V380 Oph exhibit emission-line flaring with periodicities of 20 min and 47 min, respectively. The \Ha~line of \hl~shows significant blueshifted absorption modulated at the orbital period. Similarly to the emission S-wave of the high-inclination SW Sex stars, this \textit{absorption} S-wave has its maximum blue velocity at orbital phase $\sim 0.5$. We estimate an orbital inclination for \hl~in the range $19\degr < i < 27\degr$, which is much lower than that of the emission-dominated, non-eclipsing SW Sex stars ($i \sim 60\degr-70\degr$). This gives rise to the interesting possibility of many low-inclination nova-likes actually being SW\,Sex stars, but with a very different spectroscopic appearance as they show significant absorption. The increasing blueshifted absorption with decreasing inclination points to the existence of a mass outflow with significant vertical motion.

This six new additions to the SW Sex class increase the presence of non-eclipsing systems to about one third of the whole SW Sex population, which therefore makes the requirement of eclipses as a defining criterion for SW Sex membership no longer valid. The statistics of the cataclysmic variable population in the vicinity of the upper period gap is also discussed. 
 \end{abstract}


\section{Introduction}
\label{sec_intro}

The theory of cataclysmic variable (CV) evolution predicts an abrupt
cessation of mass transfer at an orbital period of $\sim 3$\,h,
i.e. the upper boundary of the period gap, where magnetic braking is
believed to be interrupted. However, it is becoming increasingly clear
that the upper edge of the gap is overpopulated by CVs which show unusual behaviour. These systems, collectively known as the SW\,Sextantis stars \citep{thorstensenetal91-1}, show a number of common properties, including uncommon ``V''-shaped eclipse profiles,
the presence of \Line{He}{II}{4686} emission, substantial delays of the
radial velocities of the Balmer and \he{i} lines with respect to the motion of
the white dwarf, single-peaked emission lines that display central
absorption dips around orbital phases $\simeq0.4-0.7$, and high-velocity emission S-waves with maximum blueshift near phase $\sim 0.5$. These observational characteristics do not easily fit into the picture of a steady-state, hot optically-thick accretion disc which is expected to be common in intrinsically bright, weakly-magnetic CVs above the period gap.

In terms of numbers, the SW\,Sex stars currently represent half the total population of nova-likes in the $3-4$\,h orbital period range (\citealt{rodriguez-gil05-1}, \citealt{aungwerojwitetal05-1}, Rodr\'\i guez-Gil et al. \citeyear{rodriguez-giletal07-1}). Thus, their study is of critical importance to investigate the evolution of CVs as they enter the gap.

Many different ideas have been proposed to model the unusual behaviour of the SW\,Sex stars, but none of them has been able to explain the SW\,Sex phenomenon in full. Among the many mechanisms so far invoked we highlight stream
overflow \citep{hellier+robinson94-1, hellier96-1}, magnetic accretion
\citep{rodriguez-giletal01-1, hameury+lasota02-1}, magnetic propeller
anchored in the inner disc \citep{horne99-1}, and self-obscuration of
the inner disc by a flared outer disc \citep{kniggeetal00-1}.

In order to address the actual impact of the SW\,Sex phenomenon on the
nova-like population found between 3 and 4 hours, we started to search
for characteristic SW\,Sex features in poorly-studied nova-likes
of both hemispheres in that period range. Since the majority of
these properties are actually spectroscopic we will entirely focus on
phase-resolved optical spectroscopy. This paper contains the first
results of this project.

\section{Observational data and their reduction}
\label{data}

Time-resolved spectroscopy of the candidate SW\,Sex stars was carried
out with the New Technology Telescope (NTT) at the European Southern
Observatory on La Silla, and the William Herschel Telescope (WHT) at
the Roque de los Muchachos Observatory on La Palma. At the NTT the
EMMI (ESO Multi-Mode Instrument) was used in the Medium Dispersion
Spectroscopy mode with grating D/7 centered on H$\alpha$ and the $2048
\times 4096$\,pixel$^2$ MIT/LL CCD detector. A useful wavelength
range of $\lambda\lambda6170-6900$ at 0.8\,\AA~spectral resolution (full width at half maximum, FWHM) was obtained with a 1.0-arcsec slit width.

The double-armed ISIS spectrograph was used to perform the observations at the WHT. The blue and red arms were equipped with the R600B and the R316R gratings,
respectively. The wavelength ranges $\lambda\lambda3700-5045$ and
$\lambda\lambda6230-8900$ were sampled at 1.8\,\AA~and
3.3\,\AA~resolution (FWHM; 1.0-arcsec slit) on the $2048 \times 4200$\,pixel$^2$ EEV12 and the $2148 \times 4700$\,pixel$^2$ Marconi CCDs in the blue and the red, respectively. Spectra of arc lamps were regularly taken in order to achieve a proper wavelength calibration in both telescopes. See Table~\ref{obstab} for
some details of the observations.

\begin{figure}
\mbox{\epsfig{file=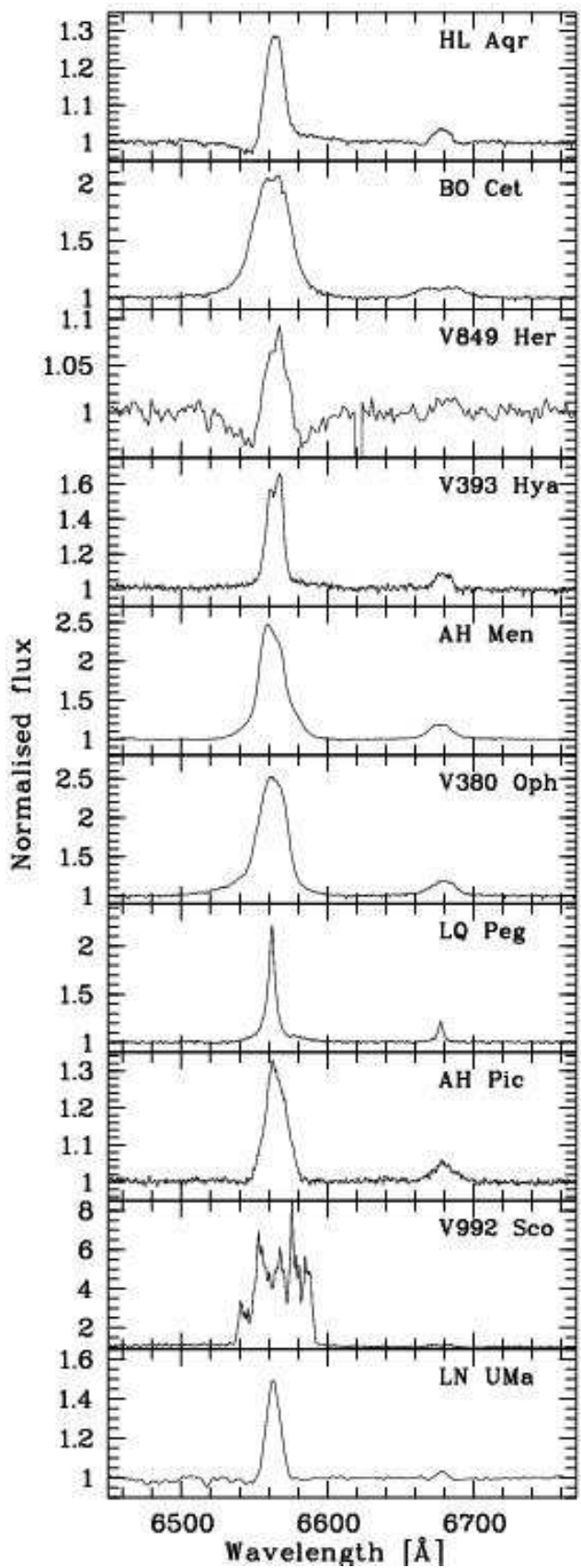,width=\columnwidth}}
\caption{\label{spec_all}
Average spectra around \Ha~of all the SW\,Sex candidates observed. The continuum has been normalised to unity.}
\end{figure}

\begin{table}
\centering
\caption{\label{obstab} Summary of the observational details. The object name,
date and UT at the start of the first exposure, the number of exposures, the
individual exposure time, and the covered orbital cycles are
given (with the exception of LQ Peg, for which $P_\mathrm{orb}$ is not known).}
 \begin{tabular}{@{}lccccc@{}}
  \hline
   Name     & Date & UT  & $\#_{\rm exp}$ & $t_{\rm exp}$ [s] & Cycles \\
 \hline
HL\,Aqr   &  2005-09-27 & 02:50:04 & 35     & 300 & 0.99 \\
          &  2005-09-28 & 00:30:19 & 37     & 300 & 1.05 \\
BO\,Cet   &  2005-09-27 & 06:25:17 & 36     & 300 & 0.99 \\
V849\,Her &  2005-05-22 & 23:42:50 & 25 blue& 300 & 0.84 \\
          &  2005-05-22 & 23:42:50 & 25 red & 300 & 0.84 \\
V393\,Hya &  2005-04-17 & 03:07:02 & 6      & 600 & 0.28 \\
          &  2005-04-18 & 03:15:44 & 15     & 600 & 0.80 \\
AH\,Men   &  2005-04-16 & 23:21:24 & 18     & 600 & 1.14 \\
V380\,Oph &  2005-04-17 & 06:05:33 & 16     & 900 & 1.08 \\
          &  2005-05-23 & 04:09:56 & 12 blue& 300 & 0.39 \\
	  &  2005-05-23 & 04:09:56 & 12 red & 300 & 0.39 \\
	  &  2005-05-24 & 01:25:02 & 67 blue& 180 & 1.03 \\
	  &  2005-05-24 & 01:25:02 & 67 red & 180 & 1.03 \\
LQ\,Peg   &  2005-09-26 & 23:32:40 & 17     & 600 & ---  \\
AH\,Pic   &  2005-04-17 & 23:26:53 & 24     & 480 & 1.03 \\
V992\,Sco &  2005-04-18 & 06:15:55 & 15     & 900 & 1.02 \\
LN\,UMa   &  2005-05-23 & 21:13:30 & 39 blue& 300 & 1.05 \\
          &  2005-05-23 & 21:13:30 & 39 red & 300 & 1.05 \\
\hline
\end{tabular}
\end{table}

The spectroscopic data were reduced using standard procedures in
{\sc iraf}\footnote{{\sc iraf} is distributued by the National
Optical Astronomy Observatories.}. After subtracting the bias level,
the images were divided by an average flat field which was normalised
by fitting Chebyshev functions to remove the detector
specific spectral response. The sky contribution was then subtracted
and the final spectra optimally extracted following the method
described by \citet{horne86-1}. For wavelength calibration, a
low-order polynomial was fitted to the pixel-wavelength arc data. The
wavelength solution for each target spectrum was obtained by
interpolating between the two nearest arc spectra. All subsequent
analysis was done with the {\sc molly}\footnote{Tom Marsh's
{\sc molly} package is available at
\texttt{http://deneb.astro.warwick.ac.uk/phsaap/software/}} package.

\section{Criteria for SW Sex status}
\label{swsex_criteria}

Before presenting the results on the individual objects it is worth clarifying which spectroscopic characteristics a nova-like must display in order to be classified as a SW Sex star. Note that not all the SW Sex stars show the whole set of features we will describe here at any one time.

The SW Sex class was initially populated by eclipsing systems only \citep{thorstensenetal91-2}. Contrary to what is expected for an accretion disc in Keplerian rotation viewed at high inclination, the emission lines observed in the SW Sex stars are single-peaked in both the eclipsing and non-eclipsing systems (compare e.g. the optical spectra of the dwarf nova HT Cas, \citealt{youngetal81-3}, and the SW Sex star V1315 Aql, \citealt{dhillonetal91-1}). Significant \hel{ii}{4686} and Bowen blend emission are also typical of the SW Sex stars. Apart from that, the Balmer line profiles are highly asymmetric with enhanced wings extending up to $\sim \pm 2000-3000$~\kms. The trailed spectra reveal that this is the consequence of a high-velocity emission S-wave which has maximum blue velocity at orbital phase $\sim 0.5$ (see e.g. \citealt{casaresetal96-1}, \citealt{hellier00-1}, and \citealt{rodriguez-giletal04-2}). In addition, an absorption component in the Balmer and \he{i} lines shows its maximum strength at the same orbital phase, transiently turning the single-peaked line profiles into ``fake'' double peaks (see \citealt{szkody+piche90-1}, \citealt{hellier96-1}, and \citealt{hoard+szkody97-1}). However, the \hel{ii}{4686} emission line is not affected by this absorption. In the eclipsing SW Sex stars, the Balmer and \he{i} radial velocity curves (and occasionally \hel{ii}{4686}) make their red to the blue crossing about $0.1-0.2$ orbital cycle \textit{later} than the phase zero defined by the eclipses \citep[e.g.][]{hoard+szkody97-1,rodriguez-giletal01-2}. The same phase delays have been observed in non-eclipsing SW Sex stars, but in this case between the \hel{ii}{4686} (believed to follow the white dwarf motion) and the Balmer (and \he{i}) radial velocity curves. As mentioned earlier, the \hel{ii}{4686} radial velocity curve occasionally shares the Balmer and \he{i} phasing. When this happens, absolute phases are usually estimated by assuming that (i) the maximum blueshift of the high-velocity S-wave occurs at orbital phase $\sim 0.5$, or (ii) the strength of the transient absorption component is maximum also at that orbital phase (e.g. \citealt{thorstensen+taylor00-1}, \citealt{thorstensen+taylor01-1}). A SW Sex nature is therefore very likely if the emission-line radial velocity curves of a given nova-like show a $\sim 0.1-0.2$ orbital cycle delay with respect to that phase convention, as it is then clear that it behaves in the way eclipsing SW Sex stars do.

As more spectroscopic studies with better time resolution (few minutes) are becoming available, it is apparent that the emission lines of some SW Sex stars show rapid variability with time-scales of the order of tens of minutes (\citealt{smithetal98-1,rodriguez-giletal01-1,rodriguez-gil+martinez-pais02-1}; also see the results on BO Cet and V380 Oph in this paper). This, however, cannot yet be considered as a common feature of the class due to the scarcity of studies with adequate time resolution.

Even though the behaviour of the emission lines in the SW Sex stars is a challenge for Doppler tomography techniques (eclipses, transient absorption, variability; see e.g. \citealt{steeghs03-1}), the Doppler maps consistently share a common feature: the bulk of Balmer emission is concentrated in the $(-V_x,-V_y)$ quadrant \citep[e.g. ][]{hellier+robinson94-1,hoard+szkody97-1,hoardetal00-1,rodriguez-giletal01-2}.

In summary, our SW Sex classification strategy can be outlined as follows:

\newcounter{clapton}
\begin{list}{\arabic{clapton})}{\usecounter{clapton}}
\item First step will be to check whether the emission line profiles are single- or double-peaked in the average spectrum.
~~\\
\item Check for the characteristic high-velocity, emission S-wave. Note that none of the systems presented in this paper shows eclipses. If no other component (as e.g. line emission from the irradiated donor star) can be used to determine absolute phases, we will assume that the maximum blue excursion of the S-wave takes place at phase $\varphi \sim 0.5$ \citep[see e.g.][]{hellier96-1,hoard+szkody97-1,hellier00-1,rodriguez-giletal01-2}, as it happens in the eclipsing SW Sex stars. To avoid confusion, we will use $\varphi_\mathrm{r}$ for relative phases and $\varphi_\mathrm{a}$ for (nearly) absolute phases. As a final note on phasing, the trailed spectrum diagrams and the folded radial velocity curves presented in the paper contain absolute phases unless otherwise stated.   
~~\\
\item Search for the transient absorption, which has to reach its maximum strength at phase $\sim 0.5$ according to the above phase definition.
~~\\
\item Check if the radial velocity curve of the line wings is delayed by $0.1-0.2$ cycle. Unfortunately, most of our spectra only cover the \Ha~and \hel{i}{6678} emission lines, thus phase lags with respect to the \hel{ii}{4686} line cannot be investigated. This line is observed in three systems, for which blue WHT spectra were obtained (V849 Her, V380 Oph, and LN UMa), but it proved too weak to perform a radial velocity study in any of the systems.
~~\\
\item Find the characteristic emission distribution in the Doppler tomogram.
\end{list}

\section{Average spectra}
\label{sec-results}

\begin{table}
\centering
\caption{\label{linetab} \Ha~and \hel{i}{6678} FWHM and equivalent width (EW) for each system. The emission line parameters were measured in the average spectrum. No Gaussian fit was possible for \Ha~in V992\,Sco.}
\begin{tabular}{@{}lcccc@{}}
\hline\noalign{\smallskip}
            &  \Ha   & \hel{i}{6678} & \Ha & \hel{i}{6678}\\
   Name     &  FWHM  &  FWHM   &  EW  &  EW  \\
            & [\kms] & [\kms] & [\AA] & [\AA]\\
\hline\noalign{\smallskip}
HL Aqr    & 620  & 530  & 4.8(2)  &  0.4(2)\\
BO Cet    & 1310 & 1620 & 35.0(2) &  3.5(2)\\
V393 Hya  & 580  & 520  & 9.6(3)  &  1.3(3)\\
AH Men    & 940  & 800  & 33.9(2) &  4.3(2)\\
V380 Oph  & 1020 & 900  & 41.1(1) &  4.4(1)\\
LQ Peg    & 310  & 210  & 10.3(1) &  1.2(1)\\
AH Pic    & 760  & 780  & 5.5(1)  &  0.9(1)\\
V992 Sco  & ---	 & --- 	& 166.8(4)&  2.6(3)\\
LN UMa    & 510  & 330  & 5.91(3) &  0.3(2)\\
\hline\noalign{\smallskip}
\end{tabular}
\end{table}

In Fig.~\ref{spec_all}, the average spectra around \Ha~for
all observed SW\,Sex candidates are plotted. All of them are dominated by
\Ha~in emission, with the exception of \vher~which shows a broad
absorption underneath the \Ha~emission. Some blueshifted absorption is also observed in \hl. Table~\ref{linetab} gives an overview on the width and strength of the \Ha~and
\hel{i}{6678} emission lines. The average \Ha~lines do not show the double-peaked profiles characteristic of line emission in a Keplerian accretion disc. Only V393 Hya has a clear double-peaked profile. In what follows we will discuss the
observed objects in detail.


\section{Results on the individual objects}

\subsection{HL\,Aqr}

\hl~\cite[=~PHL 227;][]{haro+luyten62-1} was identified as a CV by
Hunger, Heber \& Koester (\citeyear{hungeretal85-1}), who classified it as a UX UMa nova-like according to its ultraviolet and optical spectra. \cite{haefner+schoembs87-1} showed that, despite the optical spectrum of \hl~exhibited broad \Hb~and \Hg~absorptions with weak central emissions, the \Ha~line was purely in emission. Their radial velocity analyses on this line
yielded an orbital period of $0.1356 \pm 0.0006$\,d. In addition, they detected a coherent, low-amplitude oscillation in the $B$-band light curves at
19.6 s, which was later interpreted as a dwarf nova oscillation (DNO)
by \cite{warner04-1}.

\begin{figure}
\mbox{\epsfig{file=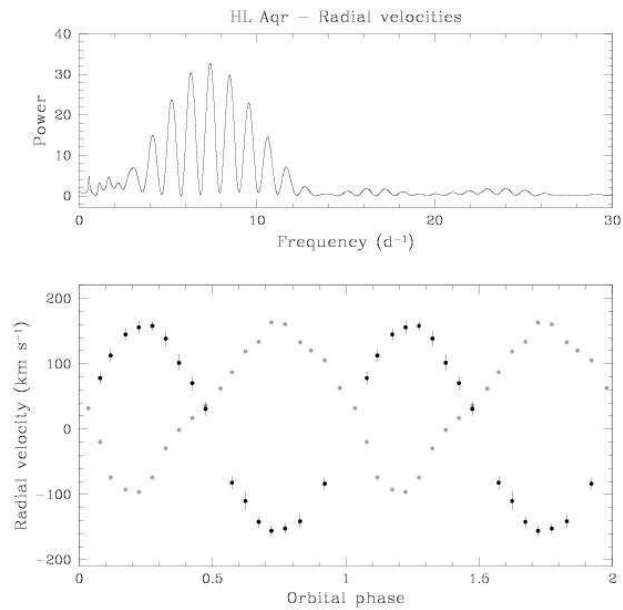,width=\columnwidth}}
\caption{\label{fig_hlaqr_rvc} \textit{Top}: Scargle periodogram computed from the \Ha~radial velocity curve of \hl~obtained by cross-correlation with a double Gaussian template (700\,\kms~separation, $\mathrm{FWHM}=300$\,\kms). \textit{Bottom}: the \Ha~wings (gray) and donor emission velocities (black) folded on $P_\mathrm{orb}=0.13557$\,d and averaged into 20 phase bins. Four clearly deviating points have been excluded from the radial velocity curve of the donor star emission. The orbital cycle has been plotted twice.}
\end{figure}

\subsubsection{\Ha~radial velocities and trailed spectrum}
\label{sec_hlaqr_rvc}

In our NTT spectra (Fig.~\ref{spec_all}) the \Ha~line is almost
entirely in emission but some absorption can be seen in
the blue wing. The profile is single-peaked and has
$\mathrm{FWHM} \simeq 620$ \kms~(Table~\ref{linetab}). We
computed the \Ha~radial velocities by applying the double Gaussian
technique of \cite{schneider+young80-2} using a 700\,\kms~separation
and $\mathrm{FWHM}=300$ \kms. The Scargle periodogram
\citep{scargle82-1} computed from the radial velocity data set
(Fig.~\ref{fig_hlaqr_rvc}) provided an improved value of the
orbital period of $P=0.13557 \pm 0.00005$ d ($=3.254 \pm 0.001$ h),
where the quoted uncertainty comes from a sine fit to the radial
velocity curves (Table~\ref{t-rvcfits}). This value is in perfect agreement with the orbital period measured by \cite{haefner+schoembs87-1}. The sine fit also provided a preliminary time of zero phase (i.e. velocity red-to-blue crossing) of $T_0(\mathrm{HJD}) = 2453640.75494 \pm 0.0003$.

\begin{table}
 \centering
  \caption{\label{t-rvcfits}\Ha~radial velocities sine fits parameters. $T_0$ refers to the time of inferior conjunction of the secondary star.}
  \begin{tabular}{@{}lcccc@{}}
  \hline
   Name     & $P_\mathrm{orb}$ & $\gamma$ & $K$  &     $T_0~(\mathrm{HJD})$      \\
            &        [h]       &  [\kms]  &[\kms]& [$2\,453\,400~+$]\\ 
 \hline
 HL\,Aqr      & 3.254(1) & 20(1)$^1$  & 113(1) & 240.7547(3)$^1$  \\
 BO\,Cet      & 3.355(1) & $-$35(2)$^1$ & 349(2) & 240.7633(1)$^1$  \\
 V849\,Her    & 3.15(48) & -11(12)$^1$ & 100(15) & 313.536(5)$^1$  \\
 V393\,Hya  & 3.23     & $41$(2)  & 46(3)  & 78.679(1) \\
 AH\,Men    & 2.95     & $-$11(1) & 209(1) & 77.5421(2)$^2$\\
V380\,Oph  & 3.69857(2) & 11.7(4) & 206.6(5) & 114.61061(8)$^2$\\
LQ\,Peg     & 2.80(2)  & $-$55.4(2)& 12.4(2) & 240.5375(3)$^3$\\
AH\,Pic     & 3.38(7)  & 32(2)$^1$& 94(5) & 78.547(1)$^1$\\
LN\,UMa     & 3.465(2)  & $-$68(2)& 62(2) & 114.4461(9)$^2$\\
 \hline
\end{tabular}\\
\begin{minipage}{10cm}
$^1)$ From donor star emission.\\
$^2)$ Corrected for phase delay.\\
$^3)$ Uncertain.\\
\end{minipage}
\end{table}

\begin{figure*}
\mbox{\epsfig{file=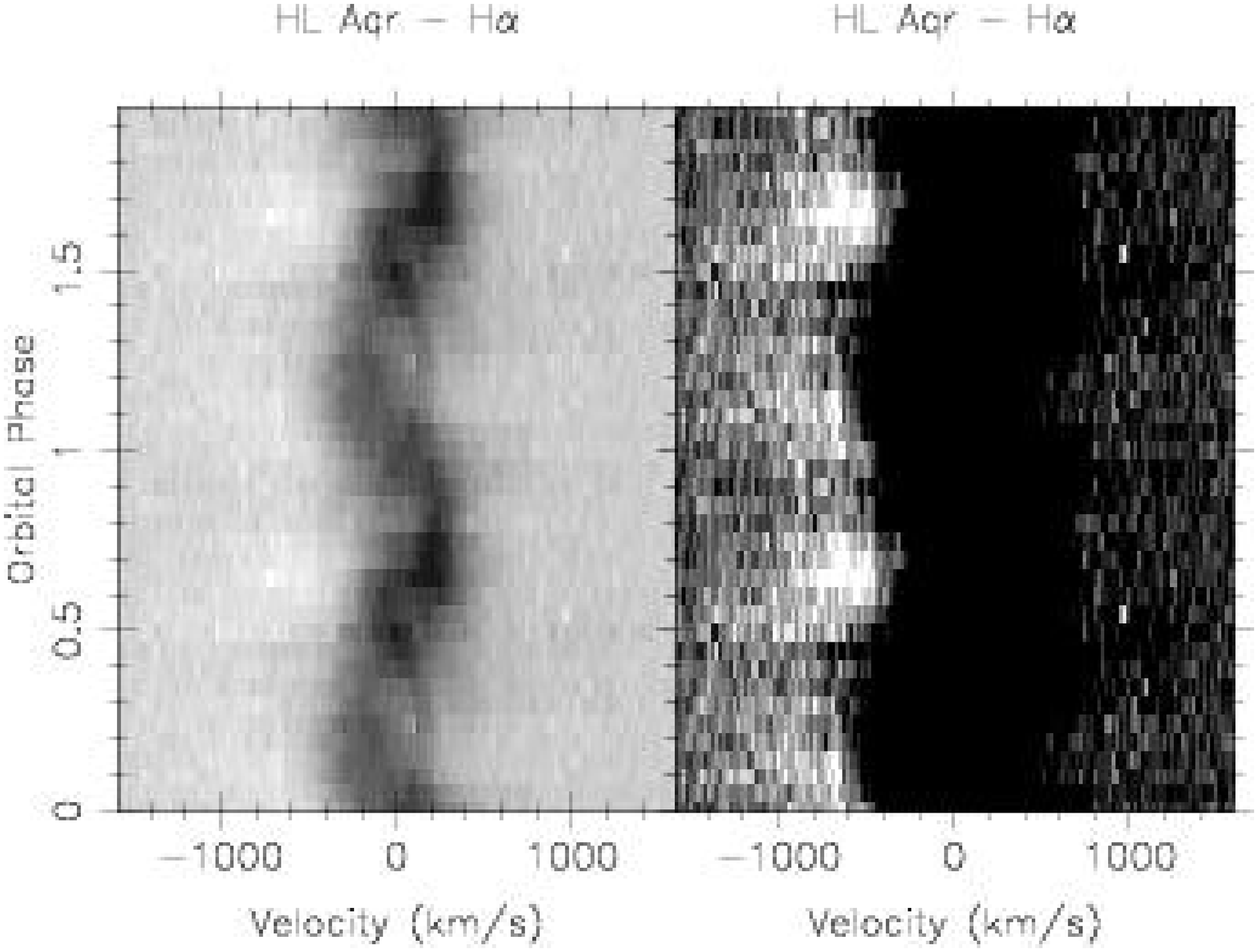,angle=-90,width=8cm}~~~~\epsfig{file=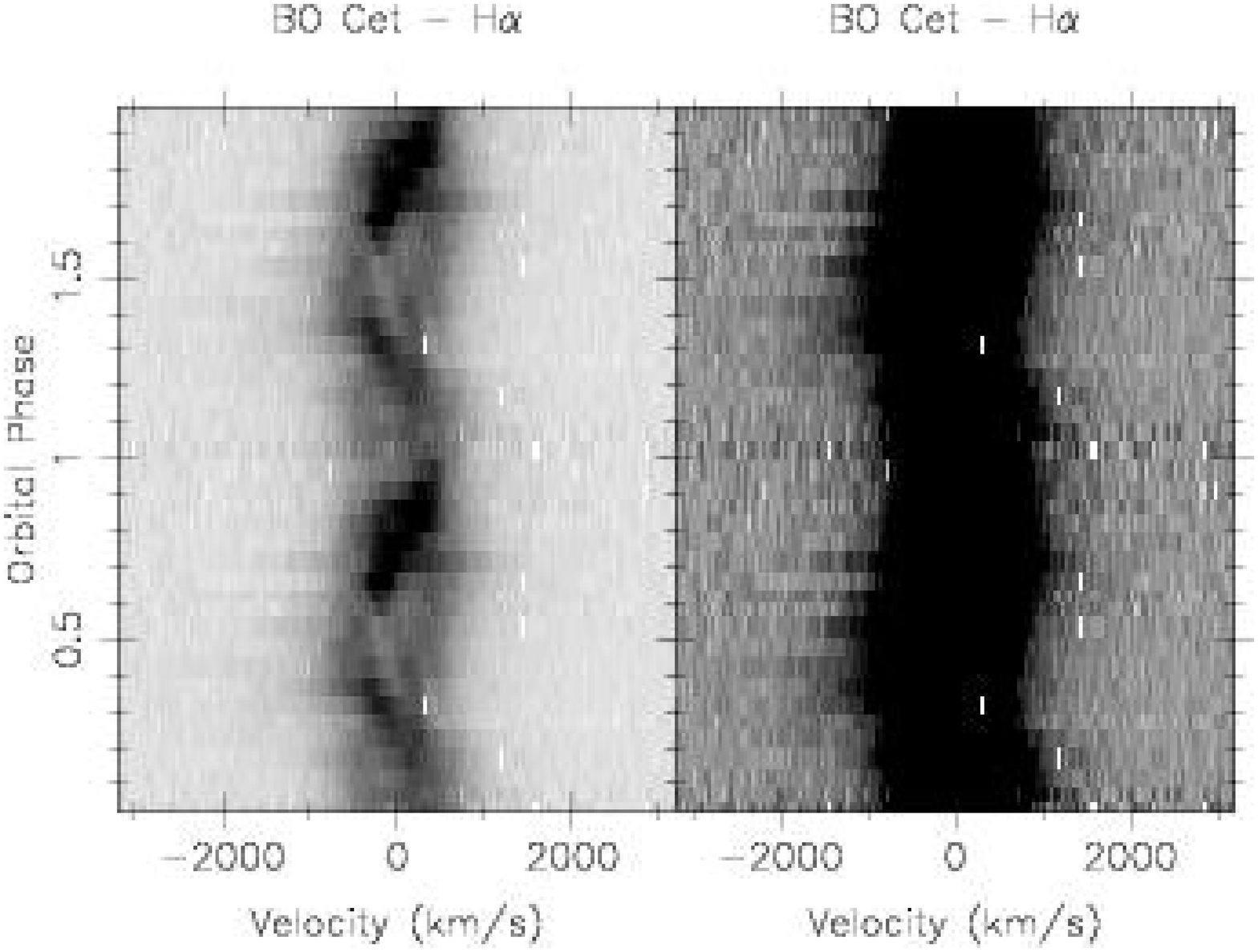,angle=-90,width=8cm}}
~~\\
~~\\
\mbox{\epsfig{file=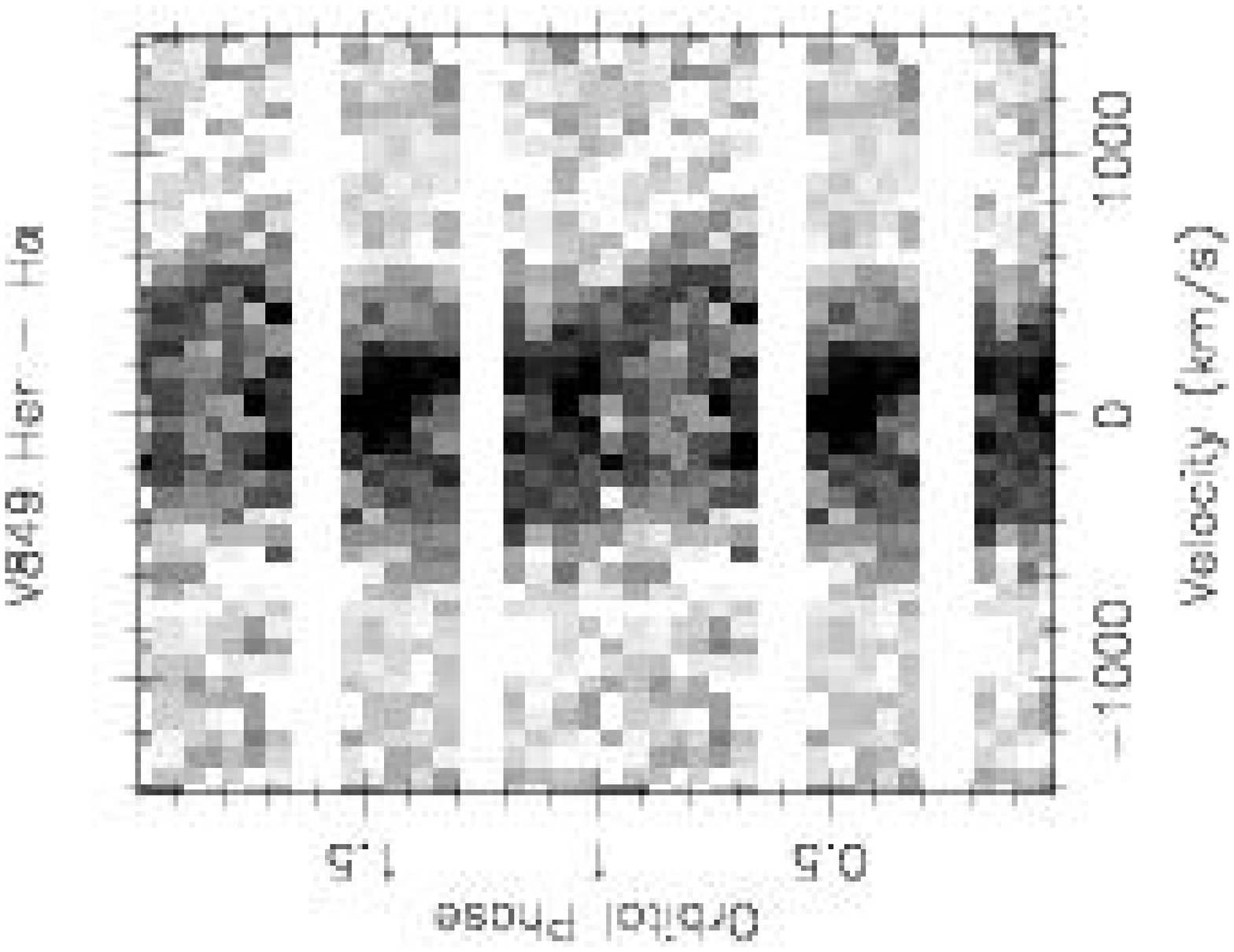,angle=-90,width=4.65cm}~~~~~~~~~~~~~~~~~~~~~~~~~~~~~~~~~~~~~~~\epsfig{file=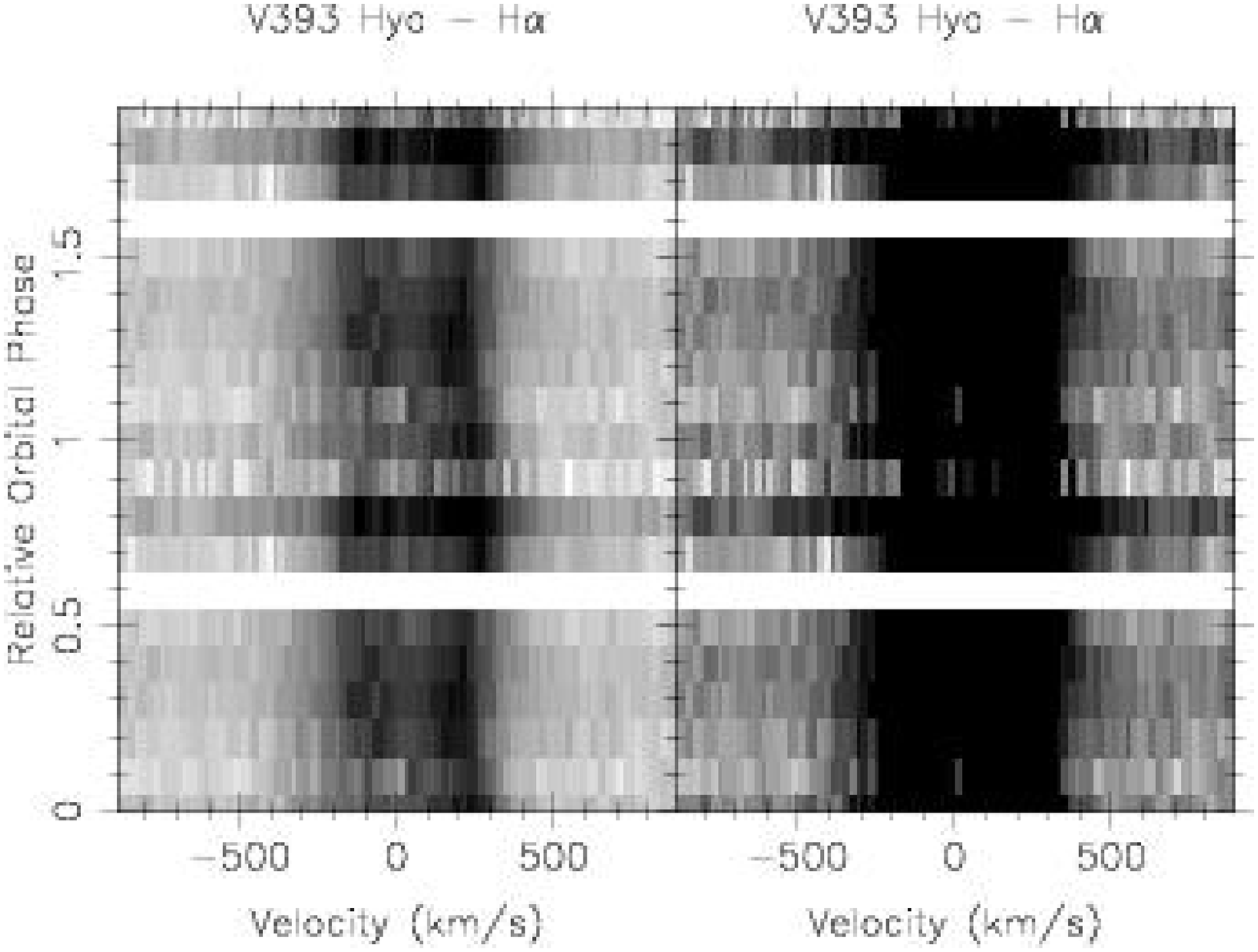,angle=-90,width=8cm}}
~~\\
~~\\
\mbox{\epsfig{file=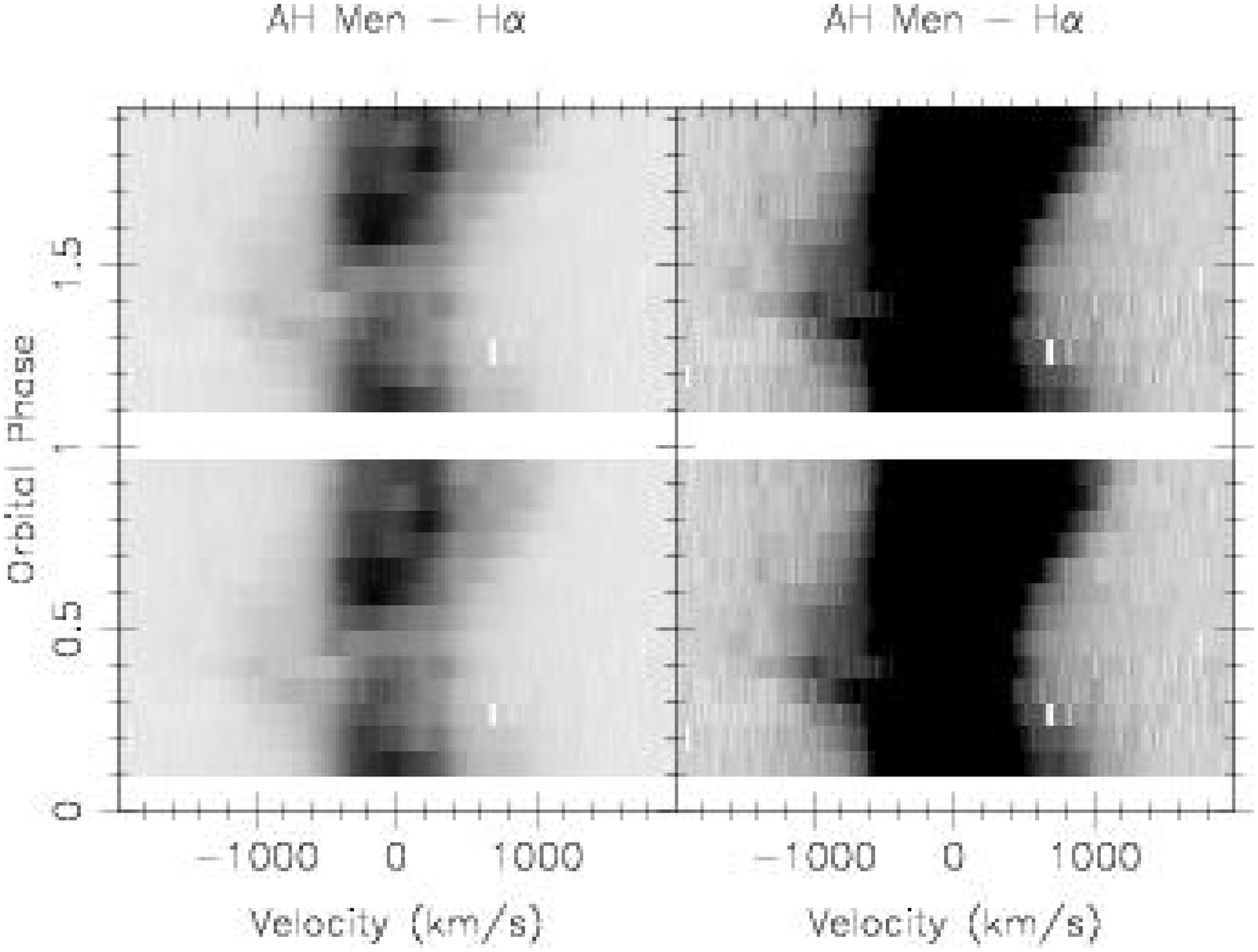,angle=-90,width=8cm}~~~~\epsfig{file=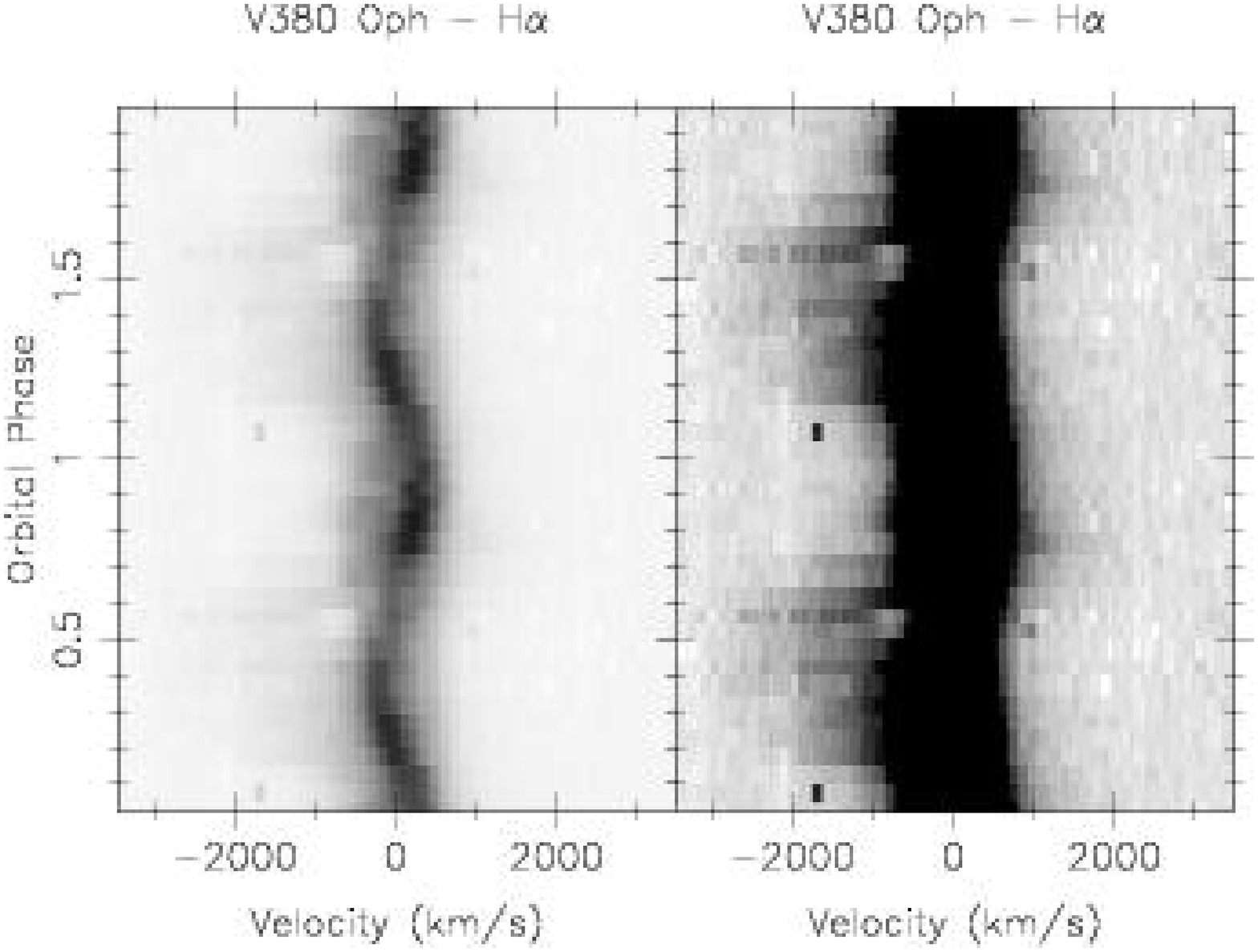,angle=-90,width=8cm}}
\caption{\label{fig_trailed} \Ha~trailed spectra. For each object (with the exception of V849 Her and V992 Sco) two panels with different constrast are shown. The left panel focuses on the line core whereas the right one emphasizes the wings.}
\end{figure*}

\begin{figure*}
\setcounter{figure}{2}
\mbox{\epsfig{file=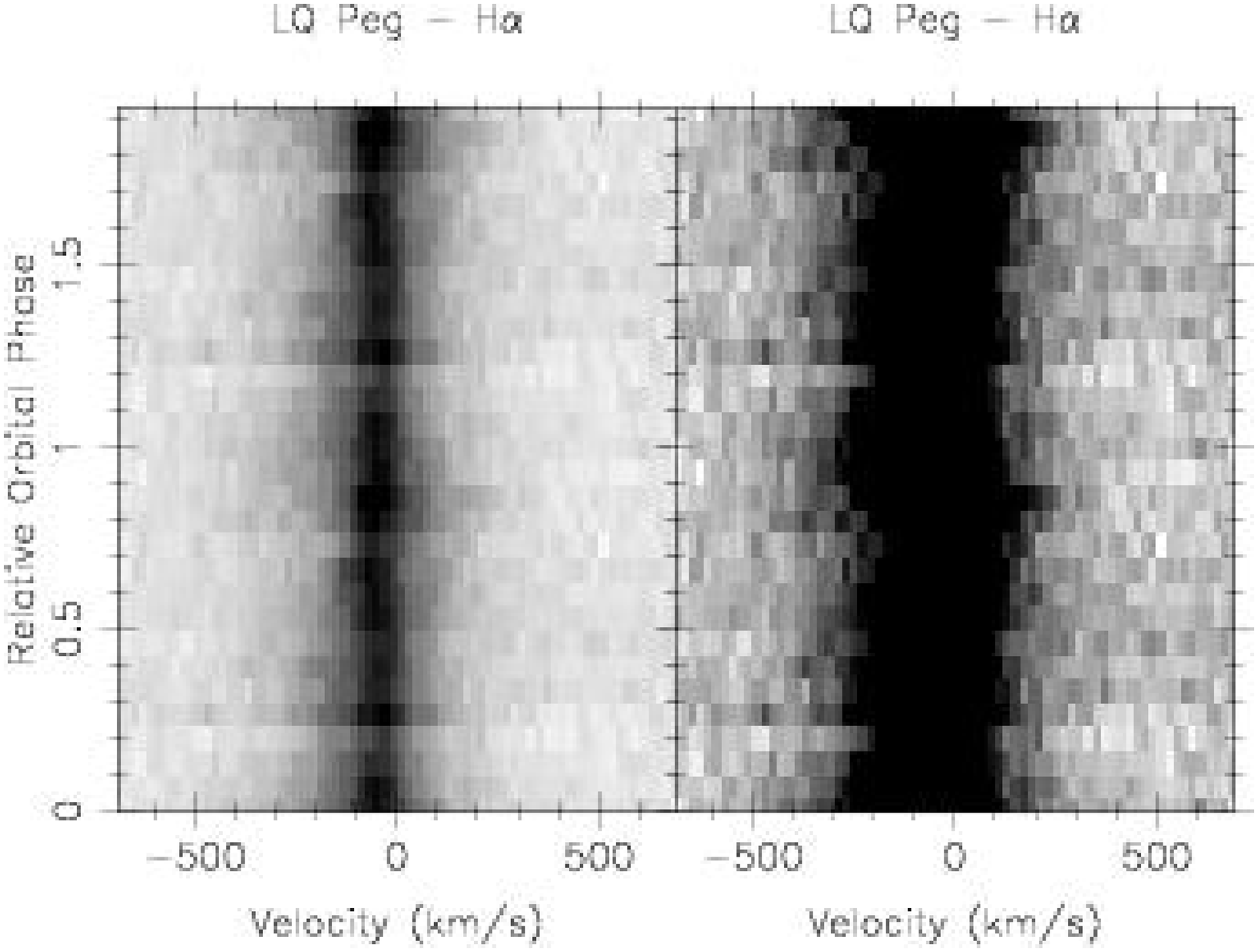,angle=-90,width=8cm}~~~~\epsfig{file=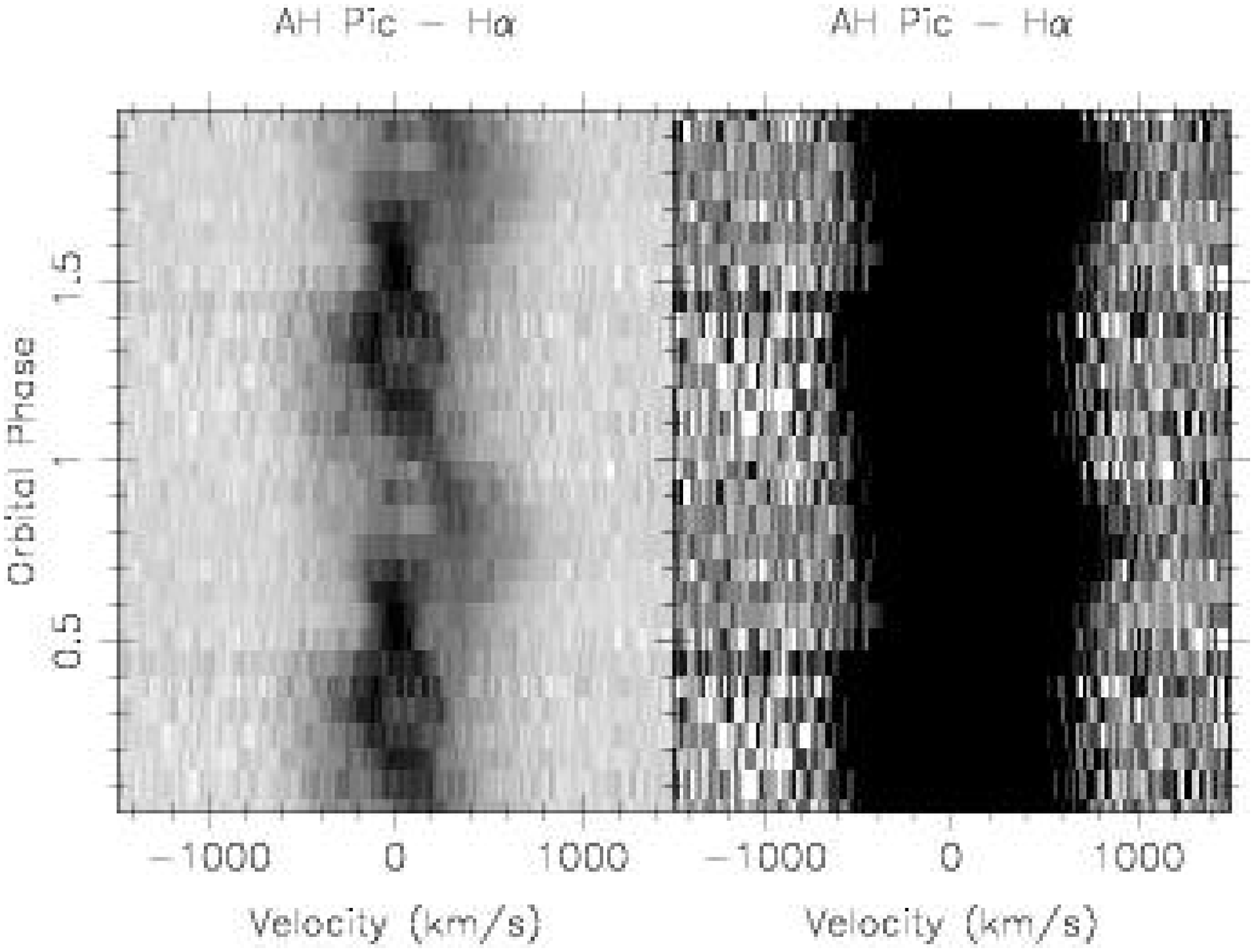,angle=-90,width=8cm}}
~~\\
~~\\
\mbox{\epsfig{file=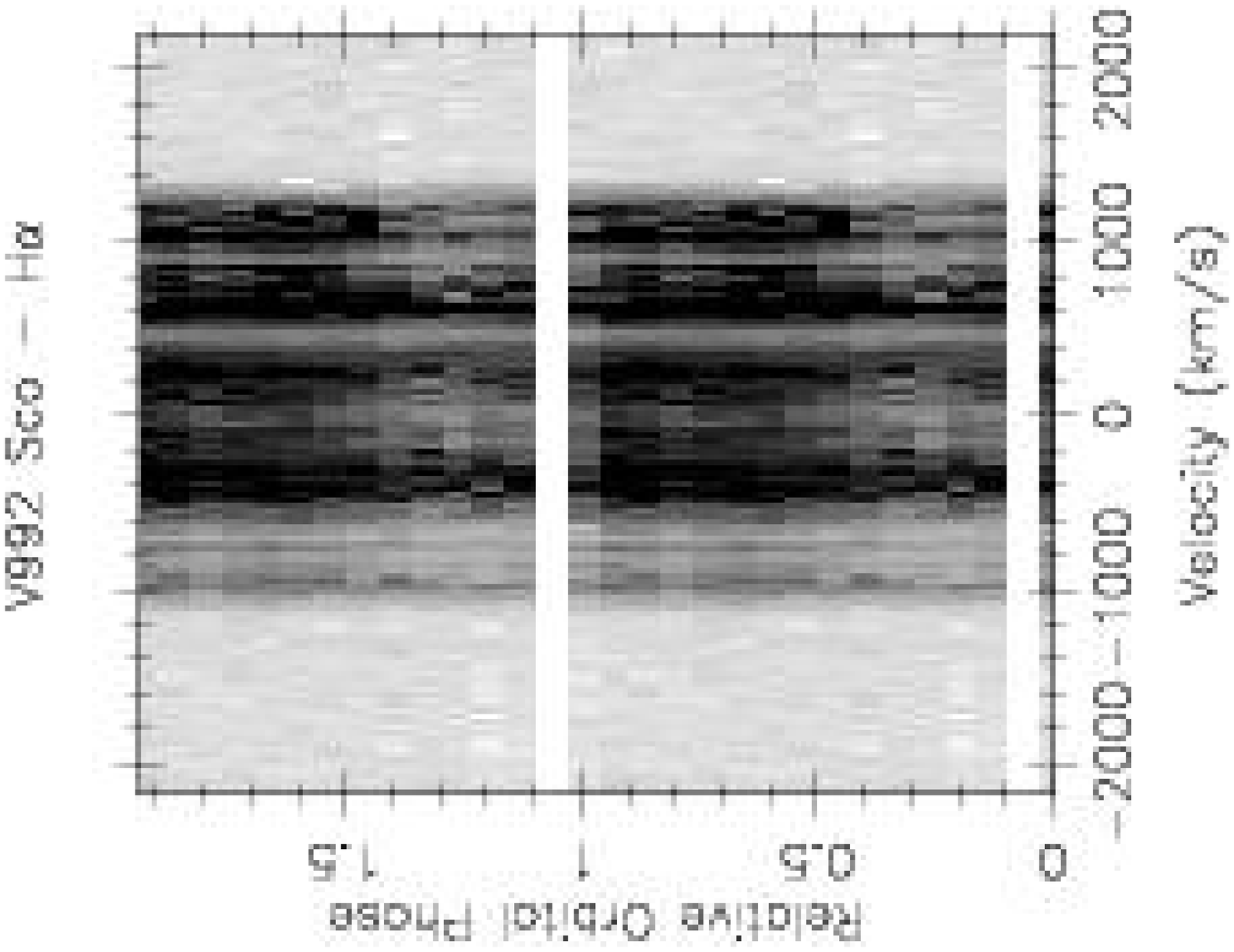,angle=-90,width=4.65cm}~~~~~~~~~~~~~~~~~~~~~~~~~~~~~~~~~~~~~~~~\epsfig{file=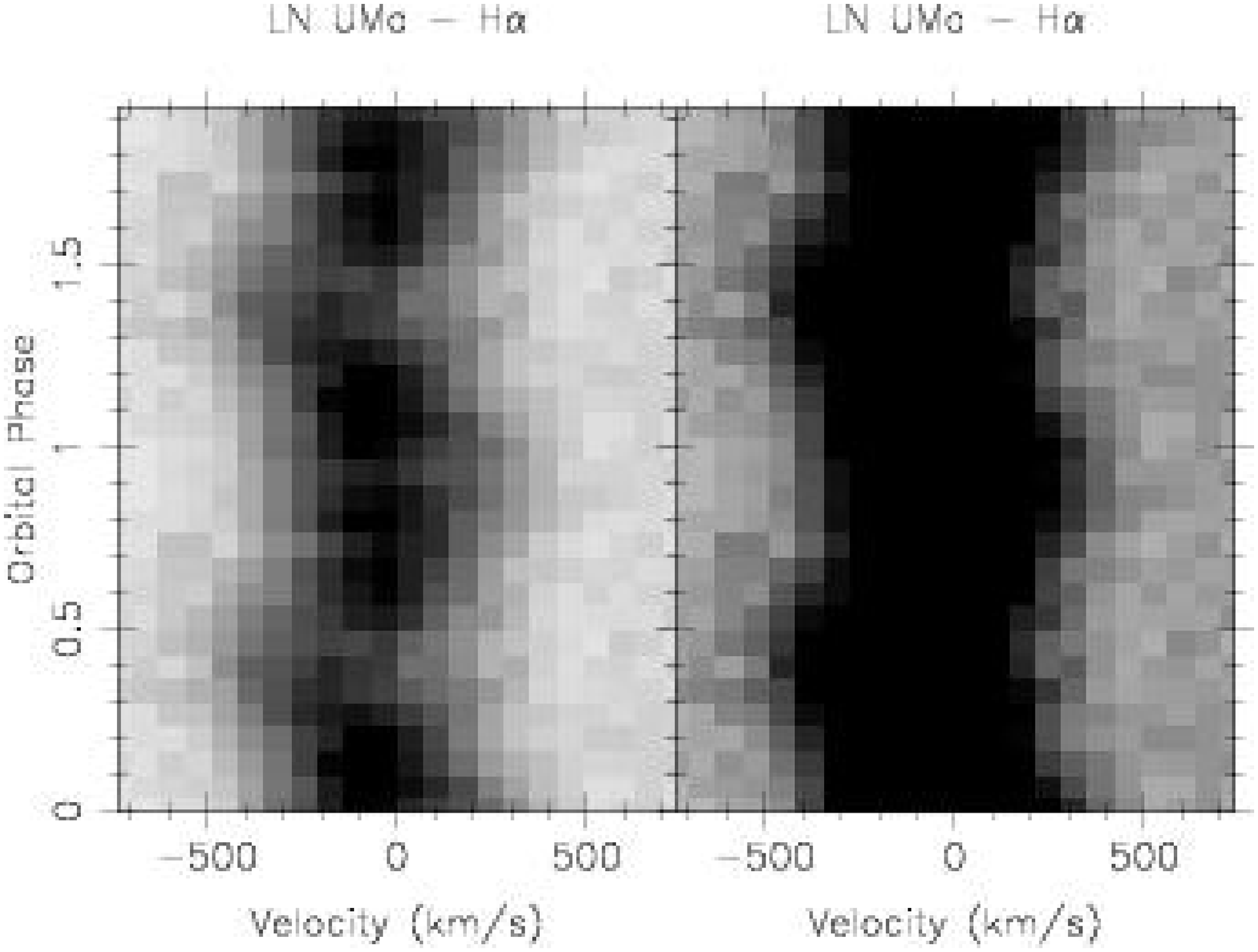,angle=-90,width=8cm}}
\caption{\label{fig_trailed} (\textit{cont.})}
\end{figure*}

\begin{figure*}
\mbox{\epsfig{file=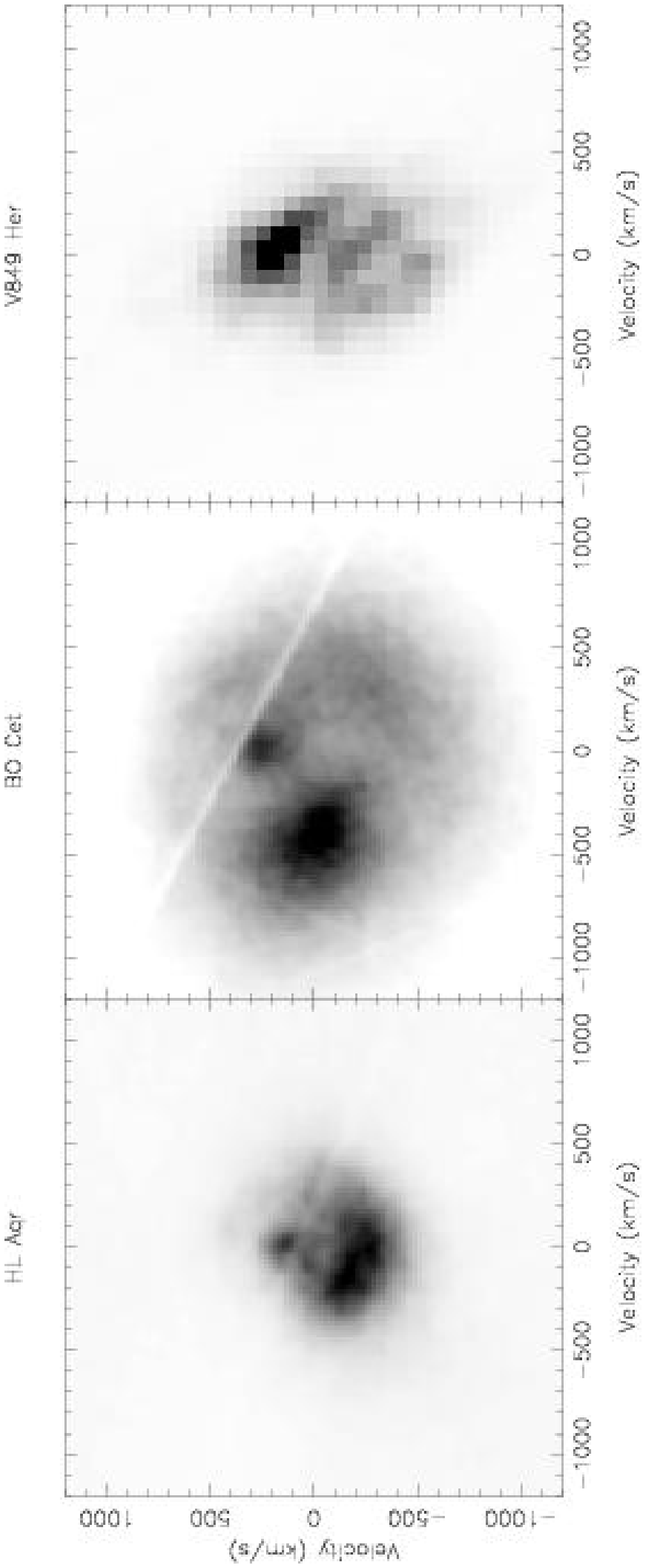,width=6.5cm,angle=-90}}
~~\\
~~\\
\mbox{\epsfig{file=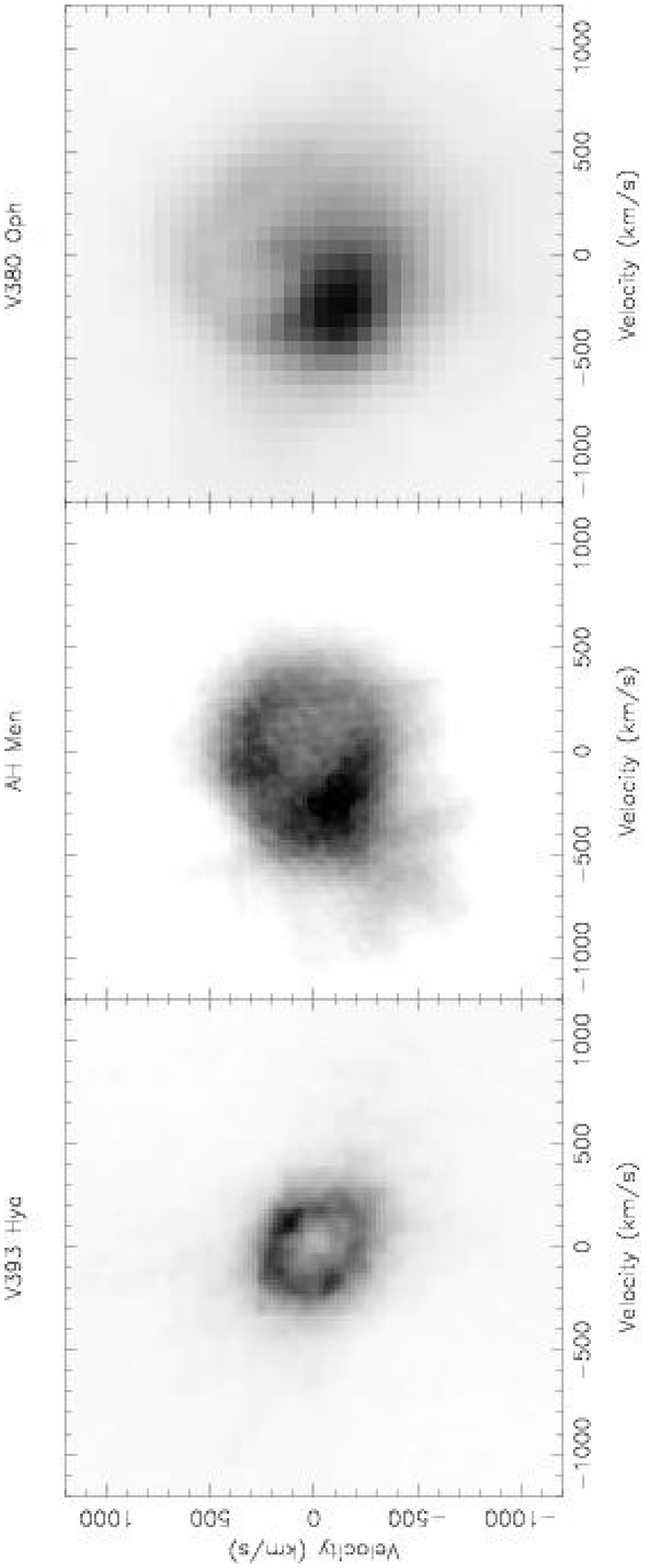,width=6.5cm,angle=-90}}
~~\\
~~\\
\mbox{\epsfig{file=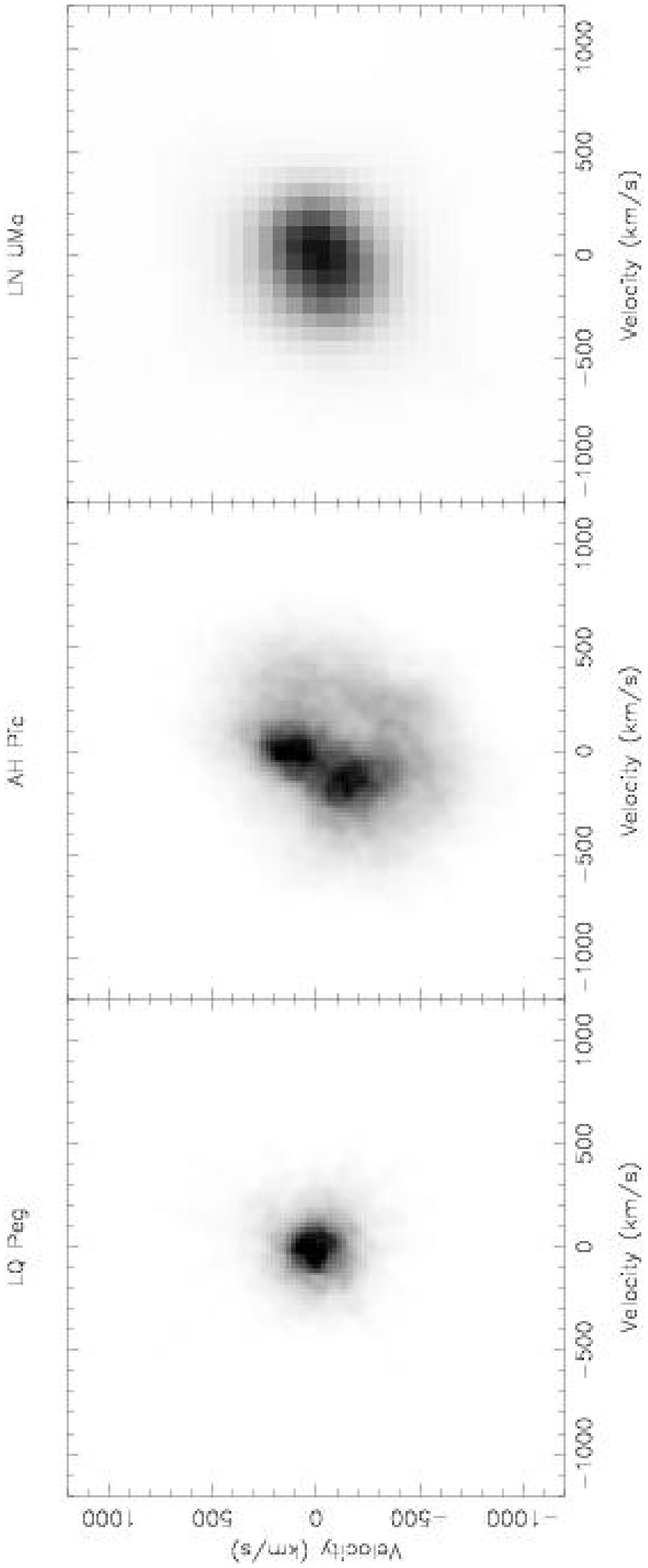,width=6.5cm,angle=-90}}
\caption{\label{fig_doppler} \Ha~Doppler tomograms. No map has been computed for V992 Sco as the \Ha~emission is still dominated by the nova shell.}
\end{figure*}

Relative orbital phases were computed by using the orbital period and the $T_0$ measured from the radial velocities. The \Ha~trailed spectrum (Fig.~\ref{fig_trailed}) of \hl~shows a narrow,
low-amplitude S-wave with blue-to-red velocity crossing at $\varphi_\mathrm{r} = 0$. This component is very likely coming from the (irradiated) donor star, which is supported by its maximum strength at $\varphi_\mathrm{r} = 0.5$. Although not shown, the \hel{i}{6678} trailed spectrum also shows narrow emission from the secondary star. In order to obtain the actual $T_0$ from the radial velocity curve of this emission, we proceeded as follows: the individual spectra were folded on the orbital period and averaged into 20 phase bins. A Gaussian-smoothed ($\mathrm{FWHM} = 200$~\kms) version of each spectrum was then subtracted from the actual one. A preliminary radial velocity curve was measured by means of a cursor on a new trailed spectrum, following by eye the path of the donor emission S-wave and clicking on its maximum flux for each phase bin. The spectra were subsequently applied a velocity shift according to that radial velocity curve in order to eliminate the velocity modulation of the donor emission, and were then averaged to construct an emission template which was centred on the rest velocity. A new radial velocity curve was finally obtained by cross-correlation of the original phase-binned spectra with this template (Fig.~\ref{fig_hlaqr_rvc}). A sine fit to this phase-folded curve provided a negligible phase shift with respect to the preliminary $T_0$, which suggests that the radial velocities of the \Ha~wings are not delayed with respect to the motion of the white dwarf (final $T_0$ values are given in Table~\ref{t-rvcfits}). In contrast with the $0.1-0.2$-cycle delays usually observed in the SW Sex stars, the radial velocity curve of the \hel{i}{6678} line wings (not shown) \textit{precedes} the white dwarf motion by 0.1 orbital cycle. Note that the phasing and amplitude of the radial velocity curve of the donor's \hel{i}{6678} emission (not shown) are the same as those of \Ha.

The \Ha~trailed spectrum shows a dominant emission S-wave with maximum excursion to the blue at $\varphi_\mathrm{a} \simeq 0.3$. No emission-line flaring is apparent. Remarkably, there is an absorption S-wave component blueshifted by $\sim 1000$\,\kms~with respect to the rest velocity which reaches maximum blue velocity at $\varphi_\mathrm{a} \simeq 0.5$, that is, 0.3 cycle later than the \Ha~emission wings. This absorption may be indicative of a mass outflow in the system. The fact that it varies with the orbital period suggests that it may originate in a wind which is at an angle with the line of sight. Although P-Cygni profiles in CVs are usually observed in the UV resonance lines, they can be also seen in the optical \citep[][ and references therein]{kafka+honeycutt04-2}. Interestingly, this absorption S-wave is bluest at $\varphi_\mathrm{a} \simeq 0.5$, just like the high-velocity emission S-wave characteristic of the SW Sex stars. A similar behaviour has been observed in the nova-like V592 Cas \citep{withericketal03-1}, where the radial velocity curve of the absorption is also delayed with respect to that of the emission by 0.3 cycle.

\subsubsection{Doppler tomography}

In order to map the \Ha~line emission sites in velocity space we constructed a Doppler tomogram from the phase-folded data using the maximum entropy technique introduced by \cite{marsh+horne88-1} (Fig.~\ref{fig_doppler}). The Doppler image does not show either a ring-like structure typical of line emission originated in an accretion disc or any feature in the path of the gas stream. Instead, the bulk of emission is distributed on an elongated spot spread over the lower left quadrant at $(V_x \sim -100, V_y \sim -200$)\,\kms, which is produced by the dominant broad S-wave observed in the trailed spectrogram. This is a common feature in the Doppler tomograms of the SW\,Sex stars. On the other hand, the narrow S-wave translates into a well-defined spot at $(V_x \sim 0,V_y \sim +150)$ \kms~in the tomogram, where emission from the (likely irradiated) donor star would be placed.

\subsubsection{Orbital inclination}
\label{sec_hlaqr_i}

The detection of an absorption S-wave in \Ha~with maximum blue velocity at $\varphi_\mathrm{a} \sim 0.5$---which may originate in a mass outflow---makes it very interesting to explore whether or not \hl~is viewed at a similar inclination as the emission-line dominated, non-eclipsing SW Sex stars ($i \sim 60\degr-70\degr$). In order to do so, estimates of the stellar masses have to be assumed. We will estimate the mass of the secondary star, $M_2$, by interpolating in the $M_2-P_\mathrm{orb}$ sequence of donor stars in CVs of \cite{knigge06-1}, and will assume a representative white dwarf mass of $M_1=0.75~\mathrm{M}_\odot$ \citep{pattersonetal05-3,knigge06-1}. Spline interpolation of Knigge's mass sequence yields $M_2=0.21~\mathrm{M}_\odot$ for the orbital period of \hl. A mass ratio of $q=M_2/M_1=0.28$ is therefore obtained. From Kepler's Third Law we get:
\begin{equation}
\frac{P_\mathrm{orb}\,{K_2}^3}{2 \pi \mathrm{G}}=M_1 \left[\frac{1}{q(1+q)}\right]\,\sin^3 i~.
\label{eq1}
\end{equation}
\noindent
In order to resolve for the orbital inclination we need an estimate of $K_2$. As mentioned earlier, the narrow \Ha~emission seen in the trailed spectrum is very likely produced in the irradiated inner hemisphere of the donor star. Thus, its radial velocity amplitude, $K_\mathrm{irr}$, is actually a lower limit to $K_2$ as the irradiation light-centre is displaced from the donor's centre of mass towards the inner Lagrangian point ($\mathrm{L_1}$). The ``$K$-correction'' that has to be applied to $K_\mathrm{irr}$ depends on the mass ratio and on a factor ($0 < f < 1$), which represents the distance from the donor star's centre of mass to the light-centre of the irradiated region \citep[][ and references therein]{munoz-dariasetal05-1} in the form:
\begin{equation}
K_2=\frac{K_\mathrm{irr}}{1-f(1+q)}~.
\label{eq2}
\end{equation}  
\noindent
Two extreme cases are: (i) emission from $\mathrm{L_1}$ only, in which case $f$ is just the distance from $\mathrm{L_1}$ to the donor star's centre of mass in units of the binary separation, $R_{\mathrm{L}_2}/a$; and (ii) emission from the limb of the irradiated hemisphere of the donor star. In the second case, and assumming a spherical Roche lobe when far from $\mathrm{L_1}$, $f \simeq (R_{\mathrm{L}_2}/a)^2$. The former case is merely theoretical, and a more realistic upper limit to $f$ can be obtained by modelling of the emission lines formed on the heated face of the secondary star \citep[see equation for $f(\alpha=0\degr)$ in][]{munoz-dariasetal05-1}. 

For \hl, a sine fit to the radial velocities of the donor emission yields $K_\mathrm{irr} \simeq 159$~\kms. On the other hand, we get $f(\alpha = 0\degr) \simeq 0.27$, and $f \simeq 0.08$ for maximum irradiation \citep[using the $R_{\mathrm{L}_2}/a$ equation given by][]{eggleton83-1}. We therefore obtain (Eq.~\ref{eq2}) $177 < K_2 < 244$~\kms, so the orbital inclination of \hl~must lie in the range $19\degr < i < 27\degr$, much lower than in the emission-dominated, non-eclipsing SW Sex stars. Note that $M_1$ and $M_2$ (and therefore $q$) have been assumed. If our mass estimates were close to the actual values, the fact that we see an \textit{absorption} S-wave with maximum blueshift at $\varphi_\mathrm{a} \simeq 0.5$ might be a direct consequence of the much lower inclination of \hl, which seems to support the hypothesis of a mass outflow with significant vertical motion in the system.

\subsubsection{SW Sex class membership}

\hl~is definitely not a ``typical'' SW Sex star. Whilst it exhibits single-peaked emission lines, the radial velocity curve of the \Ha~line wings does not show a phase delay with respect to the motion of the white dwarf. Instead, the \hel{i}{6678} emission line crosses from red to blue $\sim 0.1$ orbital cycle \textit{earlier} than expected. Moreover, a S-wave with maximum blue excursion at $\varphi_\mathrm{a} \simeq 0.5$ and blueshifted by $\sim 1000$~\kms, is observed. Remarkably, this component is seen in absorption, not in emission as is commonly observed in the SW Sex stars. However, the similar phasing and the much lower orbital inclination of \hl~may point to an interesting connection. In this regard, it is possible that the high-velocity component originates in a mass outflow. Line emission from the wind dominates at higher inclinations, whilst P-Cygni absorption is the rule in nearly face-on systems \citep[see e.g.][]{drew97-1}. In addition, the prototypical phase-0.5 absorption is absent. This is not unexpected, as the narrow emission lines already sit on significant  absorption. It is therefore plausible that the transient absorption can be only seen as transient in the high inclination SW Sex stars. Finally, the Doppler tomogram shows the bulk of emission distributed over its lower left quadrant, a typical signature of the SW Sex stars. In summary, \hl~shows features that point to a SW Sex nature. The differences with the canonical ones might be due to its much lower inclination.


\begin{figure}
\mbox{\epsfig{file=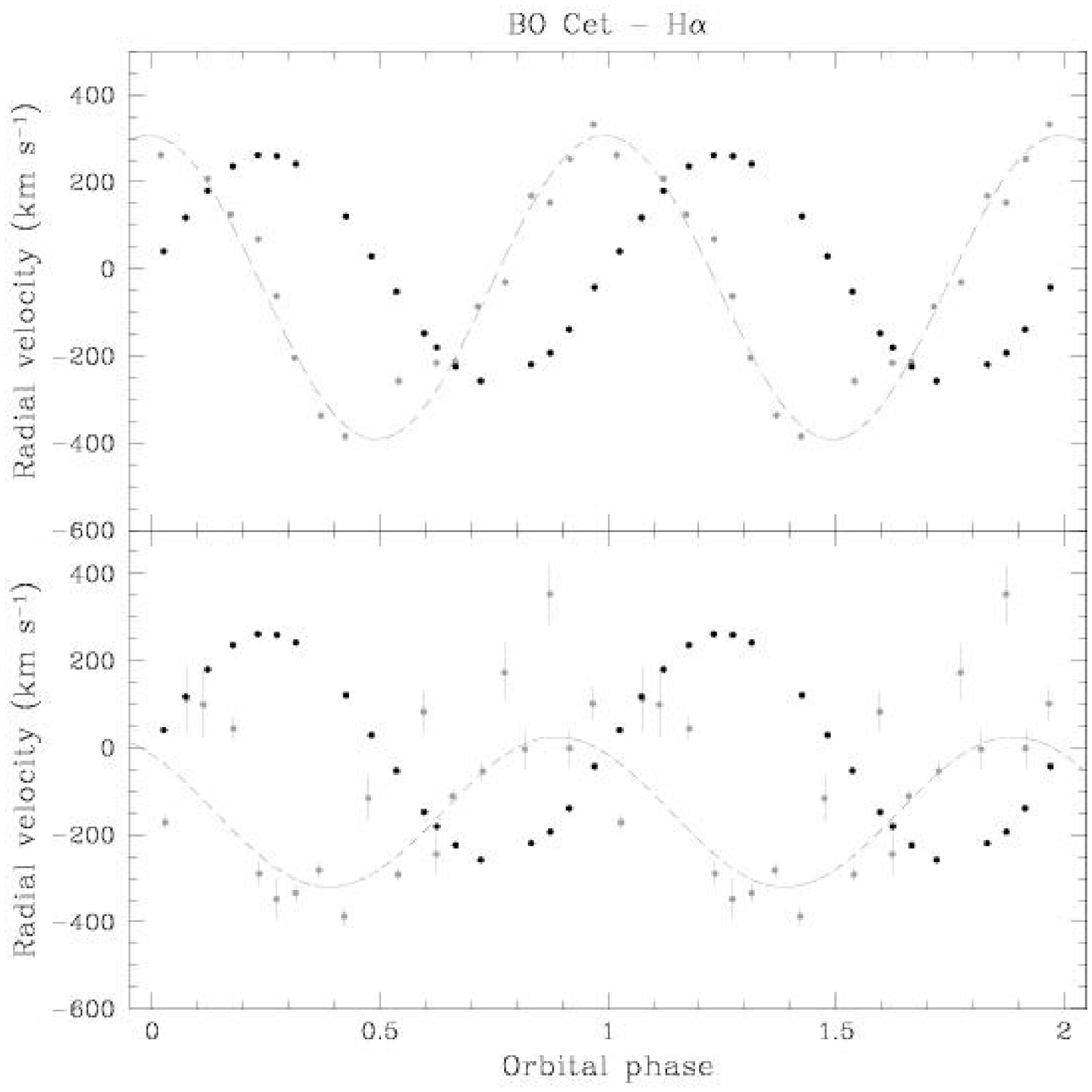,width=\columnwidth}}
\caption{\label{fig_bocet_rvcs} \textit{Top}: phase folded ($P_\mathrm{orb} = 0.1398$~d) \Ha~radial velocity curve of BO Cet obtained by cross-correlation with a single Gaussian template of $\mathrm{FWHM}=300$ \kms~(gray) and the best sine fit (dashed line); and radial velocity curve of the \Ha~emission from the donor star (black). There is a 0.24-cycle shift between the two. \textit{Bottom}: same but applying the double-Gaussian technique with $\mathrm{FWHM}=200$ \kms~and a Gaussian separation of 3500 \kms~(gray). The \Ha~wings are also delayed by 0.14-cycle with respect of the motion of the white dwarf. All the curves have been averaged into 20 phase bins. Some clearly deviant points have been removed for clarity. The orbital cycle has been plotted twice.}
\end{figure}

\subsection{BO\,Cet \label{sec_bocet}}

\bo~($=$ Cet\,4, 1H\,0204--023, PB\,6656) is listed as a nova-like CV in \cite{downesetal05-1}. The optical spectrum obtained by \cite{zwitter+munari95-1} shows a blue continuum with broad, single-peaked emission lines of H\,{\sc i} and weaker, double-peaked \he{i}. The presence of a strong \hel{ii}{4686} emission line reveals a source of ionising photons within the binary, as observed in magnetic CVs and SW Sex stars. However, there is no orbital phase-resolved spectroscopic analysis in the literature even though \bo~is a fairly bright CV ($V \sim 14-15$). \cite{downesetal05-1} quote an orbital period of 0.1398 d ($= 3.36$\,h) obtained from intensive photometric coverage compiled at the Center for Backyard Astrophysics\footnote{The news note on the orbital period of \bo~can be found at http://cba.phys.columbia.edu/communications/news/2002/oc\-to\-ber29-a.html}.

\begin{figure}
\mbox{\epsfig{file=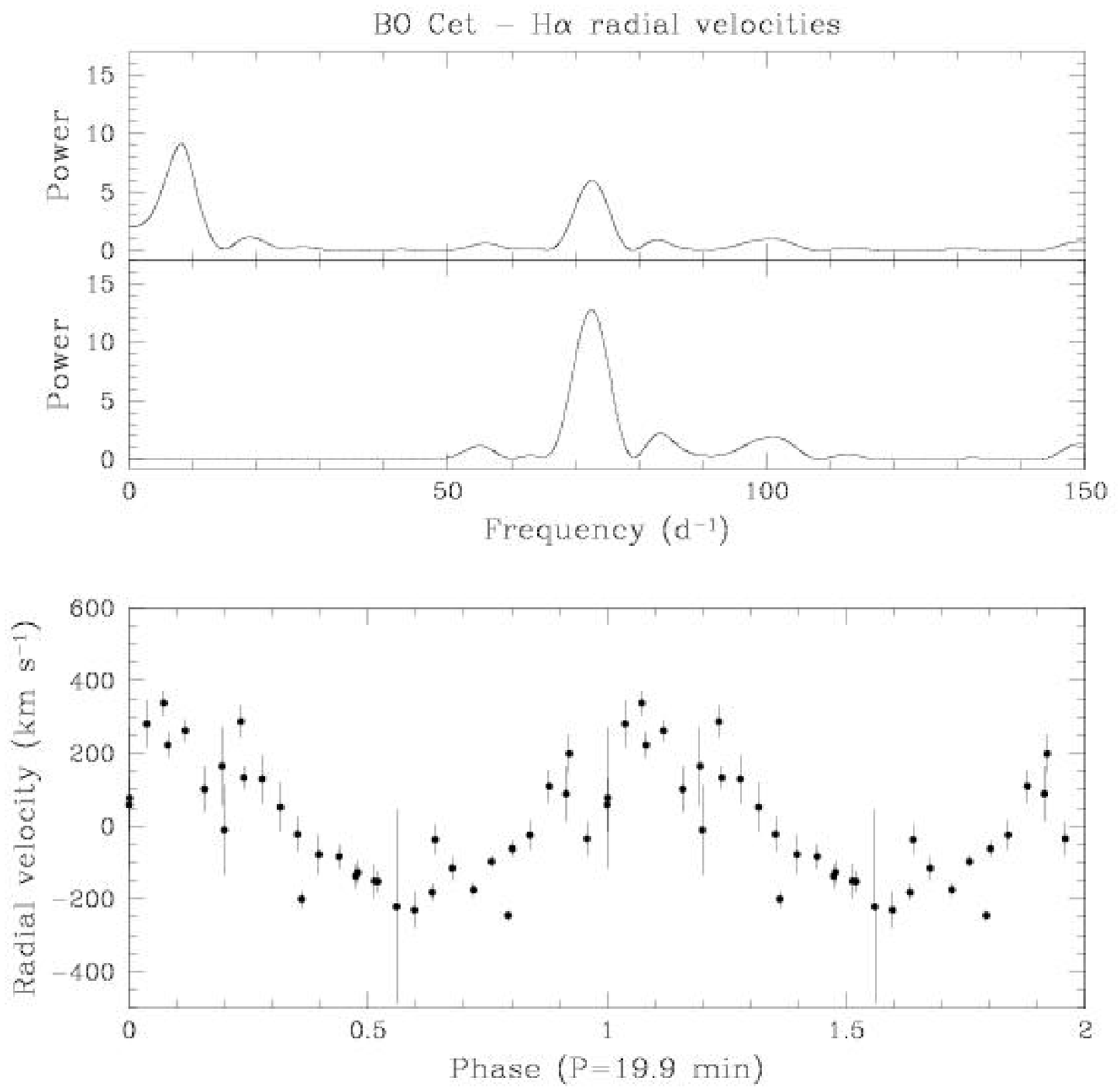,width=\columnwidth}}
\caption{\label{fig_bocet_scargle} \textit{Top panel}: Scargle
periodogram computed from the radial velocity curve of the \Ha~wings in BO Cet. Strong peaks centred at both the orbital frequency and at $\nu = 72.5$\,d$^{-1}$
are apparent. \textit{Middle panel}: Scargle periodogram after
pre-whitening the data at the orbital period. The data were detrended by subtracting a boxcar-smoothed version of the curve in order to remove the long time-scale variations. \textit{Bottom panel}:
detrended \Ha~velocities folded
on the 19.9-min period. A full cycle has been repeated for clarity.}
\end{figure}

\subsubsection{\Ha~radial velocities and trailed spectrum}

Our spectra show a relatively broad ($\mathrm{FWHM} \simeq 1300$\,\kms) \Ha~emission line, indicative of an intermediate-high orbital inclination, with a half-width at zero-intensity ($\mathrm{HWZI}$) reaching $\sim 4000$ \kms. Alas, our data only cover one orbital cycle, which is insufficient for an accurate orbital period determination. In spite of that, we measured the radial velocity variation of the \Ha~emission line by two methods: cross-correlation of the individual profiles with (i) a single $\mathrm{FWHM}=300$ \kms~Gaussian template, and (ii) two $\mathrm{FWHM}=200$ \kms~Gaussians separated by 3500 \kms~\citep[the double Gaussian technique of ][]{schneider+young80-2}. A sine fit to the core velocities yields a period of $\simeq 0.14$\,d. This is consistent with the photometric value mentioned earlier, which we adopt as the orbital period. 

An initial \Ha~trailed spectrum revealed a narrow emission S-wave with an amplitude of $\sim 200-300$ \kms~and a likely origin on the heated face of the secondary star. In order to obtain absolute phases we proceeded in the same way as with \hl~(Sect.~\ref{sec_hlaqr_rvc}), obtaining $T_0(\mathrm{HJD})=2453640.7633 \pm 0.0001$ and $K_\mathrm{irr}=262 \pm 2$~\kms. The phase folded \Ha~radial velocity curves are presented in Fig.~\ref{fig_bocet_rvcs}. The line core lags the motion of the white dwarf by 0.24 cycle. In addition, the wings are also delayed by 0.14 cycle and are significantly blueshifted, which clearly depicts the risk of adopting the amplitude of the wings radial velocity curve as the true radial velocity of the white dwarf ($K_1$) for mass measurements (a common practice as seen in the literature).

The \Ha~trailed spectrum of \bo~(Fig.~\ref{fig_trailed}) displays several interesting features. The line profile is dominated by an emission S-wave with an amplitude of $\sim 300$ \kms~and bluest velocity at $\varphi_\mathrm{a} \simeq 0.5$. This component shows maximum intensity in the $0.65-0.95$ phase interval, and probably a secondary maximum at $\varphi_\mathrm{a} \sim 0.3$. Intensity minima take place at $\varphi_\mathrm{a} \simeq 0.5$ and $\varphi_\mathrm{a} \simeq 0$, when the EW of the line core is minimum. Moreover, the line wings are shaped by a high-velocity S-wave clearly visible up to $-2000$\,\kms. This component is clearly pulsed, which is confirmed by a non-phase folded trailed spectra (not shown).

\subsubsection{Emission-line flaring}

\begin{figure}
\mbox{\epsfig{file=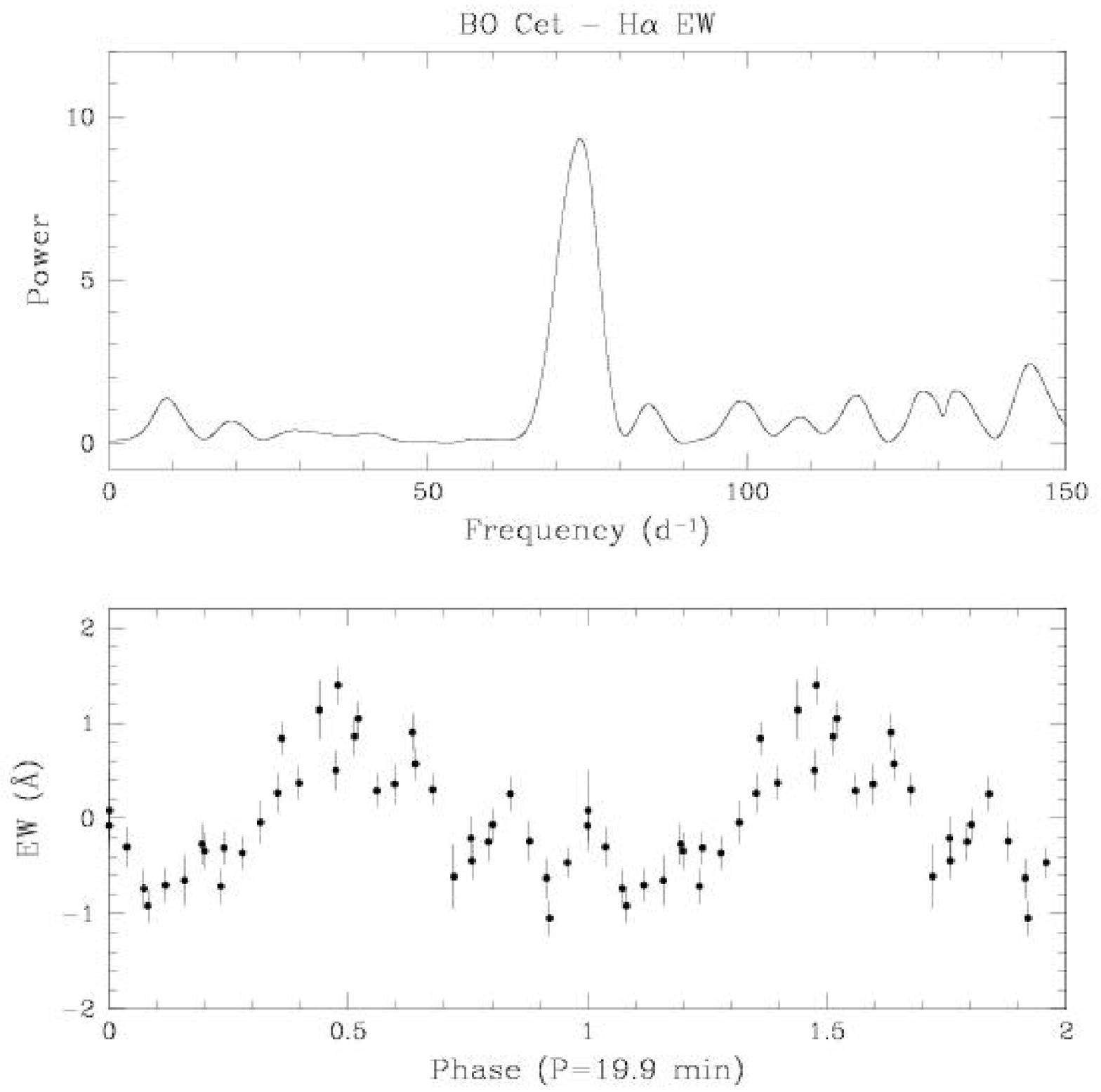,width=\columnwidth}}
\caption{\label{fig_bocet_ew_scargle} \textit{Top panel}: Scargle periodogram computed from the EW curve of the \Ha~blue wing of BO Cet pre-whitened at the orbital frequency. \textit{Bottom panel}: detrended \Ha~EWs folded on the 19.9-min period. A full cycle has been repeated for clarity.}
\end{figure}

Emission-line rapid variability is often observed in intermediate polars \cite[e.g. FO\,Aqr][]{marsh+duck96-2} and SW\,Sex stars (e.g. BT Mon, Smith, Dhillon \& Marsh \citeyear{smithetal98-1}; LS\,Peg, \citealt{rodriguez-giletal01-1}; V533\,Her, \citealt{rodriguez-gil+martinez-pais02-1}; DW\,UMa, V. Dhillon, private communication; and RX\,J1643.7+3402, Mart\'\i nez-Pais, de la Cruz Rodr\'\i guez \& Rodr\'\i guez-Gil \citeyear{martinez-paisetal07-1}).

This fast oscillation is responsible for the scatter observed in the radial velocity curve of the \Ha~wings (Fig.~\ref{fig_bocet_rvcs}, bottom panel). In fact, the non-phase folded wing velocities clearly exhibit a short time-scale oscillation at $\sim 15-20$\,min. In order to search for periodicites we computed the Scargle periodogram shown in Fig.~\ref{fig_bocet_scargle} (top panel), which contains two strong peaks: one at the orbital frequency and another one at $\nu \simeq 72.5$\,d$^{-1}$. After subtracting the orbital modulation, another periodogram was calculated, showing the main peak at a period of $19.9 \pm 0.9$\,min (the quoted uncertainty is half the FWHM of the peak). Although the peaks are significant, we can only state that this oscillation is coherent for at least 10 cycles. Therefore, a much longer spectroscopic coverage is needed in order to confirm whether this radial velocity variation is produced by a stable clock in the system. In such a case, the 19.9-min period should be related to the spin period of a magnetic white dwarf as it happens in intermediate polars like V1025\,Cen \citep{buckleyetal98-1}, DW\,Cnc \citep{rodriguez-giletal04-1}, and HS\,0943+1404 \citep{rodriguez-giletal05-2}.

With the aim of searching for rapid variations, we also measured the EW curve of the \Ha~line blue wing (the high-velocity emission S-wave is best seen in the blue) in the velocity interval $(-3500,-1200)$ \kms. After prewhitening at the orbital frequency, a Scargle periodogram was computed (Fig.~\ref{fig_bocet_ew_scargle}). A strong peak centred at $19.6 \pm 0.9$\,min, consistent with the observed periodicity in the radial velocity curves, is apparent.

\subsubsection{Doppler tomography}

The \Ha~Doppler tomogram of \bo~is presented in Fig.~\ref{fig_doppler}, where two \Ha~emission sites are clearly observed. A small, concentrated emission spot is located at $(V_x \sim 0,V_y \sim +260)$ \kms, just where emission from the donor star is expected to lie. In fact, the $V_y$ position of the spot on the map coincides with the amplitude of the radial velocity curve of the donor emission ($K_\mathrm{irr}=262$ \kms). In addition, a broad spot is seen in the lower left quadrant of the Doppler map, approximately centred on $(V_x \sim -400,V_y \sim -10)$ \kms. This is a characteristic feature of the tomograms of the SW\,Sex stars.

\subsubsection{Orbital inclination}

Adopting the component masses $M_1 = 0.75~\mathrm{M_\odot}$ and $M_2 = 0.22~\mathrm{M_\odot}$ for \bo~(see Sect.~\ref{sec_hlaqr_i}), we obtain a mass ratio of $q=0.29$. The extreme limits to the $K$-correction provide a radial velocity amplitude for the companion star in the range $292 < K_2 < 407$~\kms, which implies an orbital inclination in the interval $35\degr < i < 52\degr$.

\subsubsection{SW Sex class membership}

The above analysis qualifies \bo~as a new non-eclipsing SW\,Sex star. The emission lines are mainly single peaked, with the exception at phase 0.5, when they turn double peaked. Phase 0.5 is also the instant of minimum EW of the line core. In addition, \Ha~exhibits a clear emission S-wave with maximum blue velocity at phase 0.5, which translates into an elongated emission spot in the lower left quadrant of the Doppler tomogram. Also, both the core and wings radial velocity curves are delayed with respect of the motion of the white dwarf. Finally, \bo~displays emission-line flaring with a periodicity of 19.9 min.

A word of caution should be rised regarding the actual orbital period of \bo. In our study we have adopted a photometric period which can not be \textit{a priori}
discarded as, for example, a superhump period. Although the fact that
Patterson's analysis yielded a single periodicity supports a pure orbital origin, we should bear in mind that many SW\,Sex stars show positive and/or negative permanent superhumps \citep[see e.g.][]{patterson95-1,pattersonetal02-1,stanishevetal02-1,pattersonetal05-3}. Hence, only a more extensive spectroscopic coverage will provide the actual
orbital period.


\subsection{V849\,Her \label{sec_v849her}}

\vher~($=$ PG\,1633+115) was tentatively identified as a 15-mag CV in
the Palomar-Green survey (Green, Schmidt \& Liebert \citeyear{greenetal86-1}). Its optical spectrum
displays weak \Ha~emission and \Hb~absorption with a continuum showing
variations with a time-scale of days \citep{ringwald93-2}. Extensive
$V$-band photometry performed by \cite{misselt+shafter95-1} revealed a
0.5-mag drop in brightness and a modulation at a period of $\sim
3.5$\,h. Alas, the analysis of the light curves did not provide an
accurate orbital period. A further optical spectrum obtained by
\cite{munari+zwitter98-1} confirmed the presence of a very weak
\Ha~emission line superposed on a broader absorption. An even weaker
\Hb~emission can be seen at the bottom of a dominating \Hb~absorption
trough.

\begin{figure}
\mbox{\epsfig{file=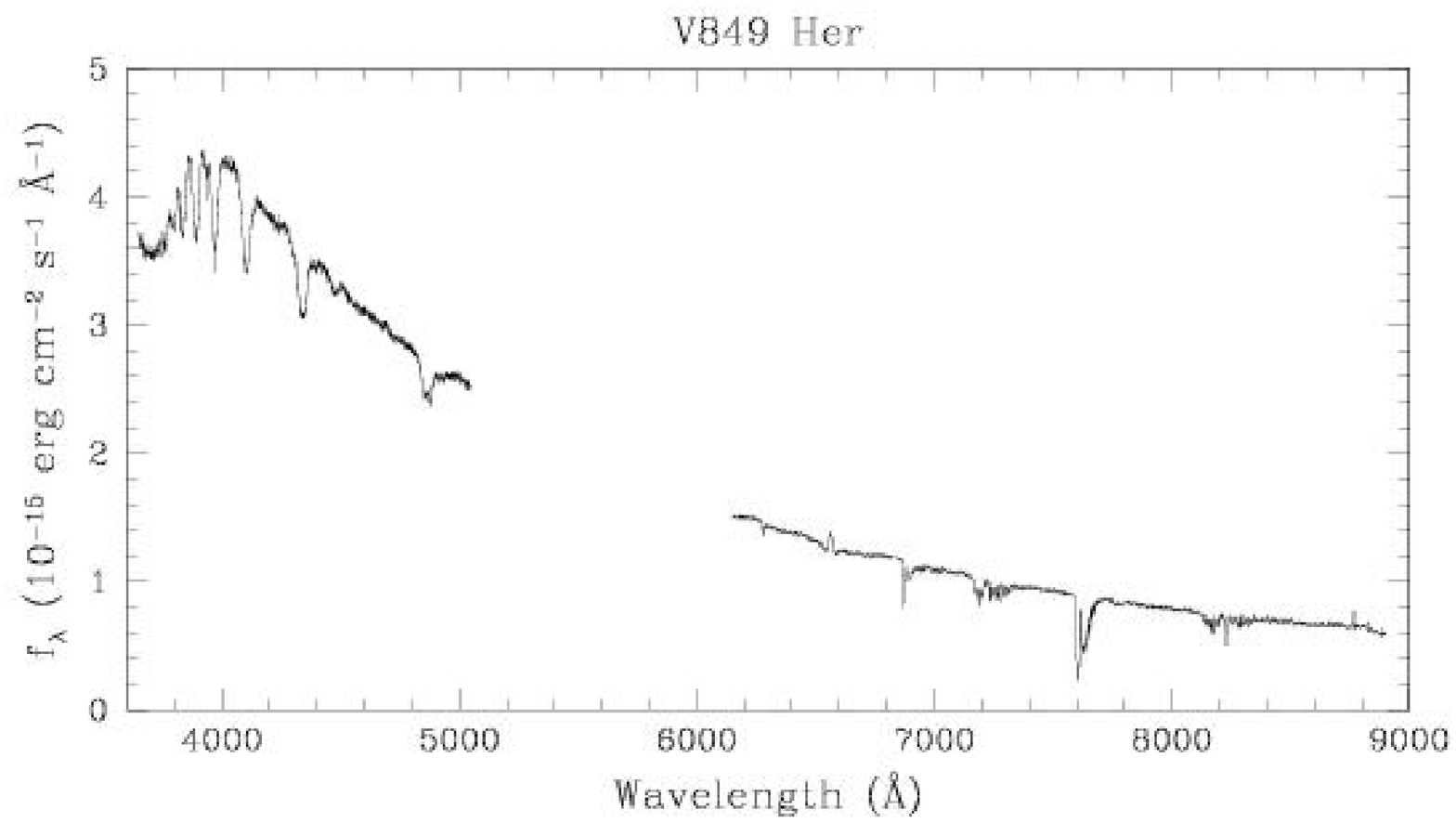,width=\columnwidth}}
\caption{\label{fig_v849her_flux} WHT flux-calibrated, average
spectrum of \vher. No telluric absorption correction has been applied.}
\end{figure}

Our WHT flux-calibrated, average spectrum is presented in
Fig.~\ref{fig_v849her_flux}. Broad absorption lines dominate blueward
of \Hb, whereas the \Ha~line is mostly in emission and shows a single-peaked profile. A weak \hel{ii}{4686} emission line is also observed. On the basis of the observed long-term brightness changes it has been argued
\citep{ringwald93-2} that \vher~might be a dwarf nova. Our average
spectrum, however, is very similar to
\citeauthor{munari+zwitter98-1}'s, suggesting that either the system
comes into outburst quite frequently or it shows on most
occasions an absorption-dominated spectrum. Thus,
\vher~is most likely a low-inclination nova-like, or possibly a Z\,Cam
dwarf nova. Long term monitoring of its visual brightness is
necessary to differentiate between the two possibilities. 

\subsubsection{\Ha~radial velocities and trailed spectrum}

With the aim of obtaining the orbital period of \vher~we measured the
radial velocities of the \Ha~emission line. The best results were
obtained using the double-Gaussian technique with
$\mathrm{FWHM}=200$\,\kms~and a Gaussian separation of 900\,\kms. A sine fit to the velocities provided a period of $3.15 \pm 0.48$\,h, but the short interval our spectroscopic data covers (only 2.64\,h) prohibits us from reaching any firm
conclusion regarding the actual orbital period. Even though the orbital period
of \vher~is not a settled issue, we folded the spectroscopic data
on the period given by the sine fit to the velocities and averaged the
spectra into 20 orbital phase bins. 

\begin{figure}
\mbox{\epsfig{file=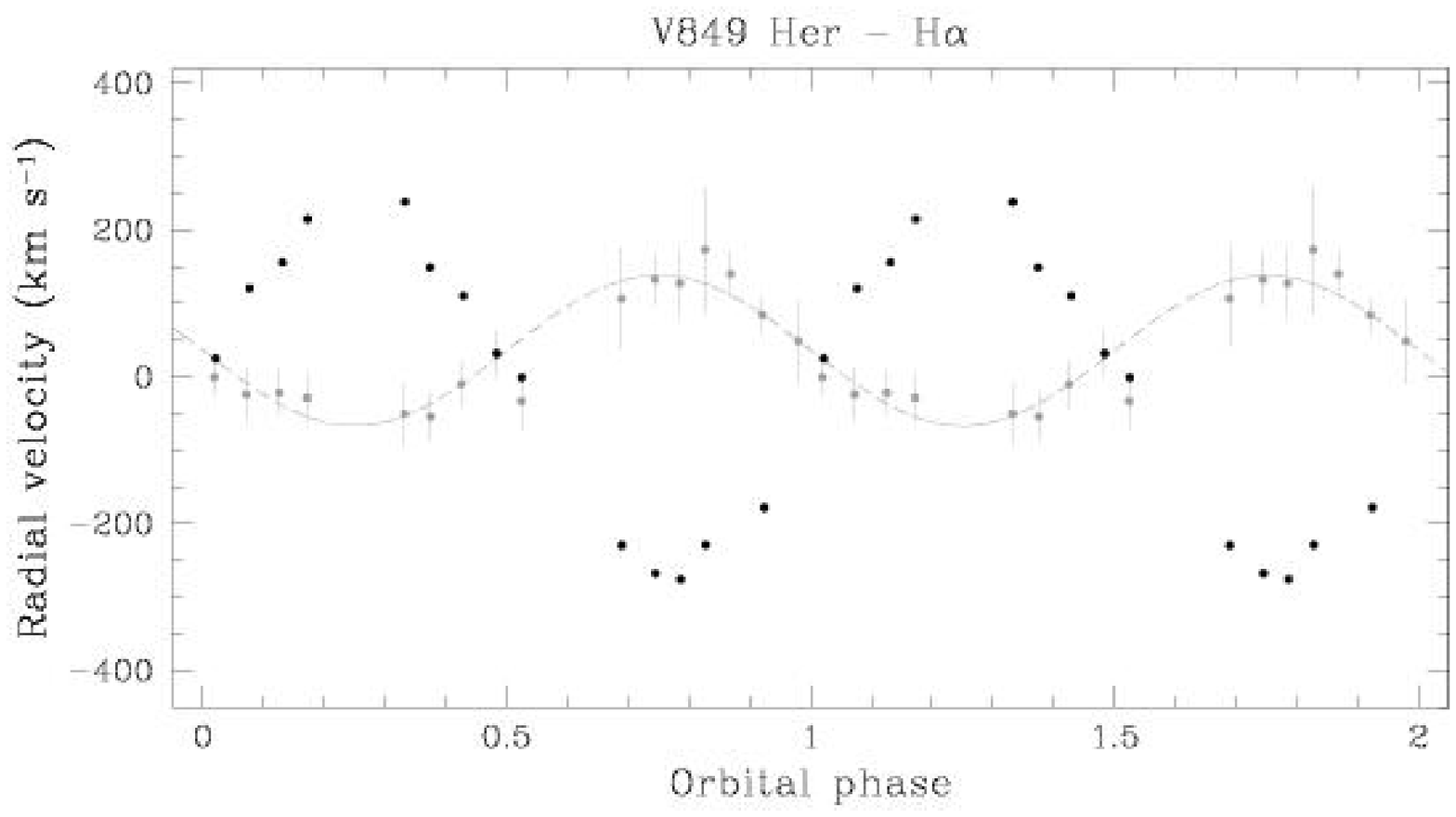,width=\columnwidth}}
\caption{\label{fig_v849her_rvc} \Ha~radial velocity curve of \vher~obtained by using a double-Gaussian template of $\mathrm{FWHM}=200$ \kms~and a Gaussian separation of $900$~\kms~(gray) and the best sine fit (dashed line); and radial velocity curve of the \Ha~emission from the donor star (black). There is no phase delay between the two. A full orbital cycle has been repeated for clarity.}
\end{figure}

A preliminary \Ha~trailed spectrogram computed from the phase-folded data (using the $T_0$ from the fit to the velocities) shows an emission S-wave extending up to $\pm 500$\,\kms~and reaching its bluest velocity at $\varphi_\mathrm{r} \sim 0.3-0.4$. There is also another emission component moving with smaller velocity amplitude ($\sim 200$~\kms) which has maximum intensity at $\varphi_\mathrm{r} \sim 0.4-0.5$. The phase-binned spectra show alternating double-peaked and single-peaked \Ha~profiles, which supports the presence of at least two emission components. The emission component with smaller velocity may originate on the heated inner face of the secondary star. Its maximum strength at $\varphi_\mathrm{r} \sim 0.5$ seems to support this origin. This will allow us to measure an absolute ephemeris so that the higher velocity component can be appropriately phased. In order to do so we applied the same technique we used with \hl~and \bo, obtaining $T_0(\mathrm{HJD})=2453513.536 \pm 0.005$ and $K_\mathrm{irr}=257 \pm 18$~\kms. The phase-binned \Ha~radial velocity curves are presented in Fig.~\ref{fig_v849her_rvc}. There is no phase delay between the radial velocities measured with the double-Gaussian technique and the white dwarf motion which, on the other hand, is true only if the assumption of emission from the donor star is correct. Alas, the signal-to-noise ratio of our data is not adequate to check whether the \hel{i}{6678} emission line is phase-shifted or not.


\subsubsection{Doppler tomography}

The \Ha~Doppler map of \vher~is shown in Fig.~\ref{fig_doppler}. Its main feature is an emission spot located at the position of the donor star $(V_x \sim 0,V_y \sim +250)$. The other elongated emissions are likely artifacts due to the poor phase sampling. The characteristic SW Sex emission at the lower left quadrant is not observed.

\subsubsection{Orbital inclination}

In this case, following our usual analysis, $M_1 = 0.75~\mathrm{M_\odot}$ and $M_2 = 0.20~\mathrm{M_\odot}$, which results in a mass ratio of $q=0.27$. The radial velocity amplitude of the secondary star should therefore lie in the interval $296 < K_2 < 389$~\kms, which translates into an orbital inclination range of $33\degr < i < 45\degr$. These numbers are very similar to the results obtained for \bo~but, clearly, both systems behave differently. If we assume a plausible $K$-correction of $K_\mathrm{irr}/K_2 \simeq 0.7$ (see Figure 4 in \citealt{munoz-dariasetal05-1}), we obtain $K_2 \simeq 367$~\kms. Identifying the amplitude of the radial velocity curve of the disc emission with $K_1$, we get $q = K_1/K_2 \simeq 0.27$, which matches the assumed value. In the SW Sex stars, the presence of the high-velocity emission S-wave tends to increase the amplitude of the wing velocities which, together with the characteristic phase shifts, suggests that the bulk of emission is not coming from an axisymetric structure around the white dwarf. This does not seem the case in \vher.

\subsubsection{SW Sex class membership}

The spectroscopic behaviour of \vher~does not seem to qualify it as a SW Sex star. Nevertheless, better quality spectra with better sampling would be desirable to validate or disprove its SW Sex nature.

As a final note, the optical spectrum of \vher~is very similar to that of a number of nova-likes found in the Hamburg Quasar Survey (e.g. HS\,0139+0559, HS\,0229+8016, and HS\,0642+5049, \citealt{aungwerojwitetal05-1}), which are dominated by absorption and also have orbital periods in the range $3-4$\,h. Because of their inconspicuous spectroscopic, photometric, and X-ray properties, this type of system may be quite frequent, with a significant number of objects still to be identified. Together with \vher, these systems are likely viewed at intermediate-low inclinations, so we are therefore looking directly at optically thick disc material. Hence, broad line absorption may dominate the spectra of nearly face-on systems. Such dominance is actually seen in the UX UMa nova-likes.


\subsection{V393\,Hya}

\vhya~($=$~EC\,10578--2935) was listed as a CV and tentatively
classified as a nova-like by \cite{kilkennyetal97-1}, an
identification later supported by the weak Balmer lines observed by
\cite{sefakoetal99-1} in its optical spectrum. The system shows a
variable emission pattern in which a spectrum with weak Balmer,
\hel{ii}{4686}, and Bowen blend emissions \citep{chenetal01-1} can
switch to a flat and featureless continuum with only \Ha~in emission
\citep{dall+schmidtobreick04-1}. The weakness of
the emission lines suggests a rather hot accretion disc and,
therefore, is indicative of the high mass transfer rate characteristic
of nova-like CVs. In addition, the emission lines are very narrow,
showing a \Ha~full width at zero intensity ($\mathrm{FWZI}$) of only
$\sim 700-900$ \kms, which indicates a low orbital inclination. In
fact, \citeauthor{chenetal01-1} were not able to extract any useful
radial velocity information from the line profiles.

Like \bo, \vhya~was also subject to extensive photometric scrutiny by the observers associated to the Center for Backyard Astrophysics, who found a periodicity of 0.1346\,d ($= 3.23$\,h) in the phase-resolved light curves\footnote{CBA news note on \vhya~at:\\
http://cba.phys.columbia.edu/communications/news/2003/\\
april27.html}.     

\subsubsection{\Ha~radial velocities, trailed spectrum, and Doppler tomogram}

The NTT average spectrum shown in Fig.~\ref{spec_all} exhibits
double-peaked \Ha~emission together with a flat-topped \hel{i}{6678}
line. Contrary to previous works, we measured $\mathrm{FWZI} \ga 2000$
\kms, a factor of two higher than reported by other authors. The radial
velocity curve of the \Ha~line was computed by cross-correlation of
the individual profiles with a double-Gaussian template with
$\mathrm{FWHM} = 200$\,\kms~and a Gaussian separation of 800\,\kms. A
Scargle periodogram of the \Ha~velocities (not shown) suggests an orbital 
period of $0.14-0.15$\,d, consistent with the photometric
periodicity. However, our limited coverage (0.9\,h and 2.6\,h on April
17 and 18, respectively) is clearly insufficient to provide an
accurate orbital period, so we use the photometric estimate.

The \Ha~trailed spectra of \vhya~(Fig.~\ref{fig_trailed}; relative orbital phases) only shows the orbital motion of the double-peaked profile and enhanced wings reaching $\sim \pm 500$\,\kms. A much better spectral resolution and signal-to-noise ratio are needed to resolve any component moving beneath the two peaks. The Doppler map (Fig.~\ref{fig_doppler}) solely reveals a ring of emission, which is the result of the double-peaked shape of the \Ha~emission line. In spite the poor quality ot our data, \vhya~does not seem to behave as a SW Sex star.  


\subsection{AH\,Men}

The bright ($V \simeq 13$) counterpart of the X-ray source 1H\,0551--819 has at first been regarded as an intermediate polar cataclysmic variable \citep{warner+wickramasinghe91-1,wu+wickramasinghe91-1}, even before being identified as a nova-like CV by \cite{buckleyetal93-1}. These authors obtained 79\,h of photometric data which revealed a periodicity of 3.34\,h. Apart from erratic flickering the light curves also showed quasi-periodic oscillations (QPOs) with preferred time-scales in the range $\sim 600-2500$\,s. The light curve of \ah~also displayed dips at maximum light which were suggested to be shallow eclipses, but this point was never confirmed. In such a case the orbital inclination of the CV would be around 70\degr.

\cite{patterson95-1} secured 190\,h of high-speed photometry of \ah~finding a stable period of 2.95\,h which is therefore believed to be the orbital period of the system. Lower coherence signals at 3.05\,h and 2.88\,h were also found, suggesting the presence of positive (apsidal) and negative (nodal) superhumps, respectively. This is supported by the detection of two much longer periodicities close to 4\,d likely associated with the corresponding precession periods of an eccentric, wobbling accretion disc. On the other hand, the high frequency range is characterised by oscillations at $\sim 1020-1320$\,s with a probable white dwarf spin period of either 1040\,s or 2080\,s.

\begin{figure}
\mbox{\epsfig{file=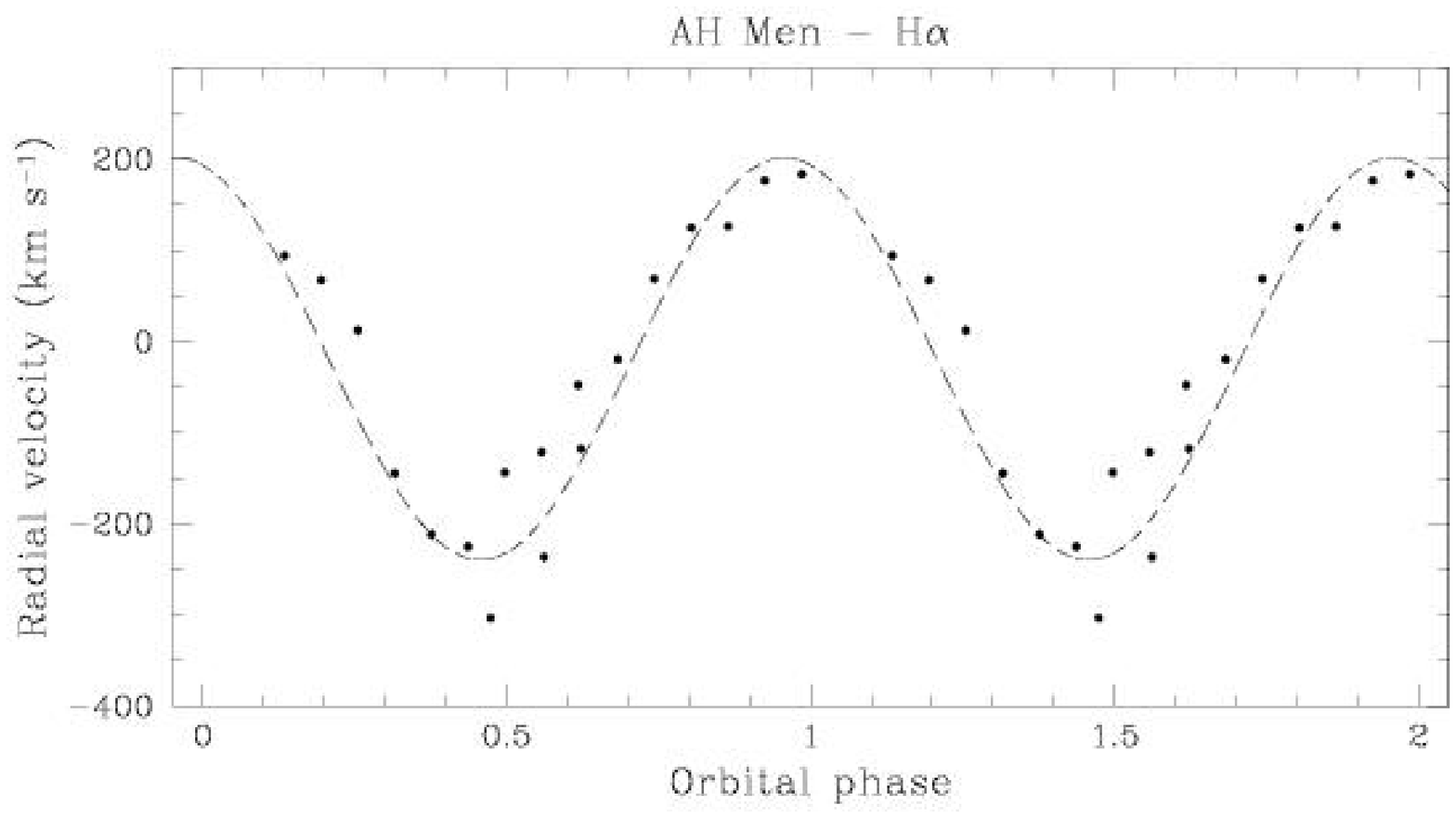,width=\columnwidth}}
\caption{\label{fig_ahmen_rvc} \Ha~radial velocity curve of \ah. A 0.2-cycle delay with respect to the expected red-to-blue crossing is seen (see text for details on the assummed ephemeris). The dashed curve is the best sine fit to the data. A full cycle has been repeated.}
\end{figure}

The average optical spectrum of \ah~reported by \cite{buckleyetal93-1} exhibited double-peaked Balmer and \he{i} emission lines and a single-peaked \hel{ii}{4686} emission, the latter indicating a moderate level of excitation of the accretion flow. \citeauthor{buckleyetal93-1} performed a radial velocity study from spectroscopic data spanning 5\,h obtaining a periodicity of $3.1 \pm 0.3$\,h and a $K$-amplitude of $\sim 140$\,\kms. This period is consistent with \citeauthor{patterson95-1}'s findings but the spectroscopic data alone were not   sufficient to unambiguously confirm the orbital period of \ah.

The UV spectrum presented by \cite{mouchetetal96-1} is similar to that
of the non-eclipsing SW\,Sex star V795\,Her
(Prinja, Drew \& Rosen \citeyear{prinjaetal92-1}). Both exhibit transient absorptions in the
C\,{\sc iv} $\lambda$1549 line proposed to be due to an accretion disc
wind but, unlike V795\,Her, the UV spectrum of \ah~shows the C\,{\sc
iv} $\lambda$1549 line predominantly in emission, suggesting a higher
inclination \citep[V795\,Her is seen at $\sim 53\degr-56\degr$,][]{casaresetal96-1,rodriguez-giletal01-2}. \cite{gaensicke+koester99-1} derived a lower limit to the distance of $d=150$\,pc from the non-detection of the donor star, and showed that the
UV spectrum of AH Men can be qualitatively modelled with an optically thick
accretion disc model at a distance of $\sim200$\,pc.

\subsubsection{\Ha~radial velocities and trailed spectrum}

The average \Ha~profile (Fig.~\ref{spec_all}) has $\mathrm{FWHM} \simeq 940$\,\kms~and displays asymmetric wings disappearing into the continuum at velocities of $\sim \pm 4000$\,\kms. We measured the radial velocity curve of the \Ha~emission line by cross correlation of the individual profiles with a double Gaussian template (1600\,\kms~separation, $\mathrm{FWHM}=200$\,\kms). Adopting the orbital period reported by \cite{patterson95-1}, $P_\mathrm{orb}=0.122992$\,d, a sine fit to the radial velocity curve yields a time of red-to-blue crossing of $2453477.5667 \pm 0.0002$\,(HJD). The subsequently phase-folded trailed spectrum showed a high-velocity emission S-wave with maximum blueshift at $\varphi_\mathrm{r} \sim 0.3$. Note that in the eclipsing SW\,Sex stars (for which accurate eclipse ephemeris are available) this happens at $\varphi_\mathrm{a} \sim 0.5$ (see Sect.~\ref{swsex_criteria}). Hence, the actual time of inferior conjunction of the donor star in \ah~should take place $\sim 0.2$ orbital cycle earlier than the $T_0$ provided by the radial velocity curve, that is, $T_0=2453477.5421 \pm 0.0002$\,(HJD).      

The trailed spectrum (Fig.~\ref{fig_trailed}) clearly shows a high-velocity S-wave with an amplitude of $\sim 1000$\,\kms~and a central absorption when the S-wave is at its bluest excursion, a hallmark of the SW\,Sex stars. In fact, the EW of the line has a minimum at $\varphi_\mathrm{a} \simeq 0.5$. It does not appear to show flaring, as can be also seen in the trailed spectrum. Contrary to \hl~and \bo, no narrow emission from the donor star is apparent.

\subsubsection{Doppler tomography}

The \Ha~Doppler map of \ah~(Fig.~\ref{fig_doppler}) shows the characteristic emission pattern of the SW\,Sex stars, with emission concentrated in the lower left quadrant of the tomogram. There might be also an emission spot close to the expected location of the secondary star at $(V_x \sim -80,V_y \sim +290)$~\kms. However, the trailed spectrum does not show an apparent emission with the phasing of the donor star.  

\subsubsection{SW Sex class membership}
\label{sec_ahmen_swsex}

The \Ha~emission of \ah~is characterised by a high-velocity S-wave which maximum blue velocity at $\varphi_\mathrm{a} \sim 0.5$. The single-peaked line profiles reach minimum EW at the same phase, which is probably due to a crossing absorption component. Additionally, the radial velocity curve of the line wings shows a phase delay of $\sim 0.2$ orbital cycle with respect to the white dwarf motion. Furthermore, the \Ha~Doppler tomogram displays the characteristic emission in its lower-left quadrant. All of this represent defining features of the SW\,Sex class. Hence, we classify \ah~as a SW\,Sex star. Finally, it would be desirable to perform phase-resolved spectroscopy with better time resolution to check whether the rapid oscillation in the light curve has a counterpart in the emission lines.


\begin{figure}
\mbox{\epsfig{file=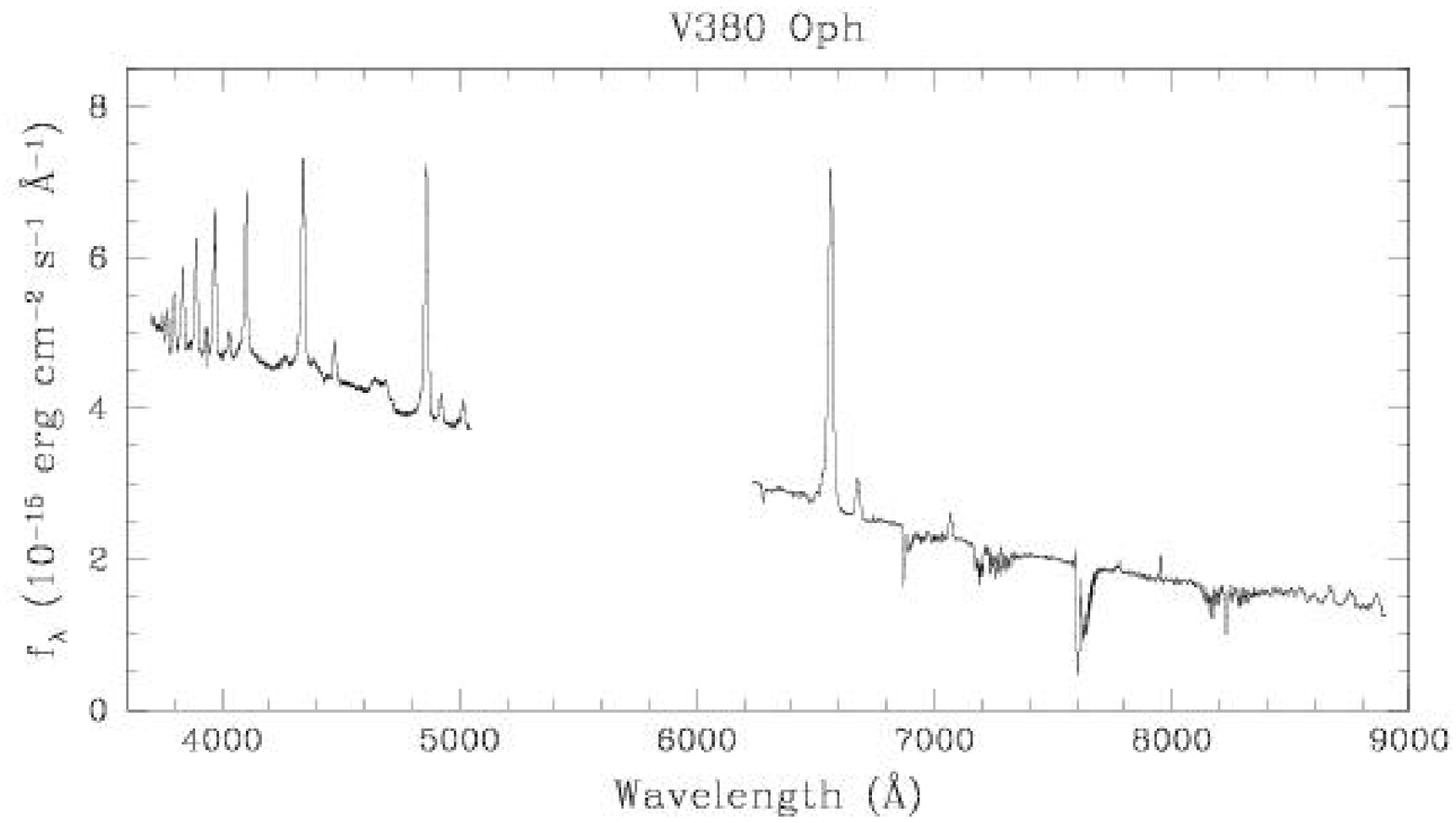,width=\columnwidth}}
\caption{\label{fig_v380oph_flux} WHT flux-calibrated, average spectrum of \voph. No telluric absorption correction has been applied.}
\end{figure}

\begin{figure}
\mbox{\epsfig{file=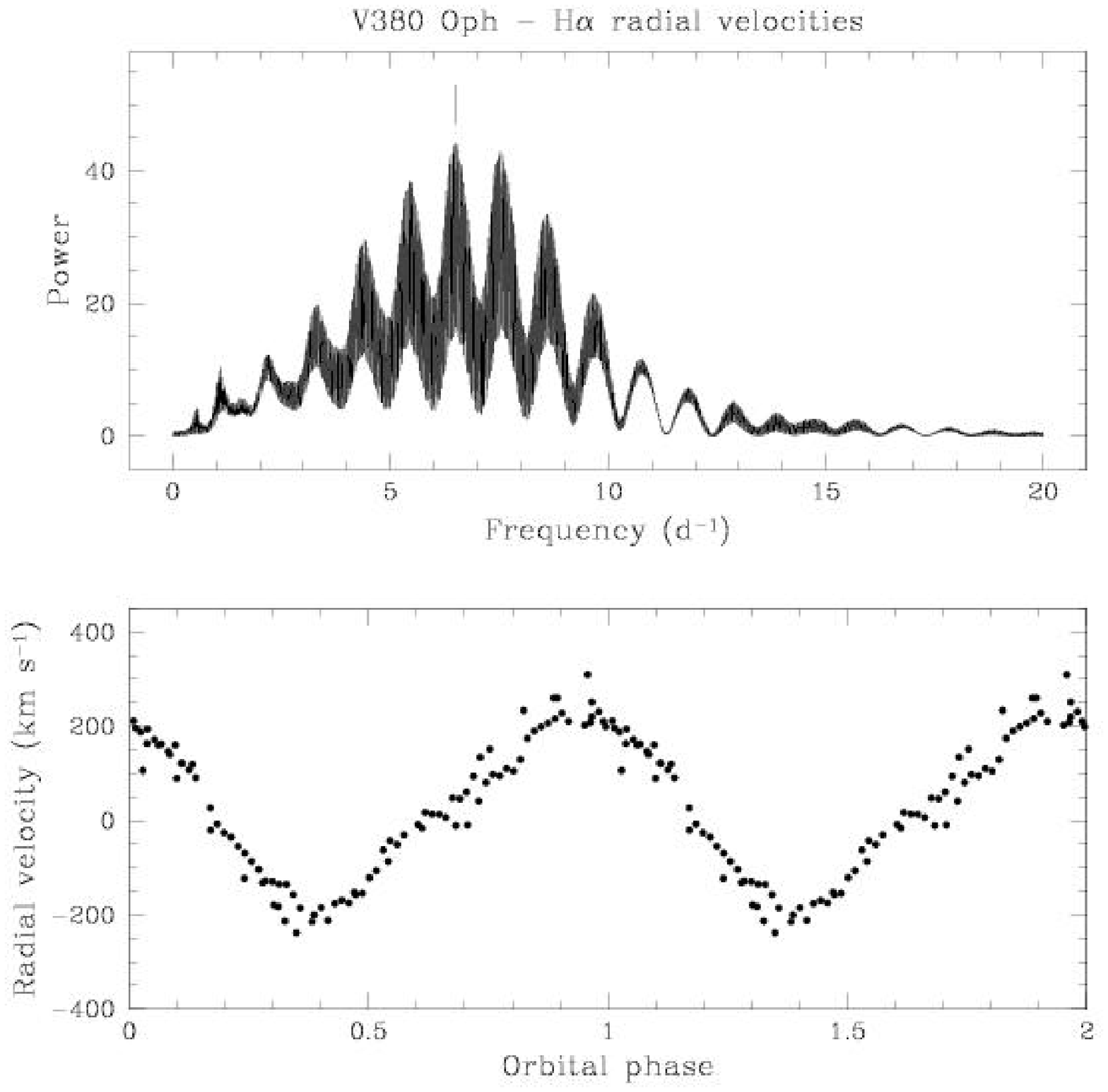,width=\columnwidth}}
\caption{\label{fig_v380oph_scargle} \textit{Top}: Scargle periodogram of the \Ha~radial velocity curves of \voph. The adopted orbital period of 0.154107\,d corresponds to the marked peak. \textit{Bottom}: phase-folded \Ha~radial velocity curve (no phase binning has been applied) showing a 0.15-cycle delay with respect to expected red-to-blue crossing time (see text for details on the assummed $T_0$). A full cycle has been repeated.}
\end{figure}

\begin{figure*}
\mbox{\epsfig{file=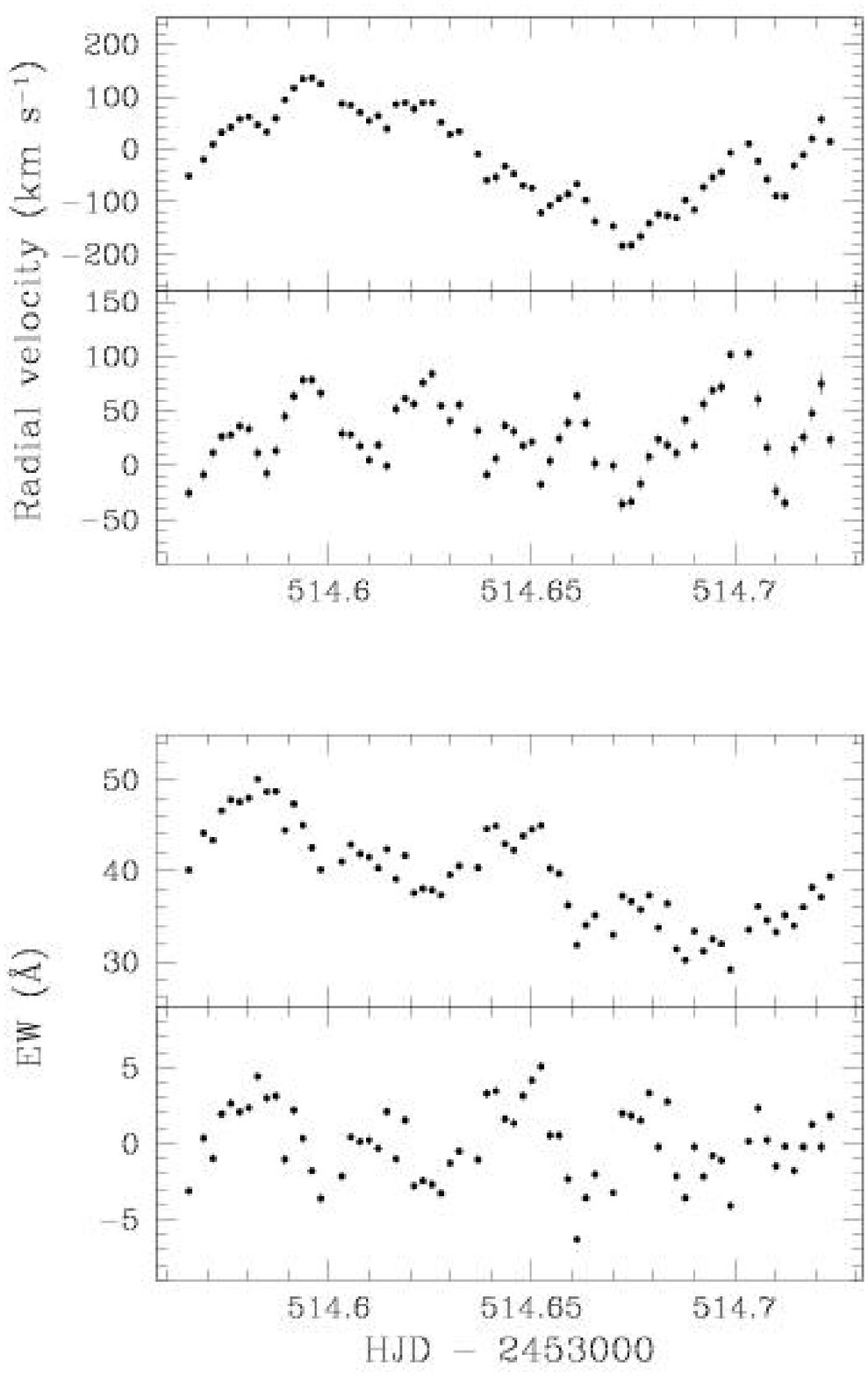,width=5.7cm}~~\epsfig{file=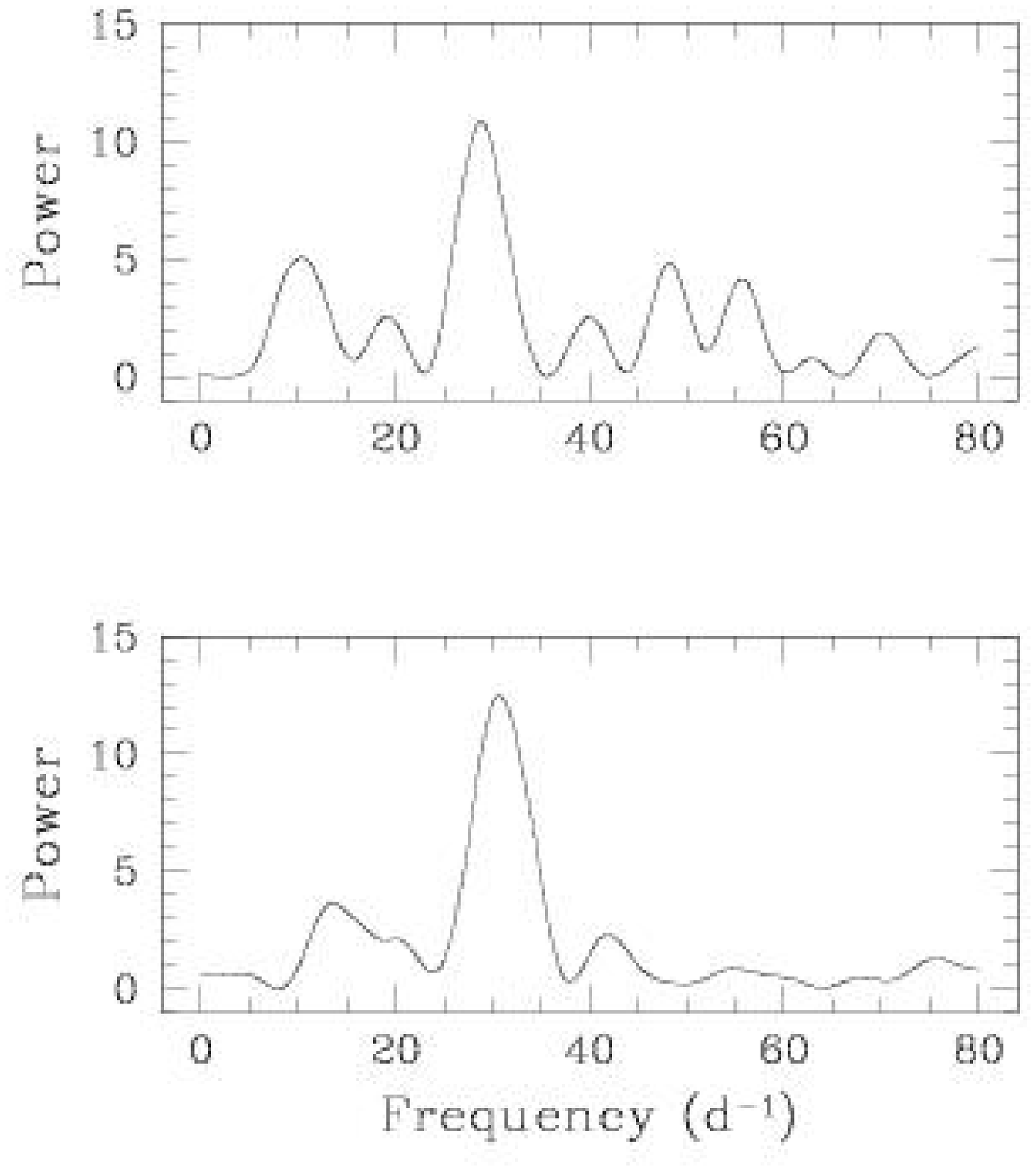,width=5.7cm}\epsfig{file=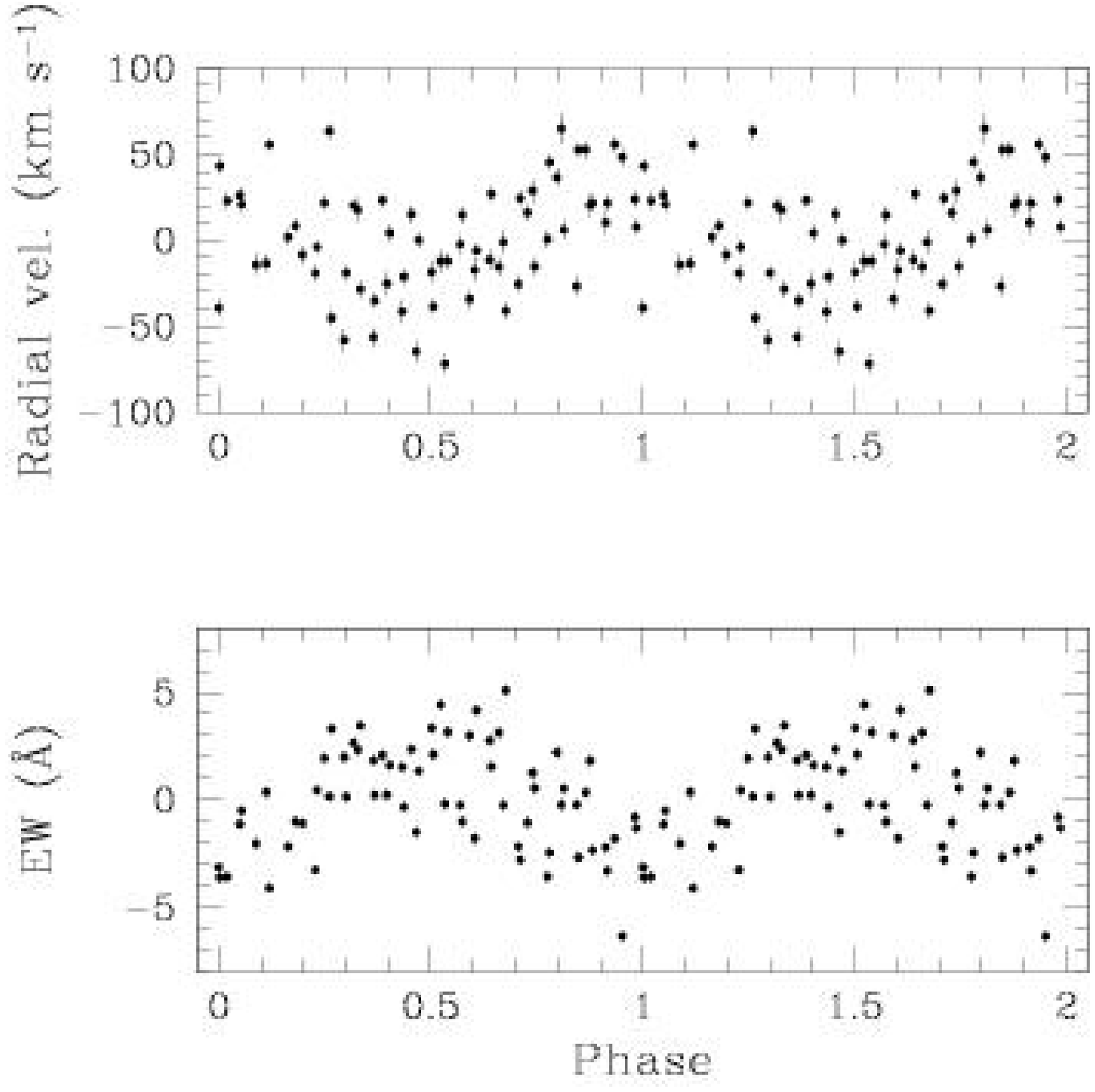,width=5.7cm}}
\caption{\label{fig_v380_flaring} \textit{Top left}: actual (\textit{top}) and detrended (\textit{bottom}) \Ha~radial velocity curves of \voph. \textit{Bottom left}: the same for the EWs. \textit{Middle panel}: Scargle periodograms from the detrended \Ha~radial velocities (\textit{top}) and EWs (\textit{bottom}). \textit{Right panel}: the detrended radial velocities (\textit{top}) and EWs (\textit{bottom}) folded on the 46.7-min period. The orbital cycle has been plotted twice.}
\end{figure*}

\subsection{V380\,Oph}

The first spectroscopic analysis of V380\,Oph showed the system to have
strong Balmer emission lines on top of a blue continuum 
\citep{shafter85-1}.
Shafter concluded that the system is a high mass transfer nova-like CV.
He also performed radial velocity studies and derived an orbital period
of $P_{\rm orb} = 3.8$\,h. Analysis of archival plates shows that the 
brightness of V380\,Oph occasionally drops by $\sim 2.5$\,mag reaching a photographic magnitude of $B_\mathrm{pg} \simeq 17$ \citep{shugarovetal05-1}. These low states seem to last for about one year and qualify \voph~as a VY\,Scl star. \citeauthor{shugarovetal05-1} also found two periodicities in their optical light curves: 0.148\,d and 4.5\,d. The former is quite close to the spectroscopic period reported by \citeauthor{shafter85-1} whilst the latter may be related to disc precession. 

The average WHT spectrum of \voph~is shown in
Fig.~\ref{fig_v380oph_flux}. It is dominated by strong, single-peaked
Balmer emission lines. Much weaker \he{i}, \hel{ii}{4686}, and the Bowen blend are also seen. The average NTT \Ha~profile (Fig.~\ref{spec_all}) has an EW of
41\,\AA~(Table~\ref{linetab}), about twice the value given by
\cite{shafter85-1}. Photometry made on our NTT acquisition images confirmed that the brightness of \voph~was almost the same as during Shafter's observations.

\subsubsection{\Ha~radial velocities and trailed spectrum}

The \Ha~radial velocity curves of \voph~were obtained by applying the
double Gaussian technique to both the NTT and WHT individual profiles,
adopting a Gaussian separation of 400\,\kms~and
$\mathrm{FWHM}=200$\,\kms. The Scargle periodogram
(Fig.~\ref{fig_v380oph_scargle}) shows the highest peak centred at
$\nu \simeq 6.49$\,d$^{-1}$, which corresponds to an orbital period of
$P=0.154107 \pm 0.000001$\,d ($=3.70\,\mathrm{h}$), where the
uncertainty comes from a sine fit to the velocities. Our period measurement
supersedes the less accurate value of 0.16\,d reported by
\cite{shafter85-1}. The
photometric period measured by \cite{shugarovetal05-1} is 3.8 per cent
\textit{shorter} than the orbital period, suggesting the occurrence of a
negative (nodal) superhump. As in the case of \ah, the detection of a
much longer period at $\sim 4$\,d strongly supports the presence of an
eccentric, wobbling accretion disc in \voph.

A preliminary trailed spectrum (not shown) computed from the phase-binned data (using the $T_0$ given by the above sine fit) shows a prominent, high-velocity S-wave with maximum blue velocity taking place at $\varphi_\mathrm{r} \sim 0.3$. Once more, as this emission feature should be bluest at $\varphi_\mathrm{a} \sim 0.5$ in the SW Sex stars, a phase delay of $\Delta \varphi \sim 0.2$ is suggested. After correcting for this offset we get a more realistic time of zero phase of $T_0(\mathrm{HJD})=2453514.61061 \pm 0.00008$. 

A new trailed spectrum is presented in Fig.~\ref{fig_trailed}. The line core is dominated by an emission S-wave with an amplitude of $\sim 200$\,\kms. In addition, there may be another similar amplitude, narrow S-wave which appears to reach maximum blue velocity at $\varphi_\mathrm{a} \sim 0.75$. If our $T_0$ is the correct one that emission component would be naturally placed on the donor star, presumably on its irradiated face. Unfortunately, it proved impossible to isolate this component in order to confirm the ephemeris. Besides, an EW curve of the line core (between $-300$\,\kms~and $+300$\,\kms) reveals a minimum at $\varphi_\mathrm{a} \simeq 0.5$, probably caused by the prototypical absorption that the SW\,Sex stars show at that orbital phase. On the other hand, a high-velocity S-wave reaching values as large as $-3000$\,\kms~produces the line wings.

\subsubsection{Emission-line flaring}

The intensity of the high-velocity S-wave does not appear to be uniform, but follows a humpy pattern that is clearly visible in the blue wing, where the S-wave is best seen. This bears close similarity with what is seen in \bo~and other SW\,Sex stars already mentioned in Sect.~\ref{sec_bocet}, and is likely caused by emission-line flaring. In order to probe for this possibility we computed a set of radial velocity curves of the \Ha~line for the longest data set (May 22), but this time choosing several Gaussian separations that entered the line wings. At a separation of 1000\,\kms~the radial velocity curve is still dominated by the orbital modulation but a variation at a much shorter time-scale ($\sim 50$\,min) is also seen (Fig.~\ref{fig_v380_flaring}). A similar oscillation is present in the EW curve of \Ha~obtained in the $(-1200,1200)$\,\kms~velocity interval (Fig.~\ref{fig_v380_flaring}).

With the aim of searching for periodicities we first detrended both the radial velocity and EW curves by subtracting a sine fit, and then computed Scargle periodograms. Both power spectra are dominated by a strong peak centred at $\sim 31$\,d$^{-1}$, which corresponds to a period of $\sim 47$\,min. A Gaussian fit to the peak in the EW periodogram gives $46.7 \pm 0.1$\,min (the error comes from a sine fit to the EW curve). At this point we have to rise a word of caution. The velocity and EW curves presented in Fig.~\ref{fig_v380_flaring} span slightly more than an orbital cycle, hence, they contain only five 47-min cycles. This alone is not clear evidence of a coherent oscillation, but the fact that the same periodicity is observed in both the velocities and EWs supports this hypothesis. A much larger data set will be needed to confirm or reject the 47-min periodicity. Unfortunately, the time resolution of our 1.1-cycle long NTT data is insufficient to sample it properly.

\subsubsection{Doppler tomography}

The \Ha~Doppler map of \voph~is presented in Fig.~\ref{fig_doppler}. The \Ha~emission is mainly concentrated in the lower-left quadrant of the tomogram, around $(V_x \sim -250,V_y \sim -100)$~\kms, a clear signature of the SW Sex stars.

\subsubsection{SW Sex class membership}

\voph~displays all the characteristic features described in Sect.~\ref{sec_ahmen_swsex} for \ah. Therefore, it fully qualifies as a SW Sex star. Apart from that, the radial velocities and the EWs show rapid variations with a likely periodicity of 46.7 min, which can be related to the rotation of a magnetic white dwarf.



\begin{figure}
\mbox{\epsfig{file=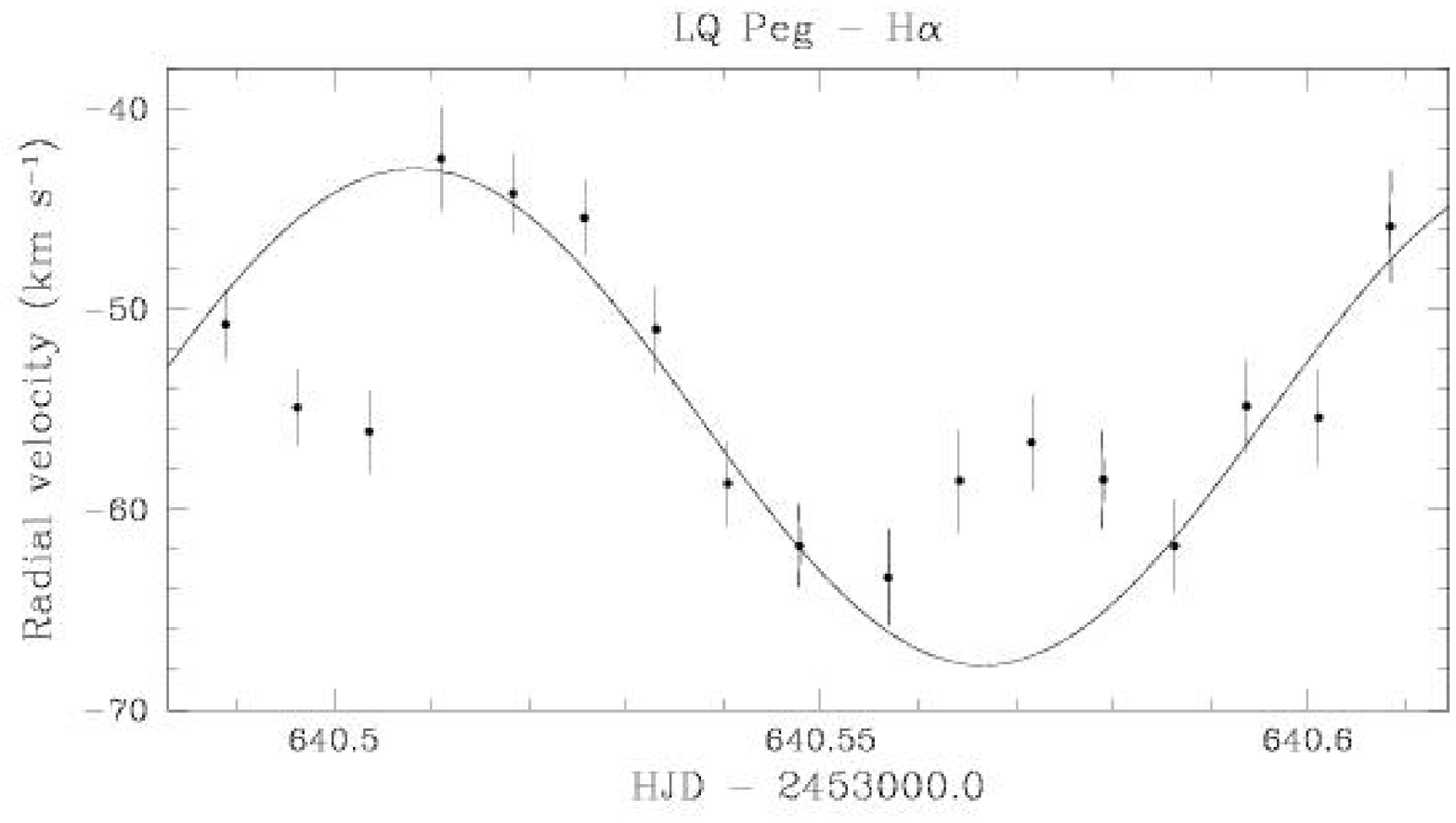,width=\columnwidth}}
\caption{\label{fig_lqpeg_rvc} \Ha~radial velocity curve of \lqpeg. The solid line is the best sine fit to the data. The two points flanking 640.5 and three points about 640.52 have been excluded from the fit.}
\end{figure}

\subsection{LQ\,Peg}

\lqpeg~($=$~PG\,2133$+$115 $=$ Peg\,6) was found as a peculiar absorption-line object in the Palomar--Green survey \citep{greenetal86-1}, and was classified as a CV of possible UX\,UMa type from follow--up
optical and ultraviolet \textit{IUE} spectroscopy (Ferguson, Green \& Liebert \citeyear{fergusonetal84-1}). \cite{ringwald93-2}
derived an uncertain orbital period of 2.9\,h from radial velocity measurements. The optical spectrum of \lqpeg~appears to be highly variable, switching from weak Balmer emission superimposed on broad absorptions to pure Balmer and \hel{ii}{4686} emission profiles (see the spectra shown in \citealt{fergusonetal84-1} and \citealt{papadakietal06-1}).

Photometrically, \lqpeg~undergoes deep ($\sim 3-4$\,mag), VY\,Scl-like low states \citep{sokolovetal96-1,watanabe99-1,kato+uemura99-3,honeycutt+kafka04-1,kafka+honeycutt05-1}, which drive the system down to $V \sim 18$. Additionally, time resolved photometric observations in both the high and low states show no coherent oscillations apart from strong flickering \citep{misselt+shafter95-1,schmidtkeetal02-1}. \cite{papadakietal06-1} found a 2.99-h wave in the optical light curve spanning about 50 hours spread over three months. Whether this modulation is the actual orbital period or a positive (or negative) superhump has still to be investigated. There might also be a QPO at $\sim 30$\,min, but no firm conclusion could be addressed.

\subsubsection{\Ha~radial velocities, trailed spectrum, and Doppler tomogram}

The average of all the spectra we obtained at the NTT (Fig.~\ref{spec_all}) shows very narrow \Ha~($\mathrm{FWHM} \simeq 310$\,\kms) and \hel{i}{6678} ($\mathrm{FWHM} \simeq 210$\,\kms) lines purely in emission, indicating a low orbital inclination. The \Ha~radial velocities were measured by applying the double-Gaussian method with a separation of 200\,\kms~and $\mathrm{FWHM}=200$\,\kms, which yielded the curve presented in Fig.~\ref{fig_lqpeg_rvc}. A sine fit to the velocities gives a period of 2.8\,h (that is highly unreliable) and an amplitude of only $\sim 12$ \kms. 

The \Ha~trailed spectrum (Fig.~\ref{fig_trailed}) does not provide any useful information either, as expected from the small amplitude of the radial velocity curve. On the other hand, the \Ha~EWs do not show periodic rapid variations. In the \Ha~Doppler tomogram (Fig.~\ref{fig_doppler}), the lack of resolved structure in the line profiles translates into a single emission spot centred on the origin. It is therefore apparent that higher resolution spectroscopic data are needed to eventually obtain the actual orbital period. Until then no speculation about the nature of the photometric period should be made.


\subsection{AH\,Pic}

\ahp~($=$ EC\,05565--5934) was identified as a CV in the
Edinburgh--Cape Survey by \cite{chenetal01-1}. They classified the
system as a UX\,UMa-type star, based on several blue spectra which at
one epoch showed the Balmer lines in weak and shallow absorption,
while two years later they were present in weak emission. \citeauthor{chenetal01-1} also performed time resolved photometry of the system and found
a clear modulation at $3.40 \pm 0.04$\,h, which they proposed as the orbital
period. In addition, the light curves also displayed rapid variations
with a time-scale of $10-20$\,min

\subsubsection{\Ha~radial velocities and trailed spectrum}

In our average spectrum (Fig.~\ref{spec_all}), the \Ha~line is clearly
in emission, although a shallow broad absorption is also seen when the
flux scale is properly adjusted. The \Ha~emission shows a
single-peaked profile with $\mathrm{FWHM} \simeq 760$\,\kms, typical
for nova-likes viewed at intermediate to low inclination.

We measured the \Ha~radial velocities by applying the double-Gaussian technique (1200\,\kms~separation, $\mathrm{FWHM}=150$\,\kms). A sine fit provided an orbital period of $P_\mathrm{orb} = 0.146 \pm 0.04$\,d. An initial \Ha~trailed spectrogram with relative orbital phases computed using the $T_0$ from the radial velocity curve shows several emission S-waves.  A close inspection of the individual profiles reveals a multi-component structure with a broad emission component giving rise to line wings extending up to $\sim \pm 800$\,\kms, and probably two other components moving closer to the line core which produce double peaks with changing relative intensity. The component with smallest velocity amplitude has maximum intensity at $\varphi_\mathrm{r} \sim 0.5$. This suggests a likely origin in the irradiated hemisphere of the donor star. In order to obtain absolute phases we isolated this component by applying the usual method and obtained $T_0(\mathrm{HJD})=2453478.547 \pm 0.001$ and $K_\mathrm{irr}=184 \pm 3$~\kms. 

\begin{figure}
\mbox{\epsfig{file=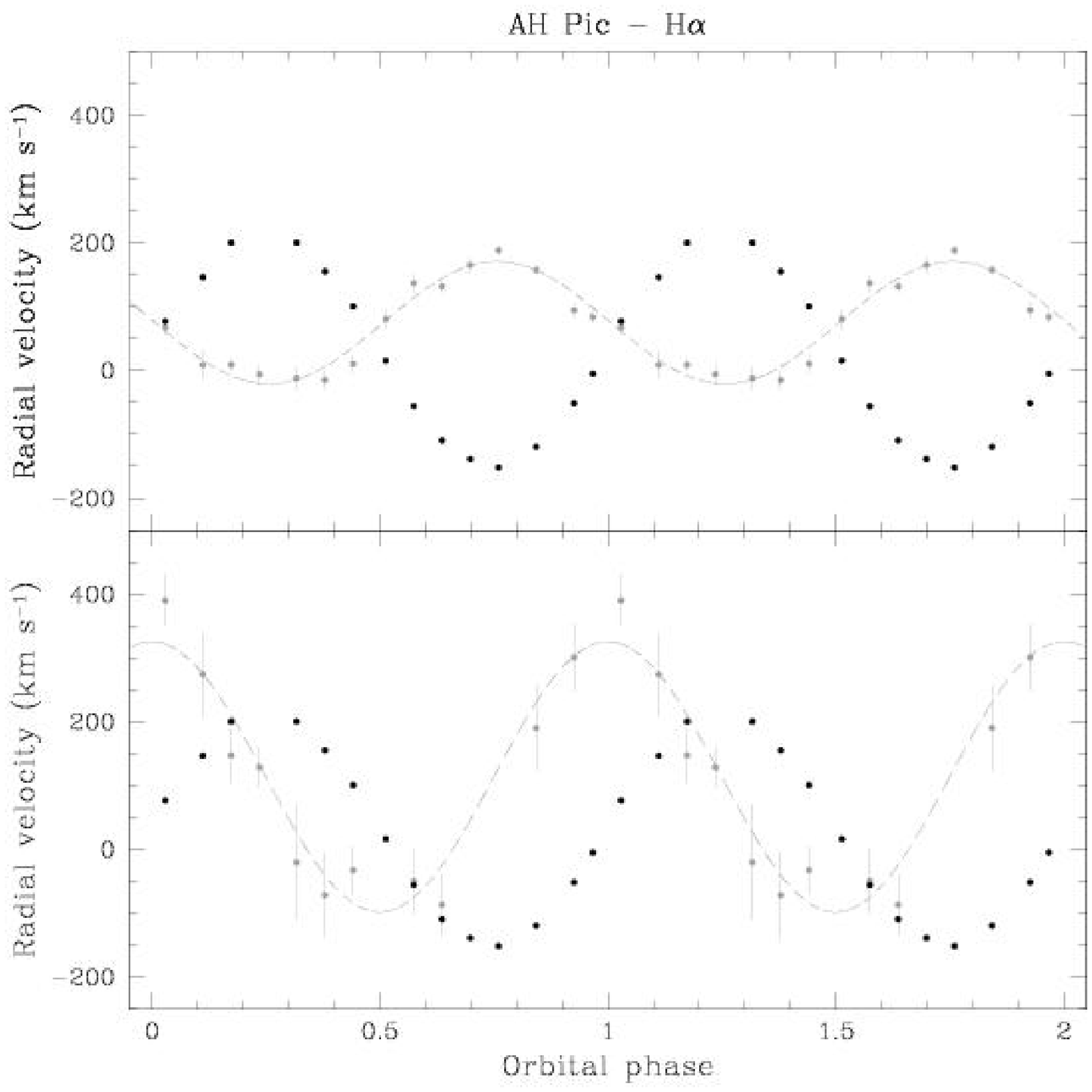,width=\columnwidth}}
\caption{\label{fig_ahpic_rvc} \textit{Top}: the \Ha~radial velocity curve of \ahp~computed by cross-correlation with a double-Gaussian template (gray; 1200\,\kms~separation, $\mathrm{FWHM}=150$\,\kms), and best sine fit (gray dashed curve). Black dots are the \Ha~radial velocity curve of the likely donor emission. A deviant point has been omitted. \textit{Bottom}: the same but with a Gaussian separation of 2000~\kms~(gray). A 0.25-orbital cycle delay with respect to the motion of the white dwarf is apparent. The orbital cycle has been plotted twice.}
\end{figure}

With the aim of searching for phase offsets, we constructed a new radial velocity curve of \Ha~by using the double-Gaussian technique, but going further into the line wings (2000\,\kms~separation, $\mathrm{FWHM}=150$\,\kms). All the radial velocity curves are presented in Fig.~\ref{fig_ahpic_rvc}. The wing velocities clearly lag the motion of the white dwarf by $\sim 0.25$ orbital cycle, which is a clear indication of a SW Sex nature.

A new \Ha~trailed spectrum with absolute phases is shown in Fig.~\ref{fig_trailed}. There are two emission components clearly moving in counter phase. One is located at the central part of the line, which peaks at a velocity of $\sim 180$\,\kms~at $\varphi_\mathrm{a} = 0.25$ (we have assumed an origin on the heated donor star), whilst the other is peaking at velocities of about 300\,\kms~at $\varphi_\mathrm{a} \simeq 0.8$. A third, much broader emission S-wave is responsible for the extended wings up to $\pm 800$\,\kms~and for the delayed velocities. Finally, the EWs of the \Ha~line have two minima, one at $\varphi_\mathrm{a} \simeq 0$ and a shallower one (likely filled by the maximum emission from the donor star) at $\varphi_\mathrm{a} \simeq 0.5$. As in the case of \hl, the trailed spectrum of \ahp~shows enhanced blueshifted absorption at $\varphi_\mathrm{a} \sim 0.4-0.5$.

\subsubsection{Doppler tomography}

The \Ha~Doppler tomogram of \ahp~can be seen in Fig.~\ref{fig_doppler}. Two emission sites are clearly seen. One is located at the expected position of the donor star, and the other at $(V_x \sim 0,V_y \sim +180)$~\kms. This is obvious, as the S-wave which produces this spot is assumed to come from the heated face of the secondary star. However, this assumption is supported by two observations: (i) the presence of another emission spot in the lower left quadrant of the Doppler map, and (ii) the 0.25-cycle phase delay seen in the radial velocity curve of the line wings. These are characteristic features of the SW Sex stars.

\subsubsection{Orbital inclination}

Proceeding in the same way as with other systems with apparent emission from the donor star already presented, we can place limits to the orbital inclination from estimates of the $K$-correction. For \ahp~we assume $M_1 = 0.75~\mathrm{M_\odot}$ and $M_2 = 0.24~\mathrm{M_\odot}$ and, therefore, $q=0.32$. From a sine fit to the donor emission radial velocities we obtain $K_\mathrm{irr} = 184 \pm 3$~\kms. Hence, the radial velocity amplitude of the donor star ranges between $207 < K_2 < 293$~\kms, and the orbital inclination between $25\degr < i < 37\degr$.

\subsubsection{SW Sex class membership}

Apart from the emission in the lower left quadrant of the tomogram mentioned above, \ahp~shows many other features typical of the SW Sex stars. The \Ha~emission has single-peaked profiles and shows a high-velocity S-wave. Contrary to the rule, this S-wave has its maximum blueshift at $\varphi_\mathrm{a} \sim 0.3$, not at $\sim 0.5$ as can be expected (the same phasing is observed in \hl). Instead, there is blueshifted absorption at $\varphi_\mathrm{a} \sim 0.5$, as in the case of \hl. The similarities with \hl~ are not unexpected, as the estimated orbital inclinations are very close. The deeper absorption seen in \hl~may therefore be a consequence of its lower inclination. Finally, the radial velocity curve measured in the line wings lags the motion of the white dwarf by 0.25 cycle. We conclude that \ahp~is very likely a SW Sex star.


\subsection{V992\,Sco}

V992\,Sco (Nova Scorpii 1992) was discovered as a possible nova in outburst by 
\cite{camilleri92-1}, and was spectroscopically confirmed by
\cite{dellavalle+smette92-1}. Although it was a slow nova, high-speed 
photometry performed 10 years after outburst showed that the nova had faded 
by about 10\,mag and revealed a fundamental period of 0.154\,d \citep[$=3.686$\,h,][]{woudt+warner03-1}. The inclination of the system was estimated as intermediate, probably $\sim 40\degr$.

\subsubsection{\Ha~radial velocities and trailed spectrum}

\begin{figure}
\mbox{\epsfig{file=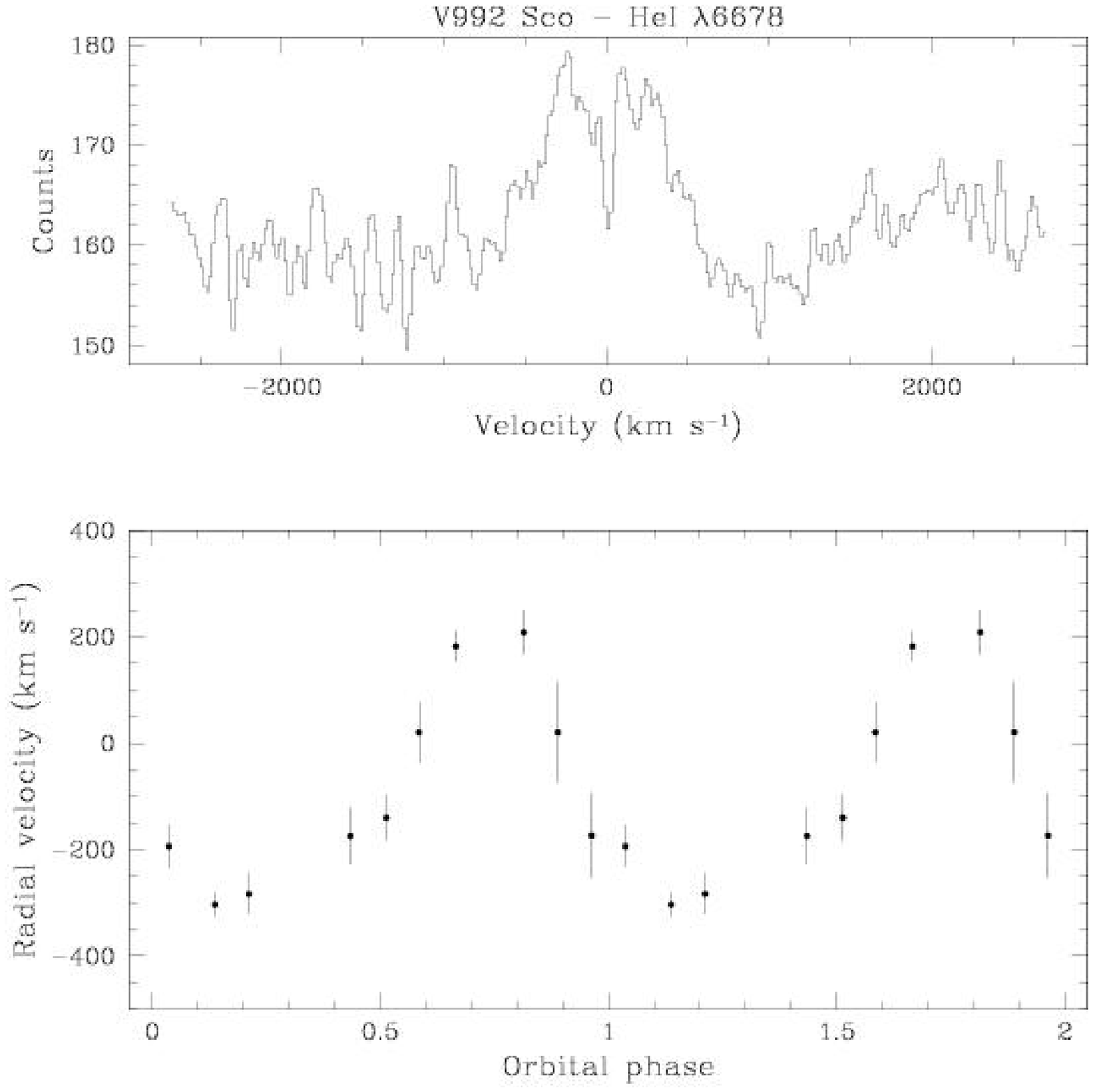,width=\columnwidth}}
\caption{\label{fig_v992sco_rvc} \textit{Top panel}: the \hel{i}{6678} average line profile of V992\,Sco. Prior to averaging, the individual spectra were smoothed by using a 1\,pixel~(FWHM) Gaussian template. A double peak may be present. \textit{Bottom panel}: the \hel{i}{6678} radial velocity curve of V992\,Sco computed by cross-correlation with a single Gaussian template ($\mathrm{FWHM}=400$\,\kms). Orbital phases are relative. A whole orbital cycle has been repeated for clarity.}
\end{figure}

Our spectrum of V992\,Sco (Fig.~\ref{spec_all}) was taken 13 years
after the nova explosion. It shows a strong, multi-peaked \Ha~emission
line. It is therefore apparent that even though the binary is
dominating in the $V$-band light curve, the nova-shell is still
the dominating \Ha~emission source. However, in order to
spectroscopically confirm \citeauthor{woudt+warner03-1}'s orbital
period we tried to find radial velocity variations in the much weaker
\hel{i}{6678} emission line (Fig.~\ref{fig_v992sco_rvc}, top panel). The radial
velocities were measured by cross-correlation with a single Gaussian template
($\mathrm{FWHM}=400$\,\kms). The period of the best fitting sine is
0.147\,d, in agreement with the photometric result. The \hel{i}{6678}
radial velocities folded on \citeauthor{woudt+warner03-1}'s period are
shown in Fig.~\ref{fig_v992sco_rvc}. Our results indicate that the
nova shell, which dominates the \Ha~emission, makes only a small contribution to \hel{i}{6678}.

The \Ha~trailed spectrum can be seen in Fig.~\ref{fig_trailed}. That of \hel{i}{6678} has not been included because of its poor quality. The analisys of better signal-to-noise ratio \hel{i}{6678} spectra can eventually provide an accurate orbital period for V992 Sco.

\subsection{LN\,UMa}

\begin{figure}
\mbox{\epsfig{file=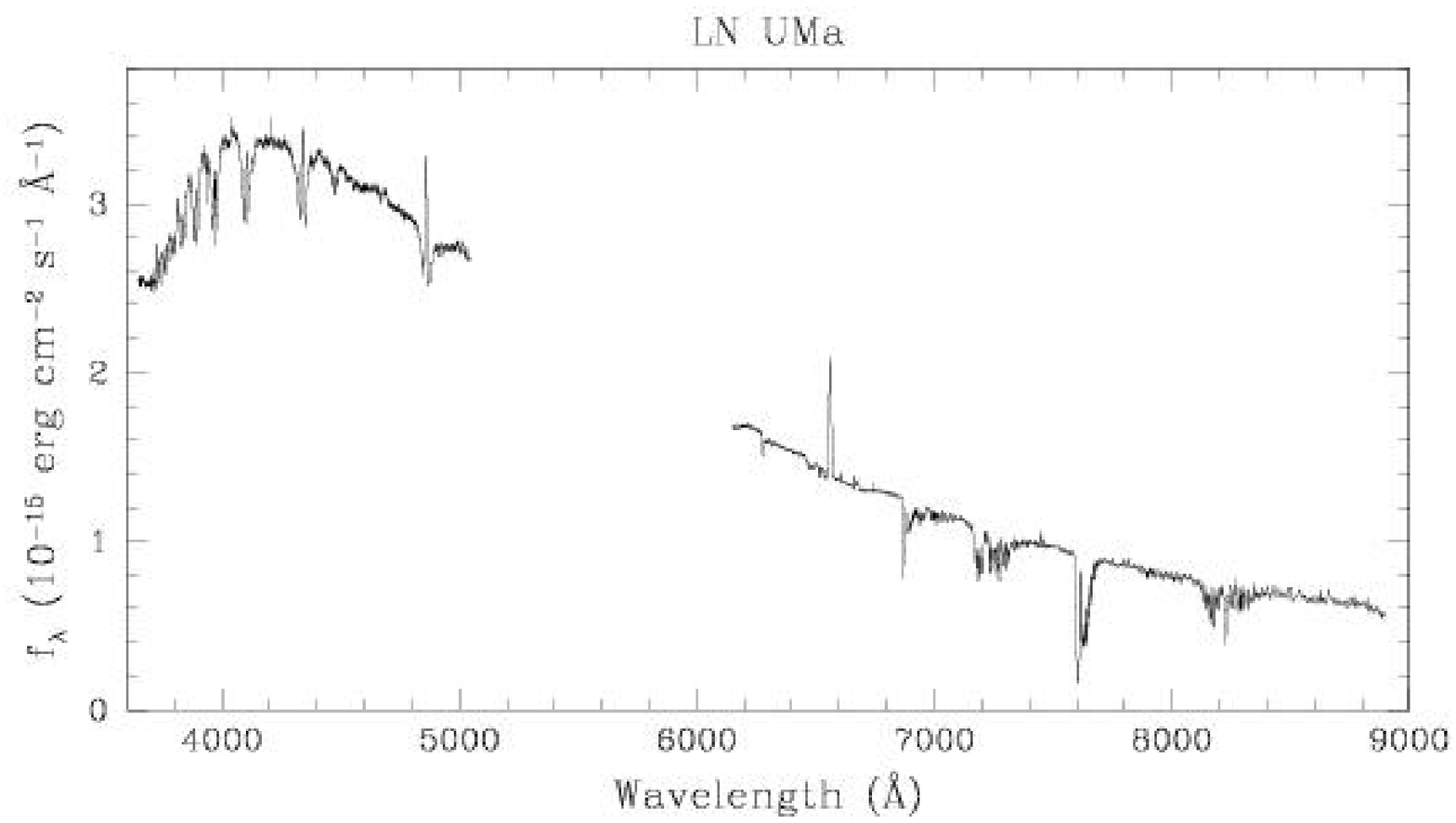,width=\columnwidth}}
\caption{\label{fig_lnuma_flux} WHT flux-calibrated, average spectrum of \lnu. No telluric absorption correction has been applied.}
\end{figure}

As many other nova-like CVs, \lnu~($=$ PG\,1000$+$667 $=$ UMa\,7) was
found in the Palomar--Green survey \citep{greenetal86-1}. The first
spectroscopic analysis was carried out by \cite{ringwald93-2}, who
first identified \lnu~as a nova-like showing narrow Balmer emission
lines with a probable orbital period of 4.06\,h. A more detailed
spectroscopic study by Hillwig, Robertson \& Honeycutt (\citeyear{hillwigetal98-1}) revealed a narrow
($\mathrm{FWHM} = 490$\,\kms) \Hb~emission line superimposed on a much
broader absorption. Their \Hb~radial velocity curve yielded an orbital
period of $0.1444 \pm 0.0001$\,d ($=3.465 \pm 0.002$\,h). In addition,
long term photometric monitoring
\citep{hillwigetal98-1,honeycutt+kafka04-1} showed \lnu~to be a
VY\,Scl star which undergoes $\sim 3$-mag fadings from its usual
brightness at $V \simeq 15.0$.

\subsubsection{\Ha~radial velocities and trailed spectrum}

\begin{figure}
\mbox{\epsfig{file=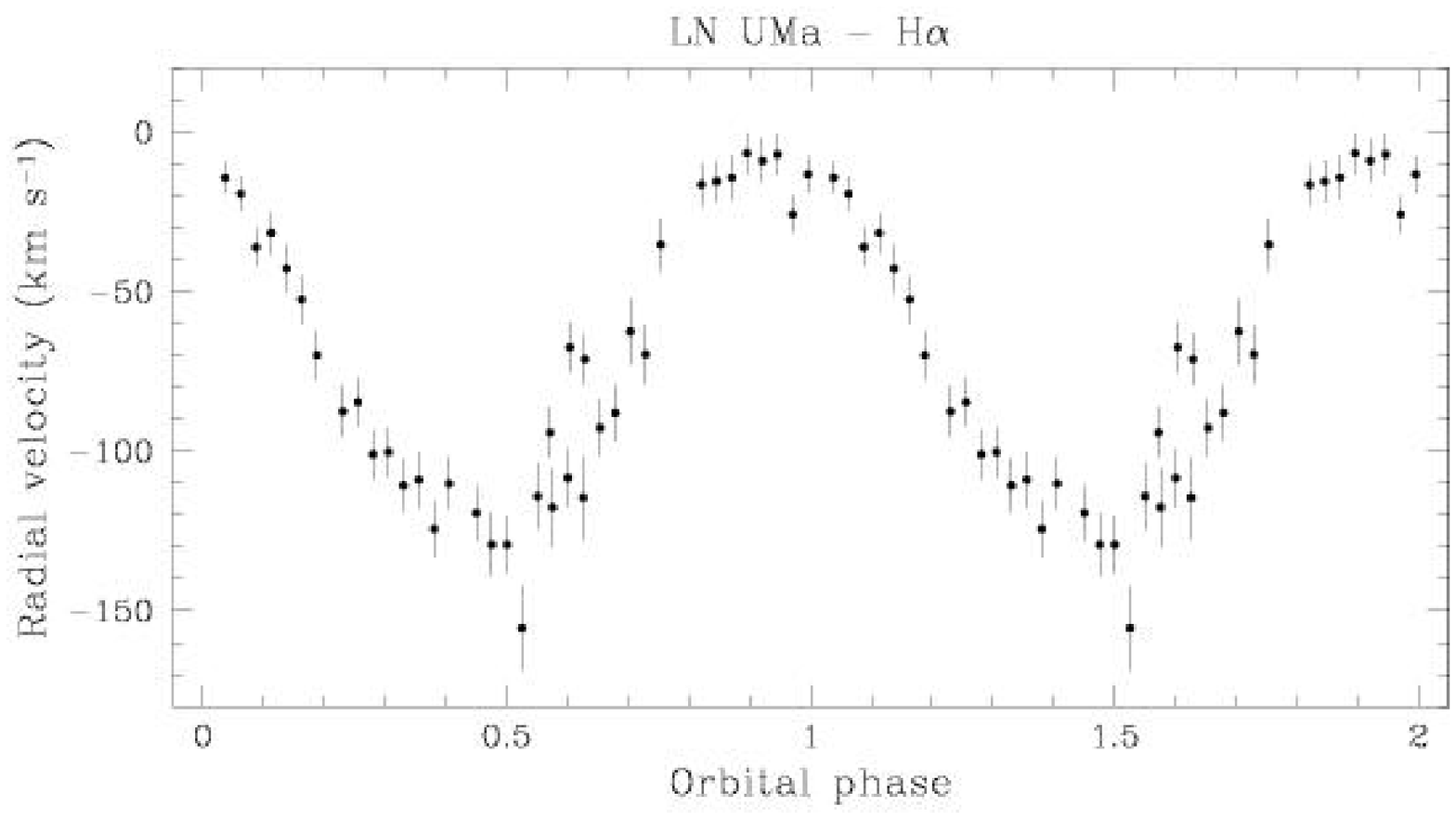,width=\columnwidth}}
\caption{\label{fig_lnuma_rvc} \Ha~radial velocity curve of \lnu~computed by cross-correlation with a double-Gaussian template (800\,\kms~separation, $\mathrm{FWHM}=200$\,\kms). The velocities lag the motion of the white dwarf by $\sim 0.2$ cycle. A whole orbital cycle has been repeated for clarity.}
\end{figure}

Our average spectrum (Fig.~\ref{fig_lnuma_flux}) displays narrow, single-peaked emission lines of the Balmer series superimposed on broad absorptions, with the exception of \Ha. Very much weaker \he{i} emission lines as well as \hel{ii}{4686} and the Bowen blend are also observed. The average \Ha~emission profile (Fig.~\ref{spec_all}) has $\mathrm{FWHM} = 510$\,\kms~and $\mathrm{EW}=6$\,\AA. We constructed the radial velocity curve of the \Ha~line by using the double-Gaussian method ($\mathrm{FWHM}=200$\,\kms) with a separation of 800\,\kms, which is presented in Fig.~\ref{fig_lnuma_rvc}. A sine fit gives a period of $0.1448 \pm 0.0008$\,d, consistent with \citeauthor{hillwigetal98-1}'s findings. We will adopt their $P_\mathrm{orb}=0.1444$\,d as they benefited from a longer spectroscopic data base.

A preliminary \Ha~trailed spectrum with orbital phases calculated using the $T_0$ from the above sine fit shows a high-velocity emission S-wave and a flux deficit close to the S--wave's bluest excursion
($\sim -700$\,\kms). Similarly to other systems already discussed in
this paper, this component reaches maximum blue velocity at $\varphi_\mathrm{r}
\sim 0.25$ when using this phase convention. If SW\,Sex-related at all it should do at $\varphi_\mathrm{a} \sim 0.5$, so we corrected the initial $T_0$ accordingly. With the new $T_0$ (Table~\ref{t-rvcfits}), the radial velocity curve of the \Ha~wings show a $\Delta \varphi \sim 0.2$ delay with respect to the actual time of zero phase. 

\begin{table*}
\centering
 \begin{minipage}{140mm}
\caption{\label{tab_summary} Summary table of SW Sex classification.}
 \begin{tabular}{@{}lccccccc@{}}
  \hline
   System     & Single & High-velocity  & 0.5-absorption & Phase & Doppler & Line    & SW Sex?\\
              & peaks  & S-wave         &                & shift &         & flaring &  \\
 \hline
HL\,Aqr   & \yes  & \yes$^1$ & \no  & \yes & \yes    & \no  & \yes \\
BO\,Cet   & \yes  & \yes     & \yes & \yes & \yes    & \yes & \yes \\ 
V849\,Her & \yes  & \no      & \no  & \no  & \no     & \no  & \no  \\ 
V393\,Hya & \no   & \no      & \no  & \no  & \no     & \no  & \no  \\ 
AH\,Men   & \yes  & \yes     & \yes & \yes & \yes    & \no  & \yes \\ 
V380\,Oph & \yes  & \yes     & \yes & \yes & \yes    & \yes & \yes \\ 
LQ\,Peg   & \yes  & \no      & \no  & \no  & \no     & \no  & \no  \\ 
AH\,Pic   & \yes  & \yes$^1$ & \no  & \yes & \yes    & \no  & \yes \\ 
V992\,Sco &  ---  & ---      & ---  & ---  & ---     & ---  &  --- \\ 
LN\,UMa   & \yes  & \yes     & \yes & \yes & \no$^2$ & \no  & \yes \\ 
\hline
\end{tabular}
\begin{minipage}{10cm}
$^1)$ In absorption.\\
$^2)$ Better spectral resolution needed.\\
\end{minipage}
\end{minipage}
\end{table*}

\subsubsection{Doppler tomography}

The \Ha~Doppler tomogram of \lnu~(Fig.~\ref{fig_doppler}) only shows an emission blob approximately centred on the origin. This, together with the small radial velocity amplitude observed $\sim 60$~\kms, suggests the need for data with much better spectral resolution in order to resolve the emission features on a Doppler map.

\subsubsection{SW Sex class membership}

The \Ha~emission line of \lnu~has a single-peaked profile throughout the whole orbit. It also shows a high-velocity S-wave with maximum blueshift at $\varphi_\mathrm{a} \sim 0.5$, and displays an absorption component crossing the line core from red to blue at the same instant. Moreover, the \Ha~line wings lag the expected instant of zero phase by $\sim 0.2$ cycle. Alas, the spectral resolution prevented us from resolving the emission components on a Doppler tomogram. \lnu~is therefore classified as a new SW Sex star.

In Table~\ref{tab_summary} we summarise the SW Sex classification criteria for all the systems presented in this paper.

\section{The impact of SW Sex stars on the nova-like population}

Remarkably, 60 per cent of the nova-likes presented in this paper are SW Sex stars, namely, \hl, \bo, \ah, \voph, \ahp, and \lnu. Spectroscopic data with better signal-to-noise ratio and/or better spectral resolution are needed to address any conclusion on the remaining three systems. Our findings increase the number of known SW Sex stars to 35 (see \citealt{rodriguez-giletal07-1} for a list), 37 per cent of which are non-eclipsers. It is therefore increasingly clear that the old requirement for eclipses is actually due to a selection effect. Hence, scenarios which rely on a high inclination, like those invoking the 0.5-absorption caused by material in the $\mathrm{L_3}$ point \citep{honeycuttetal86-1}, a disc-anchored magnetic propeller \citep{horne99-1}, and self-obscuration of the inner disc by a flared outer disc \citep{kniggeetal00-1} have to be revised. In fact, as our results on \hl~show, 
the orbital inclination may play a fundamental role in the spectroscopic behaviour of the SW Sex stars, with the lower inclination systems looking rather different than the higher inclination ones.

\begin{figure}
\mbox{\epsfig{file=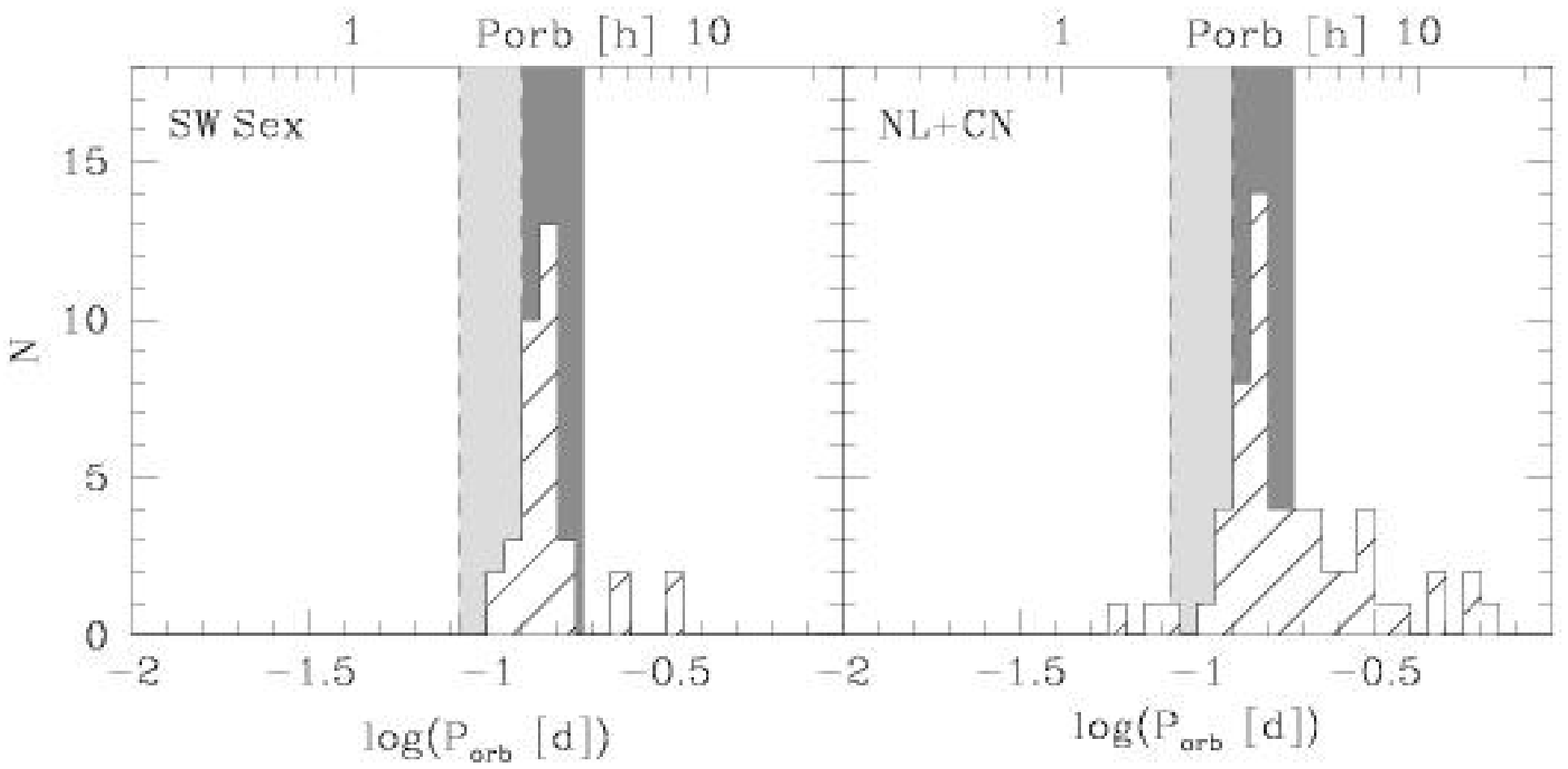,width=\columnwidth}}
\caption{\label{fig_swsex_porb} Period distributions of confirmed SW\,Sex stars (left panel) and nova-likes and classical novae that are not known to exhibit SW\,Sex behaviour so far.}
\end{figure}

The orbital period distributions of the known SW Sex stars and the non-SW Sex nova-likes and classical novae are presented in Fig.~\ref{fig_swsex_porb}. The latter was constructed from the Ritter \citep{ritter+kolb03-1}, CVcat \citep{kubeetal03-1}, and Downes \citep{downesetal05-1} catalogues, and only robust orbital period determinations were included. Remarkably, close to
40 per cent (35 out of 93) of the whole nova-like/classical nova
population are SW Sex stars. Moreover, the impact of SW Sex stars on this population can reach 50 per cent if a number of SW Sex candidates are confirmed as members of the class. In the narrow $\sim 3-4$\,h orbital period range the situation is even more striking: almost \textit{half} of the inhabitants are SW Sex stars (26 out of 53). This fraction is likely bound to increase as more SW Sex stars are found in this orbital period interval, as the similarity of the SW Sex period distribution with that of non-SW Sex nova-likes/classical novae---which also shows a spike in the $\sim 3-4$\,h range---suggests.

This pile-up of SW Sex stars is difficult to reconcile with the standard CV evolution theory. If the magnetic nature of these systems \citep{rodriguez-giletal01-1,hameury+lasota02-1} is confirmed, then the majority of CVs close to the upper boundary of the period gap will be magnetic. This is clearly at odds with the majority of CVs below the gap being non-magnetic. On the other hand, \cite{podsiadlowskietal03-1} suggested that the population of CVs longward of $P_\mathrm{orb} \simeq 4.7$\,h is dominated by systems with evolved donors. A direct consequence of this is that the dominance of SW Sex stars among CVs with unevolved secondary stars above the gap would be even higher. Yet this is very speculative as no direct measurement of the stellar masses (i.e. a full dynamical solution for both masses) in any SW Sex star has been achieved so far, which is at the same time a weakness in the $M_2-P_\mathrm{orb}$ sequence of \cite{knigge06-1} as it lacks robust dynamical $M_2$ determinations for the numerous SW Sex class. It is therefore apparent that the study of the increasing family of the SW Sex stars, with special emphasis on the determination of the stellar masses involved, is instrumental for the current theory of CV evolution \citep[see][~for a more detailed discussion]{rodriguez-giletal07-1}.

\section*{Acknowledgments}
We thank the anonymous referee for valuable comments. PRG thanks the European Southern Observatory for granting a stay within their Visiting Scientist programme. BTG was supported by a PPARC Advanced Fellowship. The use of the {\sc molly}, {\sc doppler}, and {\sc trailer} packages developed by Tom Marsh is acknowledged. We also acknowledge the use of the Simbad database operated at CDS, Strasbourg, France.
Based in part on observations collected at the European Southern Observatory, La Silla, Chile, and
on observations made with the William Herschel Telescope, which is
operated on the island of La Palma by the Isaac Newton Group in the
Spanish Observatorio del Roque de los Muchachos of the Instituto de
Astrof\'\i sica de Canarias (IAC). 

\bibliographystyle{mn2e}
\bibliography{mn-jour,aabib}

\begin{thebibliography}{}

\bibitem[\protect\citeauthoryear{{Aungwerojwit}, {G{\"a}nsicke},
  {Rodr{\'{\i}}guez-Gil}, {Hagen}, {Harlaftis}, {Papadimitriou}, {Lehto},
  {Araujo-Betancor}, {Heber}, {Fried}, {Engels} \& {Katajainen}}{{Aungwerojwit}
  et~al.}{2005}]{aungwerojwitetal05-1}
{Aungwerojwit} A., et al.,  2005, A\&A, 443, 995

\bibitem[\protect\citeauthoryear{{Buckley}, {Cropper}, {Ramsay} \&
  {Wickramasinghe}}{{Buckley} et~al.}{1998}]{buckleyetal98-1}
{Buckley} D. A.~H.,  {Cropper} M.,  {Ramsay} G.,    {Wickramasinghe} D.~T.,
  1998, MNRAS, 299, 83

\bibitem[\protect\citeauthoryear{{Buckley}, {Warner}, {Remillard}, {Tuohy} \&
  {Sullivan}}{{Buckley} et~al.}{1993}]{buckleyetal93-1}
{Buckley} D. A.~H.,  {Warner} B.,  {Remillard} R.~A.,  {Tuohy} I.~R.,
  {Sullivan} D.~J.,  1993, MNRAS, 265, 926

\bibitem[\protect\citeauthoryear{{Camilleri}}{{Camilleri}}{1992}]{camilleri92-%
1}
{Camilleri} P.,  1992, IAU Circ., 5526, 1

\bibitem[\protect\citeauthoryear{{Casares}, {Mart\'\i nez-Pais}, {Marsh},
  {Charles} \& {L\'azaro}}{{Casares} et~al.}{1996}]{casaresetal96-1}
{Casares} J.,  {Mart\'\i nez-Pais} I.~G.,  {Marsh} T.~R.,  {Charles} P.~A.,
  {L\'azaro} C.,  1996, MNRAS, 278, 219

\bibitem[\protect\citeauthoryear{{Chen}, {O'Donoghue}, {Stobie}, {Kilkenny} \&
  {Warner}}{{Chen} et~al.}{2001}]{chenetal01-1}
{Chen} A.,  {O'Donoghue} D.,  {Stobie} R.~S.,  {Kilkenny} D.,    {Warner} B.,
  2001, MNRAS, 325, 89

\bibitem[\protect\citeauthoryear{{Dall} \& {Schmidtobreick}}{{Dall} \&
  {Schmidtobreick}}{2004}]{dall+schmidtobreick04-1}
{Dall} T.~H.,  {Schmidtobreick} L.,  2004, Inf. Bull. Variable Stars, 5567, 1

\bibitem[\protect\citeauthoryear{{della Valle} \& {Smette}}{{della Valle} \&
  {Smette}}{1992}]{dellavalle+smette92-1}
{della Valle} M.,  {Smette} A.,  1992, IAU Circ., 5529, 1

\bibitem[\protect\citeauthoryear{{Dhillon}, {Marsh} \& {Jones}}{{Dhillon}
  et~al.}{1991}]{dhillonetal91-1}
{Dhillon} V.~S.,  {Marsh} T.~R.,    {Jones} D. H.~P.,  1991, MNRAS, 252, 342

\bibitem[\protect\citeauthoryear{{Downes}, {Webbink}, {Shara}, {Ritter}, {Kolb}
  \& {Duerbeck}}{{Downes} et~al.}{2005}]{downesetal05-1}
{Downes} R.~A.,  {Webbink} R.~F.,  {Shara} M.~M.,  {Ritter} H.,  {Kolb} U.,
  {Duerbeck} H.~W.,  2005, VizieR Online Data Catalog, 5123

\bibitem[\protect\citeauthoryear{{Drew}}{{Drew}}{1997}]{drew97-1}
{Drew} J.~E.,  1997, in {Wickramasinghe} D.~T.,  {Ferrario} F.,   {Bicknell}
  G.,  eds, ASP Conf. Ser. Vol. 121, Accretion Phenomena and Related Outflows. Astron. Soc. Pac., San Francisco, p.~465

\bibitem[\protect\citeauthoryear{{Eggleton}}{{Eggleton}}{1983}]{eggleton83-1}
{Eggleton} P.~P.,  1983, ApJ, 268, 368

\bibitem[\protect\citeauthoryear{{Ferguson}, {Green} \& {Liebert}}{{Ferguson}
  et~al.}{1984}]{fergusonetal84-1}
{Ferguson} D.~H.,  {Green} R.~F.,    {Liebert} J.,  1984, ApJ, 287, 320

\bibitem[\protect\citeauthoryear{{G\"ansicke} \& {Koester}}{{G\"ansicke} \&
  {Koester}}{1999}]{gaensicke+koester99-1}
{G\"ansicke} B.~T.,  {Koester} D.,  1999, A\&A, 346, 151

\bibitem[\protect\citeauthoryear{{Green}, {Schmidt} \& {Liebert}}{{Green}
  et~al.}{1986}]{greenetal86-1}
{Green} R.~F.,  {Schmidt} M.,    {Liebert} J.,  1986, ApJS, 61, 305

\bibitem[\protect\citeauthoryear{{Haefner} \& {Schoembs}}{{Haefner} \&
  {Schoembs}}{1987}]{haefner+schoembs87-1}
{Haefner} R.,  {Schoembs} R.,  1987, MNRAS, 224, 231

\bibitem[\protect\citeauthoryear{{Hameury} \& {Lasota}}{{Hameury} \&
  {Lasota}}{2002}]{hameury+lasota02-1}
{Hameury} J.~M.,  {Lasota} J.~P.,  2002, A\&A, 394, 231

\bibitem[\protect\citeauthoryear{{Haro} \& {Luyten}}{{Haro} \&
  {Luyten}}{1962}]{haro+luyten62-1}
{Haro} G.,  {Luyten} W.~J.,  1962, Bolet\'\i n de los Observatorios
  Tonantzintla y Tacubaya, 3, 37

\bibitem[\protect\citeauthoryear{{Hellier}}{{Hellier}}{1996}]{hellier96-1}
{Hellier} C.,  1996, ApJ, 471, 949

\bibitem[\protect\citeauthoryear{{Hellier}}{{Hellier}}{2000}]{hellier00-1}
{Hellier} C.,  2000, New Astronomy Reviews, 44, 131

\bibitem[\protect\citeauthoryear{{Hellier} \& {Robinson}}{{Hellier} \&
  {Robinson}}{1994}]{hellier+robinson94-1}
{Hellier} C.,  {Robinson} E.~L.,  1994, ApJ, 431, L107

\bibitem[\protect\citeauthoryear{{Hillwig}, {Robertson} \&
  {Honeycutt}}{{Hillwig} et~al.}{1998}]{hillwigetal98-1}
{Hillwig} T.~C.,  {Robertson} J.~W.,    {Honeycutt} R.~K.,  1998, AJ, 115, 2044

\bibitem[\protect\citeauthoryear{{Hoard} \& {Szkody}}{{Hoard} \&
  {Szkody}}{1997}]{hoard+szkody97-1}
{Hoard} D.~W.,  {Szkody} P.,  1997, ApJ, 481, 433

\bibitem[\protect\citeauthoryear{{Hoard}, {Thorstensen} \& {Szkody}}{{Hoard}
  et~al.}{2000}]{hoardetal00-1}
{Hoard} D.~W.,  {Thorstensen} J.~R.,    {Szkody} P.,  2000, ApJ, 537, 936

\bibitem[\protect\citeauthoryear{{Honeycutt} \& {Kafka}}{{Honeycutt} \&
  {Kafka}}{2004}]{honeycutt+kafka04-1}
{Honeycutt} R.~K.,  {Kafka} S.,  2004, AJ, 128, 1279

\bibitem[\protect\citeauthoryear{{Honeycutt}, {Schlegel} \&
  {Kaitchuck}}{{Honeycutt} et~al.}{1986}]{honeycuttetal86-1}
{Honeycutt} R.~K.,  {Schlegel} E.~M.,    {Kaitchuck} R.~H.,  1986, ApJ, 302,
  388

\bibitem[\protect\citeauthoryear{{Horne}}{{Horne}}{1986}]{horne86-1}
{Horne} K.,  1986, PASP, 98, 609

\bibitem[\protect\citeauthoryear{{Horne}}{{Horne}}{1999}]{horne99-1}
{Horne} K.,  1999, in {Hellier} C.,  {Mukai} K.,  eds, ASP Conf. Ser. Vol. 157, Annapolis Workshop on Magnetic Cataclysmic Variables. Astron. Soc. Pac., San Francisco, p.~349

\bibitem[\protect\citeauthoryear{{Hunger}, {Heber} \& {Koester}}{{Hunger}
  et~al.}{1985}]{hungeretal85-1}
{Hunger} K.,  {Heber} U.,    {Koester} D.,  1985, A\&A, 149, L4

\bibitem[\protect\citeauthoryear{{Kafka} \& {Honeycutt}}{{Kafka} \&
  {Honeycutt}}{2004}]{kafka+honeycutt04-2}
{Kafka} S.,  {Honeycutt} R.~K.,  2004, AJ, 128, 2420

\bibitem[\protect\citeauthoryear{{Kafka} \& {Honeycutt}}{{Kafka} \&
  {Honeycutt}}{2005}]{kafka+honeycutt05-1}
{Kafka} S.,  {Honeycutt} R.~K.,  2005, Inf. Bull. Variable Stars, 5597, 1

\bibitem[\protect\citeauthoryear{{Kato} \& {Uemura}}{{Kato} \&
  {Uemura}}{1999}]{kato+uemura99-3}
{Kato} T.,  {Uemura} M.,  1999, Inf. Bull. Variable Stars, 4786, 1

\bibitem[\protect\citeauthoryear{{Kilkenny}, {O'Donoghue}, {Koen}, {Stobie} \&
  {Chen}}{{Kilkenny} et~al.}{1997}]{kilkennyetal97-1}
{Kilkenny} D.,  {O'Donoghue} D.,  {Koen} C.,  {Stobie} R.~S.,    {Chen} A.,
  1997, MNRAS, 287, 867

\bibitem[\protect\citeauthoryear{{Knigge}}{{Knigge}}{2006}]{knigge06-1}
{Knigge} C.,  2006, MNRAS, in press (astro-ph/0609671)

\bibitem[\protect\citeauthoryear{{Knigge}, {Long}, {Hoard}, {Szkody} \&
  {Dhillon}}{{Knigge} et~al.}{2000}]{kniggeetal00-1}
{Knigge} C.,  {Long} K.~S.,  {Hoard} D.~W.,  {Szkody} P.,    {Dhillon} V.~S.,
  2000, ApJ, 539, L49

\bibitem[\protect\citeauthoryear{{Kube}, {G{\"a}nsicke}, {Euchner} \&
  {Hoffmann}}{{Kube} et~al.}{2003}]{kubeetal03-1}
{Kube} J.,  {G{\"a}nsicke} B.~T.,  {Euchner} F.,    {Hoffmann} B.,  2003, A\&A,
  404, 1159

\bibitem[\protect\citeauthoryear{{Marsh} \& {Duck}}{{Marsh} \&
  {Duck}}{1996}]{marsh+duck96-2}
{Marsh} T.~R.,  {Duck} S.~R.,  1996, New Astronomy, 1, 97

\bibitem[\protect\citeauthoryear{{Marsh} \& {Horne}}{{Marsh} \&
  {Horne}}{1988}]{marsh+horne88-1}
{Marsh} T.~R.,  {Horne} K.,  1988, MNRAS, 235, 269

\bibitem[\protect\citeauthoryear{{Mart{\'\i}nez-Pais}, {de la Cruz
  Rodr{\'\i}guez} \& {Rodr{\'\i}guez-Gil}}{{Mart{\'\i}nez-Pais}
  et~al.}{2007}]{martinez-paisetal07-1}
{Mart{\'\i}nez-Pais} I.~G.,  {de la Cruz Rodr{\'\i}guez} J.,
  {Rodr{\'\i}guez-Gil} P.,  2007, MNRAS, submitted

\bibitem[\protect\citeauthoryear{{Misselt} \& {Shafter}}{{Misselt} \&
  {Shafter}}{1995}]{misselt+shafter95-1}
{Misselt} K.~A.,  {Shafter} A.~W.,  1995, AJ, 109, 1757

\bibitem[\protect\citeauthoryear{{Mouchet}, {Siess}, {Drew}, {Lasota},
  {Buckley} \& {Bonnet-Bidaud}}{{Mouchet} et~al.}{1996}]{mouchetetal96-1}
{Mouchet} M.,  {Siess} L.,  {Drew} J.,  {Lasota} J.~P.,  {Buckley} D.~A.~H.,
  {Bonnet-Bidaud} J.~M.,  1996, A\&A, 306, 212

\bibitem[\protect\citeauthoryear{{Mu{\~n}oz-Darias}, {Casares} \&
  {Mart{\'{\i}}nez-Pais}}{{Mu{\~n}oz-Darias}
  et~al.}{2005}]{munoz-dariasetal05-1}
{Mu{\~n}oz-Darias} T.,  {Casares} J.,    {Mart{\'{\i}}nez-Pais} I.~G.,  2005,
  ApJ, 635, 502

\bibitem[\protect\citeauthoryear{{Munari} \& {Zwitter}}{{Munari} \&
  {Zwitter}}{1998}]{munari+zwitter98-1}
{Munari} U.,  {Zwitter} T.,  1998, A\&AS, 128, 277

\bibitem[\protect\citeauthoryear{{Papadaki}, {Boffin}, {Sterken}, {Stanishev},
  {Cuypers}, {Boumis}, {Akras} \& {Alikakos}}{{Papadaki}
  et~al.}{2006}]{papadakietal06-1}
{Papadaki} C.,  {Boffin} H. M.~J.,  {Sterken} C.,  {Stanishev} V.,  {Cuypers}
  J.,  {Boumis} P.,  {Akras} S.,    {Alikakos} J.,  2006, A\&A, 456, 599

\bibitem[\protect\citeauthoryear{{Patterson}}{{Patterson}}{1995}]{patterson95-%
1}
{Patterson} J.,  1995, PASP, 107, 657

\bibitem[\protect\citeauthoryear{{Patterson}, {Fenton}, {Thorstensen},
  {Harvey}, {Skillman}, {Fried}, {Monard}, {O'Donoghue}, {Beshore}, {Martin},
  {Niarchos}, {Vanmunster}, {Foote}, {Bolt}, {Rea}, {Cook}, {Butterworth} \&
  {Wood}}{{Patterson} et~al.}{2002}]{pattersonetal02-1}
{Patterson} J., et al.,  2002, PASP, 114, 1364

\bibitem[\protect\citeauthoryear{{Patterson}, {Kemp}, {Harvey}, {Fried}, {Rea},
  {Monard}, {Cook}, {Skillman}, {Vanmunster}, {Bolt}, {Armstrong}, {McCormick},
  {Krajci}, {Jensen}, {Gunn}, {Butterworth}, {Foote} \& {Bos}}{{Patterson}
  et~al.}{2005}]{pattersonetal05-3}
{Patterson} J., et al.,  2005, PASP, 117, 1204

\bibitem[\protect\citeauthoryear{{Podsiadlowski}, {Han} \&
  {Rappaport}}{{Podsiadlowski} et~al.}{2003}]{podsiadlowskietal03-1}
{Podsiadlowski} P.,  {Han} Z.,    {Rappaport} S.,  2003, MNRAS, 340, 1214

\bibitem[\protect\citeauthoryear{{Prinja}, {Drew} \& {Rosen}}{{Prinja}
  et~al.}{1992}]{prinjaetal92-1}
{Prinja} R.~K.,  {Drew} J.~E.,    {Rosen} S.~R.,  1992, MNRAS, 256, 219

\bibitem[\protect\citeauthoryear{{Ringwald}}{{Ringwald}}{1993}]{ringwald93-2}
{Ringwald} F.~A.,  1993, PhD thesis, Dartmouth College

\bibitem[\protect\citeauthoryear{{Ritter} \& {Kolb}}{{Ritter} \&
  {Kolb}}{2003}]{ritter+kolb03-1}
{Ritter} H.,  {Kolb} U.,  2003, A\&A, 404, 301

\bibitem[\protect\citeauthoryear{{Rodr{\'{\i}}guez-Gil}}{{Rodr{\'{\i}}guez-Gil%
}}{2005}]{rodriguez-gil05-1}
{Rodr{\'{\i}}guez-Gil} P.,  2005, in {Hameury} J.-M.,  {Lasota} J.-P.,  eds,  ASP Conf. Ser. Vol. 330, The Astrophysics of Cataclysmic Variables and Related Objects. Astron. Soc. Pac., San Francisco,  p.~335

\bibitem[\protect\citeauthoryear{{Rodr{\'{\i}}guez-Gil}, {Casares},
  {Mart{\'\i}nez-Pais}, {Hakala} \& {Steeghs}}{{Rodr{\'{\i}}guez-Gil}
  et~al.}{2001}]{rodriguez-giletal01-1}
{Rodr{\'{\i}}guez-Gil} P.,  {Casares} J.,  {Mart{\'\i}nez-Pais} I.~G.,
  {Hakala} P.,    {Steeghs} D.,  2001, ApJ, 548, L49

\bibitem[\protect\citeauthoryear{{Rodr{\'{\i}}guez-Gil}, {G{\" a}nsicke},
  {Araujo-Betancor} \& {Casares}}{{Rodr{\'{\i}}guez-Gil}
  et~al.}{2004}]{rodriguez-giletal04-1}
{Rodr{\'{\i}}guez-Gil} P.,  {G{\" a}nsicke} B.~T.,  {Araujo-Betancor} S.,
  {Casares} J.,  2004, MNRAS, 349, 367

\bibitem[\protect\citeauthoryear{{Rodr{\'{\i}}guez-Gil}, {G{\" a}nsicke},
  {Barwig}, {Hagen} \& {Engels}}{{Rodr{\'{\i}}guez-Gil}
  et~al.}{2004}]{rodriguez-giletal04-2}
{Rodr{\'{\i}}guez-Gil} P.,  {G{\" a}nsicke} B.~T.,  {Barwig} H.,  {Hagen}
  H.-J.,    {Engels} D.,  2004, A\&A, 424, 647

\bibitem[\protect\citeauthoryear{{Rodr{\'{\i}}guez-Gil}, {G{\"a}nsicke},
  {Hagen}, {Nogami}, {Torres}, {Lehto}, {Aungwerojwit}, {Littlefair},
  {Araujo-Betancor} \& {Engels}}{{Rodr{\'{\i}}guez-Gil}
  et~al.}{2005}]{rodriguez-giletal05-2}
{Rodr{\'{\i}}guez-Gil} P., et al.,  2005, A\&A, 440, 701

\bibitem[\protect\citeauthoryear{{Rodr{\'{\i}}guez-Gil} \&
  {Mart{\'{\i}}nez-Pais}}{{Rodr{\'{\i}}guez-Gil} \&
  {Mart{\'{\i}}nez-Pais}}{2002}]{rodriguez-gil+martinez-pais02-1}
{Rodr{\'{\i}}guez-Gil} P.,  {Mart{\'{\i}}nez-Pais} I.~G.,  2002, MNRAS, 337,
  209

\bibitem[\protect\citeauthoryear{{Rodr{\'{\i}}guez-Gil},
  {Mart{\'{\i}}nez-Pais}, {Casares}, {Villada} \& {van
  Zyl}}{{Rodr{\'{\i}}guez-Gil} et~al.}{2001}]{rodriguez-giletal01-2}
{Rodr{\'{\i}}guez-Gil} P.,  {Mart{\'{\i}}nez-Pais} I.~G.,  {Casares} J.,
  {Villada} M.,    {van Zyl} L.,  2001, MNRAS, 328, 903

\bibitem[\protect\citeauthoryear{{Rodr{\'{\i}}guez-Gil et
  al.}}{{Rodr{\'{\i}}guez-Gil et al.}}{2007}]{rodriguez-giletal07-1}
{Rodr{\'{\i}}guez-Gil P et al.},  2007, MNRAS, submitted

\bibitem[\protect\citeauthoryear{{Scargle}}{{Scargle}}{1982}]{scargle82-1}
{Scargle} J.~D.,  1982, ApJ, 263, 835

\bibitem[\protect\citeauthoryear{{Schmidtke}, {Ciudin}, {Indlekofer},
  {Johnson}, {Fried} \& {Honeycutt}}{{Schmidtke}
  et~al.}{2002}]{schmidtkeetal02-1}
{Schmidtke} P.~C.,  {Ciudin} G.~A.,  {Indlekofer} U.~R.,  {Johnson} D.~R.,
  {Fried} R.~E.,    {Honeycutt} R.~K.,  2002, in {G\"ansicke} B.~T.,
  {Beuermann} K.,   {Reinsch} K.,  eds,  ASP Conf. Ser. Vol. 261, The Physics of Cataclysmic Variables and Related Objects. Astron. Soc. Pac., San Francisco,  p.~539

\bibitem[\protect\citeauthoryear{{Schneider} \& {Young}}{{Schneider} \&
  {Young}}{1980}]{schneider+young80-2}
{Schneider} D.~P.,  {Young} P.,  1980, ApJ, 238, 946

\bibitem[\protect\citeauthoryear{{Sefako}, {Glass}, {Kilkenny}, {de Jager},
  {Stobie}, {O'Donoghue} \& {Koen}}{{Sefako} et~al.}{1999}]{sefakoetal99-1}
{Sefako} R.~R.,  {Glass} I.~S.,  {Kilkenny} D.,  {de Jager} O.~C.,  {Stobie}
  R.~S.,  {O'Donoghue} D.,    {Koen} C.,  1999, MNRAS, 309, 1043

\bibitem[\protect\citeauthoryear{{Shafter}}{{Shafter}}{1985}]{shafter85-1}
{Shafter} A.~W.,  1985, AJ, 90, 643

\bibitem[\protect\citeauthoryear{{Shugarov}, {Katysheva}, {Seregina} \&
  {Volkov}}{{Shugarov} et~al.}{2005}]{shugarovetal05-1}
{Shugarov} S.~Y.,  {Katysheva} N.~A.,  {Seregina} T.~M.,    {Volkov} I.~M.,
  2005, in {Hameury} J.-M.,  {Lasota} J.-P.,  eds,  ASP Conf. Ser. Vol. 330, The Astrophysics of Cataclysmic Variables and Related Objects. Astron. Soc. Pac., San Francisco,  p.~495

\bibitem[\protect\citeauthoryear{{Smith}, {Dhillon} \& {Marsh}}{{Smith}
  et~al.}{1998}]{smithetal98-1}
{Smith} D.~A.,  {Dhillon} V.~S.,    {Marsh} T.~R.,  1998, MNRAS, 296, 465

\bibitem[\protect\citeauthoryear{{Sokolov}, {Shugarov} \& {Pavlenko}}{{Sokolov}
  et~al.}{1996}]{sokolovetal96-1}
{Sokolov} D.~A.,  {Shugarov} S.~Y.,    {Pavlenko} E.~P.,  1996, in {Evans} A.,
  {Wood} J.~H.,  eds, Proc. IAU Coll. 158, Cataclysmic Variables and Related Objects. Kluwer Academic Publishers, Dordrecht, p.~219

\bibitem[\protect\citeauthoryear{{Stanishev}, {Kraicheva}, {Boffin} \&
  {Genkov}}{{Stanishev} et~al.}{2002}]{stanishevetal02-1}
{Stanishev} V.,  {Kraicheva} Z.,  {Boffin} H.~M.~J.,    {Genkov} V.,  2002,
  A\&A, 394, 625

\bibitem[\protect\citeauthoryear{{Steeghs}}{{Steeghs}}{2003}]{steeghs03-1}
{Steeghs} D.,  2003, MNRAS, 344, 448

\bibitem[\protect\citeauthoryear{{Szkody} \& {Pich\'e}}{{Szkody} \&
  {Pich\'e}}{1990}]{szkody+piche90-1}
{Szkody} P.,  {Pich\'e} F.,  1990, ApJ, 361, 235

\bibitem[\protect\citeauthoryear{{Thorstensen}, {Davis} \&
  {Ringwald}}{{Thorstensen} et~al.}{1991}]{thorstensenetal91-2}
{Thorstensen} J.~R.,  {Davis} M.~K.,    {Ringwald} F.~A.,  1991, AJ, 102, 683

\bibitem[\protect\citeauthoryear{{Thorstensen}, {Ringwald}, {Wade}, {Schmidt}
  \& {Norsworthy}}{{Thorstensen} et~al.}{1991}]{thorstensenetal91-1}
{Thorstensen} J.~R.,  {Ringwald} F.~A.,  {Wade} R.~A.,  {Schmidt} G.~D.,
  {Norsworthy} J.~E.,  1991, AJ, 102, 272

\bibitem[\protect\citeauthoryear{{Thorstensen} \& {Taylor}}{{Thorstensen} \&
  {Taylor}}{2000}]{thorstensen+taylor00-1}
{Thorstensen} J.~R.,  {Taylor} C.,  2000, MNRAS, 312, 629

\bibitem[\protect\citeauthoryear{{Thorstensen} \& {Taylor}}{{Thorstensen} \&
  {Taylor}}{2001}]{thorstensen+taylor01-1}
{Thorstensen} J.~R.,  {Taylor} C.~J.,  2001, MNRAS, 326, 1235

\bibitem[\protect\citeauthoryear{{Warner}}{{Warner}}{2004}]{warner04-1}
{Warner} B.,  2004, PASP, 116, 115

\bibitem[\protect\citeauthoryear{{Warner} \& {Wickramasinghe}}{{Warner} \&
  {Wickramasinghe}}{1991}]{warner+wickramasinghe91-1}
{Warner} B.,  {Wickramasinghe} D.~T.,  1991, MNRAS, 248, 370

\bibitem[\protect\citeauthoryear{{Watanabe}}{{Watanabe}}{1999}]{watanabe99-1}
{Watanabe} T.,  1999, VSOLJ Variable Star Bull., 34, 3

\bibitem[\protect\citeauthoryear{{Witherick}, {Prinja}, {Howell} \&
  {Wagner}}{{Witherick} et~al.}{2003}]{withericketal03-1}
{Witherick} D.~K.,  {Prinja} R.~K.,  {Howell} S.~B.,    {Wagner} R.~M.,  2003,
  MNRAS, 346, 861

\bibitem[\protect\citeauthoryear{{Woudt} \& {Warner}}{{Woudt} \&
  {Warner}}{2003}]{woudt+warner03-1}
{Woudt} P.~A.,  {Warner} B.,  2003, MNRAS, 340, 1011

\bibitem[\protect\citeauthoryear{{Wu} \& {Wickramasinghe}}{{Wu} \&
  {Wickramasinghe}}{1991}]{wu+wickramasinghe91-1}
{Wu} K.,  {Wickramasinghe} D.~T.,  1991, MNRAS, 252, 386

\bibitem[\protect\citeauthoryear{{Young}, {Schneider} \& {Shectman}}{{Young}
  et~al.}{1981}]{youngetal81-3}
{Young} P.,  {Schneider} D.~P.,    {Shectman} S.~A.,  1981, ApJ, 245, 1035

\bibitem[\protect\citeauthoryear{{Zwitter} \& {Munari}}{{Zwitter} \&
  {Munari}}{1995}]{zwitter+munari95-1}
{Zwitter} T.,  {Munari} U.,  1995, A\&AS, 114, 575

\end{thebibliography}


\end{document}